\documentclass[a4paper, 11pt]{article}
\pdfoutput=1

\usepackage{jheppub}
\usepackage{subcaption}
\usepackage{graphicx}
\usepackage{xcolor}
\usepackage{amsmath,amssymb}
\usepackage{mathtools}

\title{First Saturation Correction in High Energy Proton-Nucleus
Collisions: II. Single Inclusive Semi-Hard Gluon Production}
\author[a]{Ming Li}
\affiliation[a]{Department of Physics, North Carolina State University, Raleigh, NC 27695, USA}
\author[a,b]{and Vladimir V. Skokov}
\affiliation[b]{RIKEN/BNL Research Center, Brookhaven National Laboratory, Upton, NY 11973}

\emailAdd{mli48@ncsu.edu}
\emailAdd{vskokov@ncsu.edu}

\abstract{
Exploiting recently obtained analytic solutions of classical Yang-Mills equations 
for higher order perturbations in the field of the dilute object (proton),
we derive the complete first saturation correction to the single inclusive semi-hard gluon production
in high energy {proton-nucleus} collisions by applying the Lehmann-Symanzik-Zimmermann reduction
formula.
We thus finalize the program started by Balitsky (see Ref.~\cite{Balitsky:2004rr}) and  independently by Chirilli, Kovchegov and Wertepny (see Ref.~\cite{Chirilli:2015tea})
albeit using a very different approach to carry out our calculations.
We extracted the functional dependence of gluon spectrum on the color charge densities of the colliding objects; thus our results can be used to evaluate  complete first saturation correction to the double/multiple inclusive gluon productions.
}

\begin{document}
\maketitle
\flushbottom

\section{Introduction}
In Ref.~\cite{Li:2021zmf}, in the dilute-dense regime of the Color Glass Condensate (CGC) effective theory,
we analytically solved the classical Yang-Mills equations  to  next-to-leading and  next-to-next-to-leading orders in the field of the dilute object (the proton in proton-nucleus collisions).
In this and follow-up papers, we turn our attention to computing observables; in particular, this paper is dedicated to using these solutions to derive a complete analytical expression
for the first saturation correction to the single inclusive gluon  production. This problem was previously considered in Refs.~\cite{Balitsky:2004rr, Chirilli:2015tea}
for the single inclusive gluon production and in Refs.~\cite{McLerran:2016snu,Kovchegov:2018jun} for the double inclusive gluon production. However, these studies provided only incomplete results. Specifically, only order-$g^3$ single gluon production amplitude was computed in Ref.~\cite{Chirilli:2015tea}, while the order-$g^5$ production amplitude has never been computed in previous publications. 
In this paper, we give the complete first saturation correction to the single gluon production  for the first time.

In the CGC framework, the leading contribution to particle production is due to classical gluon field. The quantum corrections, albeit important for certain aspects \cite{Romatschke:2005pm, Gelis:2013rba, Berges:2013eia},
are considered to be negligible in the first approximation. We follow this paradigm and work explicitly in the classical regime.
It was shown that the classical approximation resums the powers of the saturation momenta of the projectile and the target \cite{Kovchegov:1997pc}.
The saturation momentum $Q_s$ is proportional to $\alpha_s \mu$, where $\mu^2$ is the hadron color charge squared per unit transverse area \cite{Lappi:2007ku}.
Therefore, the single inclusive gluon productions in high energy nuclear collisions can be expressed as a
function of two dimensionless parameters  $Q_{s,P}^2/k_{\perp}^2$
and $Q_{s,T}^2/k_{\perp}^2$, involving the saturation scales of the projectile and the target respectively. In the notations of Ref.~\cite{Chirilli:2015tea},
\begin{equation}
\frac{dN}{d^2\mathbf{k} dy}=\frac{1}{\alpha_s}  f\left(\frac{Q_{s,P}^2}{k_{\perp}^2}, \frac{Q_{s,T}^2}{k_{\perp}^2}\right) \,.
\end{equation}
Here the overall factor of $\alpha_s^{-1}$ manifests that we work in the classical approximation.
The function $f$ describes the most general situation of  nucleus-nucleus scattering and is not known analytically. The series expansion of the function $f$ in one of the arguments is termed
the dilute-dense expansion. It is customary to consider the projectile as a dilute object and perform the expansion in powers of $Q_{s,P}^2/k_{\perp}^2$:
\begin{equation} \label{eq:ExpansionQs}
f\left(\frac{Q_{s,P}^2}{k_{\perp}^2}, \frac{Q_{s,T}^2}{k_{\perp}^2}\right) =
	 \sum_{n=1}^\infty  \left(\frac{Q_{s,P}^2}{k_{\perp}^2}\right)^n f_{n} \left(\frac{Q_{s,T}^2}{k_{\perp}^2}\right)\,.
\end{equation}
The coefficients $f_n$ are analytically computable and can be non-perturbatively evaluated to all order in  $Q_{s,T}^2/k_{\perp}^2$, at least in principle.
In practice, however, only the leading term of this expansion was computed, the so-called dilute-dense approximation, see Refs.~\cite{Kovchegov:1998bi,Dumitru:2001ux}:
\begin{equation}
\frac{dN}{d^2\mathbf{k} dy} \approx \frac{1}{\alpha_s}  \frac{Q_{s,P}^2}{k_{\perp}^2}   f_1\left(\frac{Q_{s,T}^2}{k_{\perp}^2}\right) \,.
\end{equation}
This truncation is justified as long as one is interested in the momentum region $k_{\perp} \gg Q_{s,P}$ of the gluon spectrum. As $k_{\perp}$ approaches $Q_{s,P}$, higher order corrections have to be taken into account. For $k_{\perp}\leq Q_{s,P}$, an all-order resummation is needed (see e.g. Refs.~\cite{Chen:2015wia, Li:2016eqr}, and numerical simulations of Refs.~\cite{Krasnitz:1999wc,Krasnitz:2000gz,Schenke:2015aqa, Schlichting:2019bvy}). We are mainly interested in the production of semi-hard gluons with transverse momentum $k_{\perp} >Q_{s,P}$. Thus our goal is to compute the next-to-leading correction  $f_{2}$ in the expansion~\eqref{eq:ExpansionQs};
the partial result for $f_{2}$ was presented in Ref.~\cite{Chirilli:2015tea} where the contributions with amplitudes of order ${\cal O} (g^5)$ were not included. In Ref.~\cite{Chirilli:2015tea}, a diagrammatic approach based on the light-cone perturbation theory was used.
In this paper,  to extract single inclusive gluon production, we instead use the solutions of classical Yang-Mills equations derived in Ref.~\cite{Li:2021zmf}. To this end we apply
properly modified  Lehmann-Symanzik-Zimmermann formula for Milne metric.

Our approach, besides the computational transparency, also allows us to go beyond configuration-averaged quantities;
that is, we can perform ``even-by-event'' studies by evaluating particle production for a fixed distribution of the
valence charge densities  in both the projectile $\rho_P$  and the target   $\rho_T$. We aim to provide the functional
$\frac{dN}{d^2\mathbf{k} dy} \left[ \rho_P, \rho_T \right]$.

The functional form of our result enables to study fluctuations of multiplicity; moreover having the functional form   is important for two additional  reasons.
To appreciate the first reason,   consider the configuration-averaged $\frac{dN}{d^2\mathbf{k} dy}$; it is often obtained from
$\frac{dN}{d^2\mathbf{k} dy} \left[ \rho_P, \rho_T \right]$
by using the Gaussian McLerran-Venugopalan model \cite{McLerran:1993ni,McLerran:1993ka}. This may miss some interesting effects potentially relevant for the observables.
Indeed, we will demonstrate  that, in general, the functional $\frac{dN}{d^2\mathbf{k} dy} \left[ \rho_P, \rho_T \right]$ is not an even function of $\rho_P$, e.g.  it contains  a product of three color charge densities   $\rho_P(\mathbf{x})
\rho_P(\mathbf{y})  \rho_P(\mathbf{z})$. The terms with odd powers of $\rho_P$  would vanish  if averaged with a Gaussian distribution.
However, for a proton projectile, a Gaussian model  might not be the most appropriate in certain kinematic regions and a more elaborated approach might include non-trivial color charge correlations \cite{Jeon:2005cf,Dumitru:2020fdh,Dumitru:2021tvw}. For instance, the proton model of Ref.~\cite{Dumitru:2020fdh} includes a nonzero C-odd part of the correlator of three color charges. Our resulting functional $\frac{dN}{d^2\mathbf{k} dy} \left[ \rho_P, \rho_T \right]$ will explicitly keep all even and odd powers of $\rho_P$ to the appropriate order and thus can be easily extended to account for non-zero three color charge correlations.
The second reason is that   the functional form of $\frac{dN}{d^2\mathbf{k} dy} \left[ \rho_P, \rho_T \right]$ allows us to evaluate multiparticle production in a very straightforward way~\cite{Gelis:2008rw,Gelis:2008ad,Gelis:2008sz} -- by finding an average of the product of the functionals, e.g. for two-particle inclusive we simply get
\begin{equation}
 \frac{d^2N}{d^2\mathbf{k}_1 dy_1 d^2\mathbf{k}_2 dy_2} =
\left \langle \frac{dN}{d^2\mathbf{k}_1 dy_1} \left[ \rho_P, \rho_T \right] \frac{dN}{d^2\mathbf{k}_2 dy_2} \left[ \rho_P, \rho_T \right] \right \rangle_{P,T}
\end{equation}
and analogously for $n$-gluon production. Although, the formulae provided in this paper can be easily applied to compute multigluon production; we will restrict this paper to evaluating the full functional including first saturation correction. We defer
configuration-averaged gluon productions  and correlations to a future work.

We also want to comment that our calculation properly includes   interactions between multiple gluons at the same rapidity, e.g. the processes when the gluon from the proton wave function merges with the gluon emitted from the proton after the scattering off the target. The formalism used in Ref.~\cite{Kovner:2010xk,Kovner:2016jfp} (see also reference therein for earlier works) does not account for such interactions.

\begin{figure}[t]
\centering % \begin{center}/\end{center} takes some additional vertical space
\includegraphics[width = 0.75\textwidth]{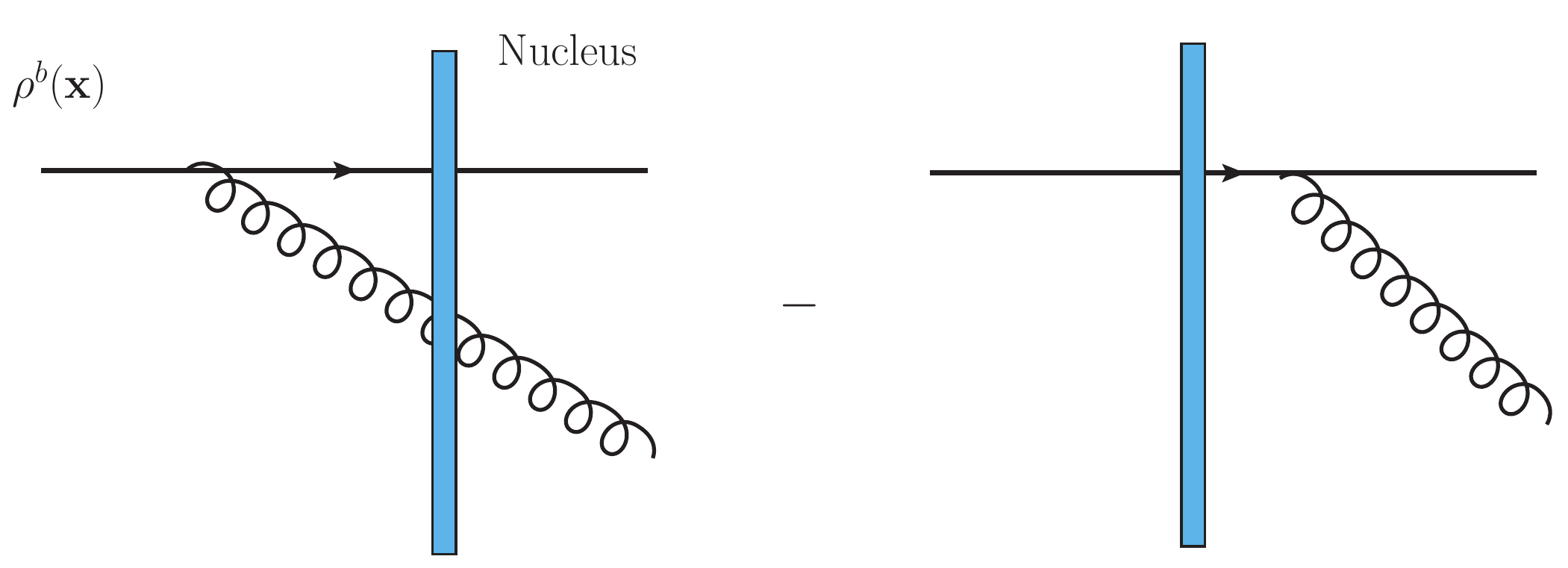}
\caption{Schematic diagrams showing the leading order single gluon production in high energy proton-nucleus collisions. The $\rho^b(\mathbf{x})$ is the valence color charge density of the proton. The colored solid rectangular represents the highly Lorentz contracted nucleus. }
\label{fig:order_g_production}
\end{figure}

Before moving on to the technical calculations, we would like to using Feynman diagrams to illustrate the first saturation corrections to single gluon production in high energy proton-nucleuse collisions.  Fig.~\ref{fig:order_g_production} shows diagram for the leading order single gluon production. The gluon could be emitted either before or after the interaction with the nucleus.  Since there is only one gluon emission vertex and it only involves one color charge density, the amplitude is proportional to $g\rho^b(x)$. It is important to point out that at the leading order each valence color charge independently emits gluons and scatters on the nucleus. They do not interact among themselves. 

\begin{figure}[t]
\centering 
\includegraphics[width = 0.9\textwidth]{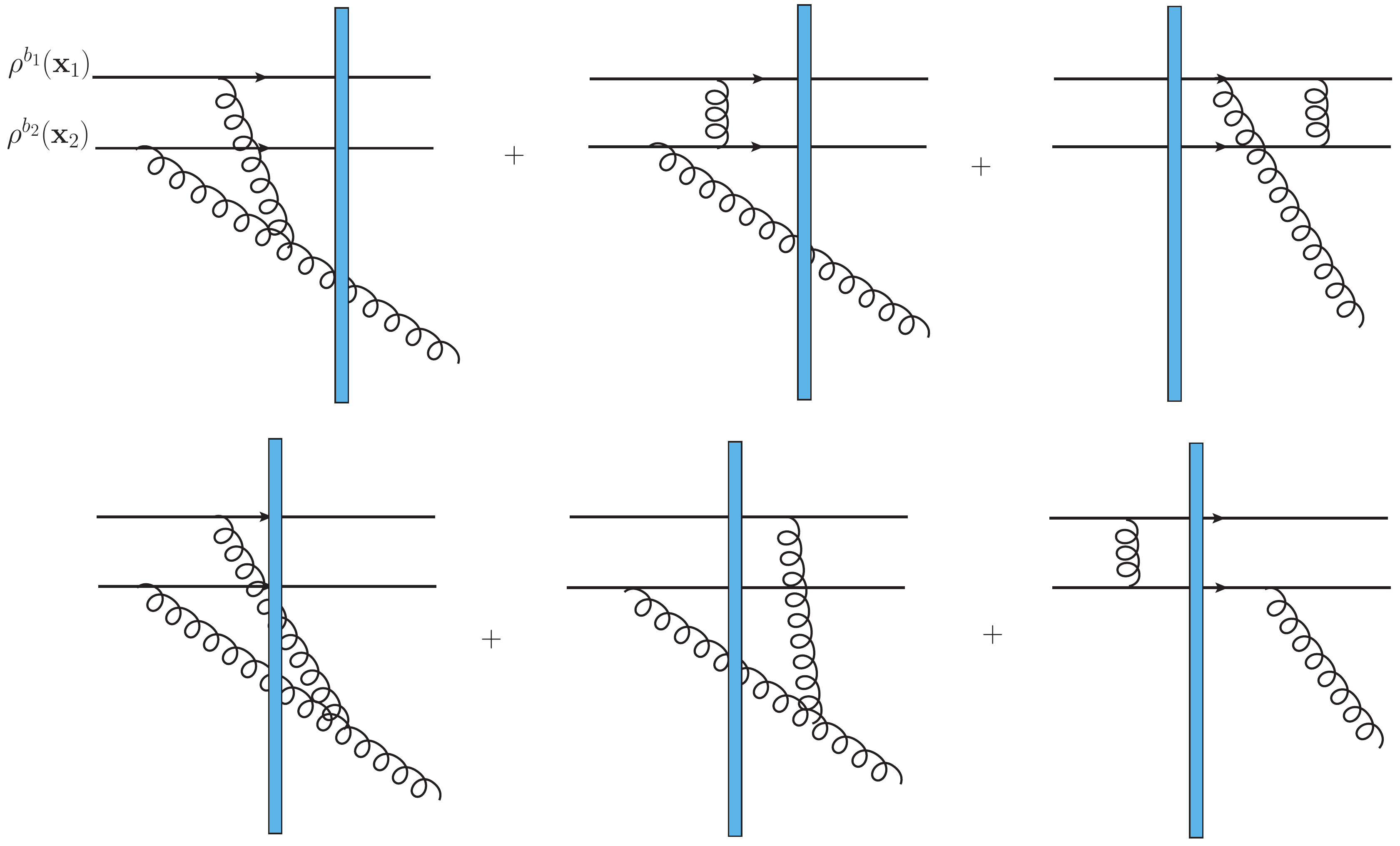}
\caption{Partial diagrams representing the order-$g^3$ single gluon production amplitude. It involves interactions of two color charge densities $\rho^{b_1}(\mathbf{x}_1)$ and $\rho^{b_2}(\mathbf{x}_2)$. }
\label{fig:order_g3_production}
\end{figure}

Fig.\ref{fig:order_g3_production} shows part of the diagrams contributing to the order-$g^3$ gluon production amplitude. This is not meant to be exhaustive and there are other diagrams due to different vertex orderings and different light-cone time flow directions, see Ref.~\cite{Chirilli:2015tea} for the complete set of diagrams.  As can be seen, these diagrams contain three gluon vertices and involve interactions of two color charge densities, thus proportional to $g^3\rho^{b_1}(\mathbf{x}_1)\rho^{b_2}(\mathbf{x}_2)$. Order-$g^3$ amplitude is the first part contributing to the first saturation correction to single gluon production. Not only the valence color charges could interact among themselves either before or after the collisions with the nucleus, the emitted gluons have final state interactions, as shown in the fourth and fifth diagrams in Fig.~\ref{fig:order_g3_production}. Note that  diagrams in Fig.~\ref{fig:order_g3_production} can be drawn from two copies of Fig.~\ref{fig:order_g_production} with the appropriate attachments of gluon lines before or after collisions with the nucleus.
\begin{figure}[t]
\centering 
\includegraphics[width = 0.9\textwidth]{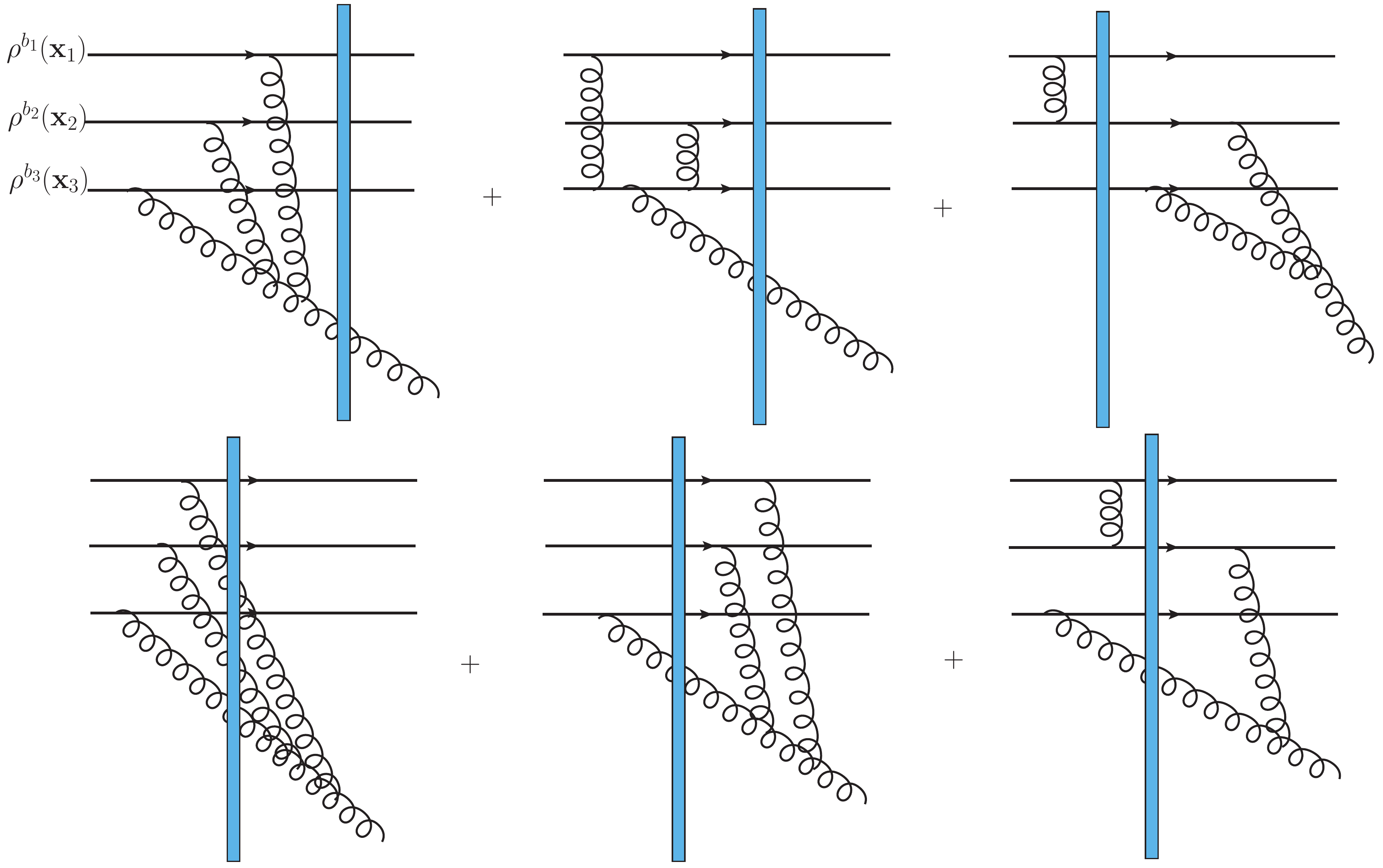}
\caption{Partial diagrams showing the order-$g^5$ single gluon production amplitude. It involves interactions of three color charge densities $\rho^{b_1}(\mathbf{x}_1)$, $\rho^{b_2}(\mathbf{x}_2)$ and $\rho^{b_3}(\mathbf{x}_3)$. }
\label{fig:order_g5_production}
\end{figure}

Fig.~\ref{fig:order_g5_production} are six of the many diagrams contributing to order-$g^5$ gluon production amplitude. These diagrams contain five gluon vertices and they involve interactions of three color charge densities, thus proportinal to $g^5 \rho^{b_1}(\mathbf{x}_1)\rho^{b_2}(\mathbf{x}_2)\rho^{b_3}(\mathbf{x}_3)$. These diagrams could be drawn from three copies of the order-$g$ diagrams in  Fig.~\ref{fig:order_g_production} with the appropriate attachments of gluon lines. 

Both order-$g^3$ and order-$g^5$ amplitudes contribute to the first saturation correction to the single inclusive gluon production. One can symbolically represent the single gluon production amplitude $M =M_{(1)} + M_{(3)} + M_{(5)}+\ldots$. Here the subscript indicates the power of coupling constant $g$. However, it should be pointed out that this is not exactly a perturbative expansion in terms of coupling constant because we only take into account diagrams that are enhanced by the color charge density of the proton at each order. Therefore $M_{(1)} \sim g\rho_P$, $M_{(3)}\sim g^3\rho_P^2$ and $M_{(5)} \sim g^5\rho_P^3$.  The single gluon production is proportional to the amplitude squared
\begin{equation}
|M|^2 = |M_{(1)}|^2 + M_{(1)}M_{(3)}^{\ast} + M_{(1)}^{\ast}M_{(3)} +|M_{(3)}|^2+ M_{(1)}M_{(5)}^{\ast} + M_{(1)}^{\ast}M_{(5)}  +\ldots
\end{equation}
The crossing terms between order-$g$ and order-$g^3$ amplitudes are proportional to odd powers of the color charge density $M_{(1)}M_{(3)}^{\ast} + M_{(1)}^{\ast}M_{(3)} \sim g^4 \rho_P^3$. In McLerran-Venugopalan model  (see also Ref.~\cite{Kovchegov:1996ty}), $\rho_P$ satisfies a Gaussian distribution; therefore $\rho_P^{3}$ terms vanish upon averaging over all possible configurations of color charge density.  The first saturation correction therefore starts from 
\begin{equation}
\frac{dN}{d^2\mathbf{k} dy}\Big|_{FSC} \sim |M_{(3)}(\mathbf{k})|^2+ M_{(1)}(\mathbf{k})M_{(5)}^{\ast} (\mathbf{k})+ M_{(1)}^{\ast}(\mathbf{k})M_{(5)}(\mathbf{k}).
\end{equation}
Both the order-$g^3$ amplitude squared and the crossing terms between order-$g$ and order-$g^5$ terms are proportional to $g^6 \rho_P^4$. In a Gaussian model for the color charge density, one has $\langle g^4\rho_P^2\rangle \sim g^4\mu_P^2  \sim Q_{s,P}^2$ so that the first saturation correction is proportional to $\frac{1}{\alpha_s} Q_{s,P}^4$. This should be compared with the leading order term $|M_{(1)}|^2 \sim \frac{1}{\alpha_s} Q_{s,P}^2$.  This power counting in terms of the saturation scale has already been formally shown in Eq.~\eqref{eq:ExpansionQs}. The goal of the paper is to derive $M_{(3)}$ and $M_{(5)}$ and finish the computation of the first saturation correction to single gluon production (previous studies in the literature only computed $M_{(3)}$). 

The paper is organized as follows. In Sec.~\ref{Sec:Review}, we briefly review the key results of the previous paper~\cite{Li:2021zmf}. In particular, we set up an initial value problem for the classical field in the forward light cone and list the required solutions. In Sec.~\ref{Sect:DefSI}, in order to extract the single inclusive gluon production, we detail the Lehmann-Symanzik-Zimmermann reduction formula in the Milne coordinate system.
We finally turn to genuinely novel material in Secs.~\ref{Sect:CalcAmp} and \ref{Sect:CalcSI}, where for the first time we compute the functional  $\frac{dN}{d^2\mathbf{k} dy} \left[ \rho_P, \rho_T \right]$ including the first saturation correction in the projectile field. We finish with conclusions in Sec.~\ref{Sect:Conclusions}. For the sake of completeness, we provide more technical details in the appendices.

\section{General Setup and Solutions of Classical Yang-Mills Equations }
\label{Sec:Review}

We briefly recapitulate the general setup of our calculations and the results obtained in Ref.~\cite{Li:2021zmf}.
%\subsection{Classical Yang-Mills equations  in the forward light cone}
In the CGC framework, the classical gluon fields produced in high energy proton-nucleus collisions are determined by the classical Yang-Mills equations $
D_{\mu} F^{\mu\nu} = J^{\nu}$. 
Here the color charge current for a right-moving projectile and a left-moving target $J^{\nu} = \delta^{\nu +} \rho_P(x^-, \mathbf{x}) + \delta^{\nu -} \rho_T(x^+, \mathbf{x})$. In the forward light cone $x^+>0$ and  $x^->0$, the color charge current vanishes and the effects of the sources $\rho_P, \rho_T$ are reflected in the initial conditions. 
We adopt the procedure described in Refs.~\cite{Kovner:1995ja, Dumitru:2001ux} and consider
the Fock-Schwinger gauge $x^-A^+ + x^+ A^- =0$. This gauge suggests a convenient parameterization for  the solutions, which are explicitly assumed to be boost invariant, $A^+ = A_- = x^+ \alpha(\tau, \mathbf{x})$, $A^- = A_+ = -x^- \alpha(\tau, \mathbf{x})$,  $A^i = \alpha^i(\tau, \mathbf{x})$ with $\tau= \sqrt{2x^+x^-}$. In terms of $\alpha(\tau, \mathbf{x}), \alpha^i(\tau, \mathbf{x})$, the classical Yang-Mills equations read
\begin{equation}\label{eq:ym_alpha_alphai}
\begin{split}
& \partial_{\tau}^2 \alpha +\frac{3}{\tau} \partial_{\tau} \alpha  - [D_i, [D_i, \alpha]] =0\,, \\
&-ig[ \alpha, \tau\partial_{\tau} \alpha] + [D_i, \frac{1}{\tau} \partial_{\tau} \alpha_i]=0\,, \\
&\frac{1}{\tau}\partial_{\tau} \alpha_i +\partial^2_{\tau}\alpha_i -ig \tau^2 [ \alpha, [D_i, \alpha]-D_jF_{ji}=0 \\
\end{split}
\end{equation}
with the  initial conditions~\cite{Kovner:1995ja, Gyulassy:1997vt}:
\begin{equation}\label{eq:initial_conditions}
\begin{split}
&\alpha(\tau=0,\mathbf{x}) = \frac{ig}{2} [\alpha_P^i(\mathbf{x}), \alpha_T^i(\mathbf{x})],\\
&\alpha^i(\tau=0, \mathbf{x}) = \alpha_P^i(\mathbf{x}) + \alpha_T^i(\mathbf{x}).\\
\end{split}
\end{equation}
The $\alpha_P^i(\mathbf{x})$ and $\alpha_T^i(\mathbf{x})$ are the Weizsacker-Williams (WW) gluon fields of the projectile and target, respectively; they  depend on the transverse coordinate and are related to $\rho_{P(T)}$ through the static Yang-Mills equations $\partial^i \alpha_{P(T)}^i(\mathbf{x}) = g \rho_{P(T)}$. Since each field $\alpha_{P(T)}^i(\mathbf{x})$ is a pure gauge,  it is convenient to use the following parametrization in terms of Wilson lines $V$ for the projectile  and $U$ for the target $\alpha_{P}^i(\mathbf{x}) = \frac{i}{g} V(\mathbf{x})  \partial^i V^\dagger(\mathbf{x})$, $\alpha_{T}^i(\mathbf{x}) = \frac{i}{g} U(\mathbf{x})  \partial^i U^\dagger(\mathbf{x})$. 
In general, the WW field of the projectile (target)  contains all powers in $\rho_P$ ($\rho_T$); for our purpose only a power series expansion in  $\rho_P$ is required:
\begin{equation}
\alpha_P^i(\mathbf{x}) = \sum_{n=0}^{\infty} g^{2n+1} \alpha_{P}^{{(2n+1)},i}(\tau,\mathbf{x})\,.
\end{equation}
The first three expansion coefficients are
\begin{align}
\alpha_P^{(1),i} &= \partial^i \phi, \\
\alpha_P^{(3),i} &= \frac{1}{2} i \left(\delta^{ij} - \frac{\partial^i \partial^j}{\partial^2}\right) [\phi, \partial^j \phi], \\
\alpha_P^{(5),i}
&=\frac{1}{2} \left(\delta^{ij} -\frac{\partial^i\partial^j}{\partial^2}\right)\left ([\frac{1}{\partial^2}[\phi, \partial^2 \phi], \partial^j\phi]-\frac{1}{3}[\phi, [\phi, \partial^j \phi]] \right) ,
\end{align}
where the gluon exchange potential
\begin{equation}
 \phi(\mathbf{x}) = \frac{1}{\partial^2} \rho_P \equiv \frac{1}{2\pi} \int d^2\mathbf{y} \ln{\left(|\mathbf{x}-\mathbf{y}| m \right)} \rho_P(\mathbf{y})\,
\end{equation}
with $m$ being some appropriate IR cutoff. 

The initial value problem can be solved  order-by-order in $\rho_P$  while preserving non-perturbative dependence to all orders in  $\rho_T$. In order to facilitate this, we consider a subgauge transformation defined by the condition $\partial^i \beta^i (\tau=0, \mathbf{x})  =0$ to which we will refer to as the \textit{initial time Coulomb gauge}, see e.g. Ref.~\cite{Blaizot:2010kh, Berges:2013fga}. The fields $\beta$ and $\beta^i$  are non-linearly related to $\alpha$ and $\alpha_i$ through a gauge transformation 
\begin{equation}
\begin{split}
&\beta^i  = \Omega^+ \alpha^i  \Omega  + \frac{i}{g}  \Omega^+  \partial^i  \Omega,\\
&\beta =   \Omega^+ \alpha  \Omega.\\
\end{split}
\end{equation}
The $\Omega = \Omega (\mathbf{x})$ is fixed  to satisfy  $\partial^i \beta^i (\tau=0, \mathbf{x})  =0$. A closed form of  $\Omega$ is not known but a perturbative expansion in coupling constant is obtained in Ref.~\cite{Li:2021zmf}.
The subgauge transformations, not affecting the FS gauge, preserve the form of the Yang-Mills equations and only modify the initial conditions.

Below we will present solutions of the classical Yang-Mills equations as power series expansion in the coupling constant $g$ in the initial time Coulomb subgauge (keeping all power of $g \rho_T$). Due to the structure of the classical Yang-Mills equations, only odd powers of $g$ are non-vanishing, that is
\begin{equation}
	\label{Eq:mom_space_general}
\begin{split}
&\beta(\tau,\mathbf{x}) = \sum_{n=0}^{\infty}g^{2n+1} \beta^{{(2n+1)}}(\tau,\mathbf{x}),\\
&\beta_i(\tau, \mathbf{x}) = \sum_{n=0}^{\infty} g^{2n+1} \beta_i^{{(2n+1)}} (\tau, \mathbf{x}).\\
\end{split}
\end{equation}
The order-$g$ and order-$g^3$ solutions in momentum space are
\begin{equation}\label{eq:sol_order_g}
\begin{split}
&\beta^{(1)}(\tau, \mathbf{k} ) = b_{\eta}(\mathbf{k}) \frac{J_1(k_{\perp}\tau)}{k_{\perp}\tau},\\
&\beta^{(1)}_i (\tau, \mathbf{k}) = \frac{-i\epsilon^{il}\mathbf{k}_l}{k_{\perp}^2} b_{\perp}(\mathbf{k}) J_0(k_{\perp}\tau) \\
\end{split}
\end{equation}
and 
\begin{equation}\label{eq:sol_order_g3_first}
\begin{split}
\beta^{(3)}(\tau,\mathbf{q}) 
=&2\beta^{(3)}(\tau=0,\mathbf{q}) \frac{J_1(q_{\perp}\tau) }{q_{\perp}\tau}+i \int \frac{d^2\mathbf{p}}{(2\pi)} \frac{\mathbf{q}\times \mathbf{p}}{p_{\perp}^2|\mathbf{q}-\mathbf{p}|^2}\Big[b_{\perp}(\mathbf{p}), b_{\eta}(\mathbf{q}-\mathbf{p})\Big] \\
&\qquad\times\int_{-\pi}^{\pi}\frac{d\phi}{2\pi}\frac{\mathbf{q}\cdot(\mathbf{q}-2\mathbf{p})+w_{\perp}^2}{q_{\perp}^2-w_{\perp}^2}\left(\frac{J_1(w_{\perp}\tau)}{w_{\perp}\tau}-\frac{J_1(q_{\perp}\tau)}{q_{\perp}\tau}\right), \\
\end{split}
\end{equation}
\begin{equation}\label{eq:sol_order_g3_second}
\begin{split}
\beta^{(3)}_{\perp}(\tau, \mathbf{q}) = &\beta_{\perp}^{(3)}(\tau=0, \mathbf{q})J_0(q_{\perp}\tau) +\frac{i}{q_{\perp}} \int \frac{d^2\mathbf{p}}{(2\pi)^2}\frac{ \mathbf{q}\times \mathbf{p}}{p^2_{\perp}|\mathbf{q}-\mathbf{p}|^2} \Big(\mathbf{p}\cdot(\mathbf{q}-\mathbf{p})\Big[b_{\eta}(\mathbf{p}), b_{\eta}( \mathbf{q}-\mathbf{p})\Big] \\
&+(\mathbf{p}\cdot\mathbf{q}-p_{\perp}^2-q_{\perp}^2) \Big[b_{\perp}(\mathbf{p}), b_{\perp}(\mathbf{q}-\mathbf{p})\Big]\Big) \int_{-\pi}^{\pi} \frac{d\phi}{2\pi}\frac{1}{w_{\perp}^2-q_{\perp}^2} (J_0(w_{\perp}\tau)-J_0(q_{\perp}\tau))\\
&+\frac{i}{q_{\perp}} \int \frac{d^2\mathbf{p}}{(2\pi)^2}\frac{ \mathbf{q}\times \mathbf{p}}{2p^2_{\perp}|\mathbf{q}-\mathbf{p}|^2} \Big[b_{\eta}(\mathbf{p}), b_{\eta}( \mathbf{q}-\mathbf{p})\Big]\Big(J_0(p_{\perp}\tau)J_0(|\mathbf{q}-\mathbf{p}|\tau) -J_0(q_{\perp}\tau)\Big), 
\end{split}
\end{equation}
\begin{equation}\label{eq:sol_order_g3_third}
\begin{split}
\Lambda^{(3)}(\tau, \mathbf{q}) 
=& - \frac{i}{q_{\perp}^2} \int\frac{d^2\mathbf{p}}{(2\pi)^2}  \frac{\mathbf{q}\cdot(\mathbf{q}-2\mathbf{p})}{4p_{\perp}^2 |\mathbf{q}-\mathbf{p}|^2} \Big(2\mathbf{p}\cdot(\mathbf{q}-\mathbf{p})\left[b_{\perp}(\mathbf{p}), b_{\perp}(\mathbf{q}-\mathbf{p})\right]\\\
&   -(p_{\perp}^2 + |\mathbf{q}-\mathbf{p}|^2 )\left[b_{\eta}(\mathbf{p}), b_{\eta}(\mathbf{q}-\mathbf{p})\right] \Big)\int_{-\pi}^{\pi}\frac{d\phi}{2\pi}\frac{1}{w^2_{\perp}}(1- J_0(w_{\perp}\tau))\\
& - \frac{i}{q_{\perp}^2} \int\frac{d^2\mathbf{p}}{(2\pi)^2}  \frac{\mathbf{q}\cdot(\mathbf{q}-2\mathbf{p})}{4p_{\perp}^2 |\mathbf{q}-\mathbf{p}|^2} \left[b_{\eta}(\mathbf{p}), b_{\eta}(\mathbf{q}-\mathbf{p})\right] \Big(1-J_0(p_{\perp}\tau)J_0(|\mathbf{q}-\mathbf{p}|\tau)\Big).
\end{split}
\end{equation}
Here $w^2_{\perp}=p^2_{\perp} + |\mathbf{q}-\mathbf{p}|^2 -2 p_{\perp}|\mathbf{q}-\mathbf{p}|\cos{\phi}$.
The decompositions of transverse field  in momentum space is
\begin{equation}
\beta_i^{(3)}(\tau,\mathbf{q}) = \frac{-i\epsilon^{ij}\mathbf{q}_j}{q_{\perp}} \beta_{\perp}^{(3)}(\tau, \mathbf{q}) -i\mathbf{q}_i \Lambda^{(3)}(\tau, \mathbf{q}).
\end{equation}

The above solutions are expressed in terms of order-$g$ initial conditions $b_{\eta}(\mathbf{k})$, $b_{\perp}(\mathbf{k})$ and order-$g^3$ initial conditions $\beta^{(3)}(\tau=0, \mathbf{q})$, $\beta_{\perp}^{(3)}(\tau=0, \mathbf{q})$. 
To present expressions for the initial conditions, we introduce an auxiliary quantity which will repeatedly appear in our calculations:
\begin{equation}
\zeta^i(\mathbf{x})=U^{\dagger}(\mathbf{x}) \alpha_{P}^i(\mathbf{x}) U(\mathbf{x})= \alpha_P^{a,i}(\mathbf{x})U^{ac}(\mathbf{x})T^c\,.
\end{equation}
The physical meaning of this combination is clear: it represents the eikonally color rotated projectile gluon field by the target Wilson line. We will also use the expansion of $\zeta^i(\mathbf{x})$ in terms of coupling constant $g$ with the coefficients
\begin{equation}
\zeta^{ i}_{(n)}(\mathbf{x})=U^{\dagger}(\mathbf{x}) \alpha_{P}^{(n),i}(\mathbf{x}) U(\mathbf{x})\,.
\end{equation}
 For example, the explicit expression for $\zeta^i(\mathbf{x})$ at order-$g$ is
\begin{equation}
\zeta^{i}_{(1)} (\mathbf{k})= \int\frac{d^2\mathbf{p}}{(2\pi)^2} \frac{i\mathbf{p}_i}{p_{\perp}^2} \rho_P^a(\mathbf{p}) U^{ac}(\mathbf{k}-\mathbf{p})T^c\,.
\end{equation}
By analogy with $\zeta^i(\mathbf{x})$,  we also introduce  the eikonally rotated projectile color charge density
\begin{equation}
\rho_R(\mathbf{x}) = U^{\dagger} (\mathbf{x}) \partial^i \alpha_{P}^{(1),i} U (\mathbf{x}) = g\, U^{\dagger} (\mathbf{x}) \rho_P(\mathbf{x}) U(\mathbf{x})\,.
\end{equation}
In terms of these auxiliary quantities, we have order-$g$ initial conditions
\begin{equation}\label{eq:beta_bperp_coulombgauge}
\begin{split}
&b_{\eta}(\mathbf{k}) = 2\beta^{(1)}(\tau=0, \mathbf{k}) =-i \mathbf{k}^i \zeta^{(1),i}(\mathbf{k}) - g\rho_R(\mathbf{k}),\\
&b_{\perp}(\mathbf{k}) = -i \epsilon_{ij}\mathbf{k}_i \beta_j^{(1)}(\tau=0, \mathbf{k}) = -i\epsilon^{ij}\mathbf{k}^i \zeta^{(1),j}(\mathbf{k}).\\
\end{split}
\end{equation}
In order to make the further notation uniform  we also introduce
\begin{equation}
b_{\|}(\mathbf{k}) = \frac{i\mathbf{k}^i}{k_{\perp}^2} \zeta^{(1),i}(\mathbf{k}).
\end{equation}
In other words we can expand the two dimensional vector $\zeta^{(1),i}$ into the longitudinal and transverse parts
\begin{equation}
\zeta^{i}_{(1)}(\mathbf{k}) = \frac{-i\epsilon^{ij} \mathbf{k}^j}{k_{\perp}^2} b_{\perp}(\mathbf{k}) - i\mathbf{k}^i  b_{\|}(\mathbf{k}).
\end{equation}

The order-$g^3$ initial conditions are
\begin{equation}
\begin{split}
\beta_{\perp}^{(3)}(\tau=0, \mathbf{k}) 
=&\frac{i\epsilon_{ih} \mathbf{k}_h}{k_{\perp}}\zeta_{(3)}^i(\mathbf{k}) -i\int \frac{d^2\mathbf{p}}{(2\pi)^2} \frac{\mathbf{k}\cdot\mathbf{p}}{k_{\perp} p_{\perp}^2} \left[b_{\|}(\mathbf{k}-\mathbf{p}), b_{\perp}(\mathbf{p})\right]\\
& -i\int \frac{d^2\mathbf{p}}{(2\pi)^2} \frac{\mathbf{p}\times\mathbf{k}}{2k_{\perp}}\left[b_{\|}(\mathbf{k}-\mathbf{p}), b_{\|}(\mathbf{p})\right],
\end{split}
\end{equation}
\begin{equation}
2\beta_{(3)}(\tau=0, \mathbf{k}) = -i\mathbf{k}^i \zeta_{(3)}^i(\mathbf{k}) -i \int\frac{d^2\mathbf{p}}{(2\pi)^2} \left[b_{\|}(\mathbf{k}-\mathbf{p}), b_{\eta}(\mathbf{p})\right] .
\end{equation}
Although we will not need the order-$g^5$ solutions of the classical Yang-Mills equations when computing order-$g^5$ gluon production amplitude using the LSZ reduction formula, we do need the order-$g^5$ initial conditions. 
The initial conditions at order-$g^5$  are 
\begin{equation}
\begin{split}
\beta_{\perp}^{(5)}(\tau=0, \mathbf{k}) 
=&\frac{i\epsilon_{lh}\mathbf{k}_h}{k_{\perp}} \zeta^l_{(5)}(\mathbf{k})+ \frac{\epsilon^{lh}\mathbf{k}^h}{k_{\perp}}\int \frac{d^2\mathbf{p}}{(2\pi)^2} \left[ b_{\|}(\mathbf{k}-\mathbf{p}), \zeta^l_{(3)}(\mathbf{p})\right]\\
&+\int \frac{d^2\mathbf{p}}{(2\pi)^2}\frac{\mathbf{k}\cdot(\mathbf{k}-\mathbf{p})}{k_{\perp} |\mathbf{k}-\mathbf{p}|^2}\frac{\mathbf{p}^l}{p_{\perp}^2}  \left[\zeta^l_{(3)}(\mathbf{p}), b_{\perp}(\mathbf{k}-\mathbf{p})\right]\\
&-\int \frac{d^2\mathbf{p}}{(2\pi)^2}\frac{d^2\mathbf{q}}{(2\pi)^2}\frac{\mathbf{k}\cdot(\mathbf{k}-\mathbf{p})}{k_{\perp} |\mathbf{k}-\mathbf{p}|^2}\frac{\mathbf{p}\times \mathbf{q}}{p_{\perp}^2q_{\perp}^2}\left[\left[b_{\|}(\mathbf{p}-\mathbf{q}), b_{\perp}(\mathbf{q})\right], b_{\perp}(\mathbf{k}-\mathbf{p})\right]\\
&-\int \frac{d^2\mathbf{p}}{(2\pi)^2}\frac{d^2\mathbf{q}}{(2\pi)^2}\frac{\mathbf{k}\cdot(\mathbf{k}-\mathbf{p})}{k_{\perp} |\mathbf{k}-\mathbf{p}|^2}\frac{\mathbf{p}\cdot\mathbf{q}}{2p_{\perp}^2}\left[\left[b_{\|}(\mathbf{p}-\mathbf{q}), b_{\|}(\mathbf{q})\right], b_{\perp}(\mathbf{k}-\mathbf{p})\right]\\
&-\frac{1}{2} \int \frac{d^2\mathbf{p}}{(2\pi)^2} \frac{d^2\mathbf{q}}{(2\pi)^2}\frac{\mathbf{k}\cdot\mathbf{q}}{k_{\perp}q_{\perp}^2} \left[ b_{\|}(\mathbf{k}-\mathbf{p}), \left[b_{\|}(\mathbf{p}-\mathbf{q}),b_{\perp}(\mathbf{q})\right]\right]\\
&-\frac{1}{3} \int \frac{d^2\mathbf{p}}{(2\pi)^2} \frac{d^2\mathbf{q}}{(2\pi)^2}\frac{\mathbf{q}\times \mathbf{k}}{k_{\perp}} \left[ b_{\|}(\mathbf{k}-\mathbf{p}), \left[b_{\|}(\mathbf{p}-\mathbf{q}),b_{\|}(\mathbf{q})\right]\right].\\
\end{split}
\end{equation}
\begin{equation}
\begin{split}
&2\beta_{(5)}(\tau=0, \mathbf{k}) \\
=& -i\mathbf{k}^i \zeta_{(5)}^{i}(\mathbf{k}) - \int \frac{d^2\mathbf{p}}{(2\pi)^2} \left[ b_{\|}(\mathbf{k}-\mathbf{p}), \mathbf{p}^i \zeta_{(3)}^i(\mathbf{p})\right] + \int \frac{d^2\mathbf{p}}{(2\pi)^2}\frac{\mathbf{p}^l}{p_{\perp}^2}\left[\zeta_{(3)}^l(\mathbf{p}), b_{\eta}(\mathbf{k}-\mathbf{p})\right]\\
&-\int \frac{d^2\mathbf{p}}{(2\pi)^2}\frac{d^2\mathbf{q}}{(2\pi)^2} \frac{\mathbf{p}\times\mathbf{q}}{p_{\perp}^2q_{\perp}^2}\left[\left[b_{\|}(\mathbf{p}-\mathbf{q}), b_{\perp}(\mathbf{q})\right],b_{\eta}(\mathbf{k}-\mathbf{p})\right]\\
&-\int \frac{d^2\mathbf{p}}{(2\pi)^2}\frac{d^2\mathbf{q}}{(2\pi)^2}  \frac{\mathbf{p}\cdot\mathbf{q}}{2p_{\perp}^2} \left[\left[b_{\|}(\mathbf{p}-\mathbf{q}), b_{\|}(\mathbf{q})\right], b_{\eta}(\mathbf{k}-\mathbf{p})\right]\\
&-\frac{1}{2} \int \frac{d^2\mathbf{p}}{(2\pi)^2}\frac{d^2\mathbf{q}}{(2\pi)^2} \left[b_{\|}(\mathbf{k}-\mathbf{p}), \left[b_{\|}(\mathbf{p}-\mathbf{q}), b_{\eta}(\mathbf{q})\right]\right].\\
\end{split}
\end{equation}

%%%%%%%%%%%%%%%%%%%%%%%%%%%%%%%%%%%%%%%%%%%%%%%%%%%
%%%%%%%%%%%%%%%%%%%%%%%%%%%%%%%%%%%%%%%%%%%%%%%%%%%
%%%%%%%%%%%%%%%%%%%%%%%%%%%%%%%%%%%%%%%%%%%%%%%%%%%
%%%%%%%%%%%%%%%%%%%%%%%%%%%%%%%%%%%%%%%%%%%%%%%%%%%
%%%%%%%%%%%%%%%%%%%%%%%%%%%%%%%%%%%%%%%%%%%%%%%%%%%

\section{Defining the Semi-Hard Gluon Productions}
\label{Sect:DefSI}
In this section we conform the conventional four-dimensional Minkowski LSZ procedure (see the textbook \cite{Itzykson:1980rh}) to the problem at hand, that is   2+1 Milne coordinate system with the initial conditions defined at $\tau=0$. There are two main sources of apparent differences from the conventional LSZ: the first is due to a  non-trivial metric modifying  the free-field modes and the second is due to a non-vanishing field at the initial hypersurface $\tau=0$.

\subsection{The LSZ reduction formula in Milne coordinate}
We begin by constructing free-field modes for the dynamical degrees of freedom.
In our case, for the gluon field there are two dynamical degrees of freedom; they are represented by two fields $\beta(\tau, \mathbf{x})$ and $\beta_{\perp}(\tau, \mathbf{x})$. The remaining field $\Lambda(\tau, \mathbf{x})$ is non-dynamical and does
not contribute to the particle spectrum.
The free-field equations  for $\tilde{\beta} (\tau, \mathbf{x})\equiv \tau \beta(\tau,\mathbf{x}) $ and $\beta_{\perp}(\tau, \mathbf{x})$ are
\begin{equation}\label{eq:LSZ_free_field_eqs}
\begin{split}
& \tau^2\partial_{\tau}^2 \tilde{\beta} +\tau\partial_{\tau} \tilde{\beta}-\tilde{\beta}  - \tau^2\partial^2_i \tilde{\beta} =0\,, \\
&\tau^2\partial^2_{\tau}\beta_{\perp}+\tau\partial_{\tau} \beta_{\perp}- \tau^2\partial^2  \beta_{\perp} =0 .\\
\end{split}
\end{equation}
%For $\tilde{\beta}(\tau, \mathbf{x})$ the two mode functions are
%\begin{equation}
%H_1^{(1)}(k_{\perp} \tau) e^{-i\mathbf{k}\cdot \mathbf{x}}  , \qquad  H_1^{(2)} (k_{\perp}\tau) e^{+i\mathbf{k}\cdot \mathbf{x}}
%\end{equation}
%For $\beta_{\perp}(\tau, \mathbf{x})$, the two mode functions are
%\begin{equation}
%H_0^{(1)}(k_{\perp} \tau) e^{-i\mathbf{k}\cdot \mathbf{x}}  , \qquad  H_0^{(2)} (k_{\perp}\tau) e^{+i\mathbf{k}\cdot \mathbf{x}}
%\end{equation}
%These are Hankel functions of first and second kinds as indicated by the superscript ``(1)'' and ``(2)''. Here only zero order and first order are used. We choose Hankel functions instead of Bessel functions as mode functions because they have the asymptotics at $x\rightarrow \infty$.
%\begin{equation}
%\begin{split}
%&H_{\nu}^{(1)}(x) \sim \sqrt{\frac{2}{\pi x}} e^{i(x-\frac{1}{2} \nu x - \frac{1}{4}\pi)},\\
%&H_{\nu}^{(2)}(x) \sim \sqrt{\frac{2}{\pi x}} e^{-i(x-\frac{1}{2} \nu x - \frac{1}{4}\pi)}.\\
%\end{split}
%\end{equation}
The free-field solutions can be easily found in the momentum space. We choose the solutions in a such a way as to resemble the plane-wave behaviour at asymptotically large $\tau$.
This fixes the form of the functions -- they are
Hankel functions of first and second kind (see Ref.~\cite{abramowitz+stegun} for mathematical definition); the first (zeroes) order Hankel function solves the first (second) equation in Eqs.~\eqref{eq:LSZ_free_field_eqs}.
The normalization can be fixed by requiring the consistency between  the equal time  commutation relations for the  fields
and the canonical commutation relations for creation and annihilation operators, as we demonstrate below.
We thus have two sets of mode functions
\begin{equation}
f(k_{\perp}\tau) = \sqrt{\frac{\pi}{4}} H_1^{(1)}(k_{\perp}\tau), \qquad f^{\ast}(k_{\perp}\tau)=\sqrt{\frac{\pi}{4}} H_1^{(2)}(k_{\perp}\tau)
\end{equation}
and
\begin{equation}
g(k_{\perp}\tau) = \sqrt{\frac{\pi }{4}} H_0^{(1)}(k_{\perp}\tau), \qquad g^{\ast}(k_{\perp}\tau)=\sqrt{\frac{\pi }{4}} H_0^{(2)}(k_{\perp}\tau).
\end{equation}
They satisfy the Wronskian identities
\begin{equation}
\begin{split}
&f(k_{\perp}\tau) \partial_{\tau} f^{\ast}(k_{\perp}\tau) - \partial_{\tau} f(k_{\perp}\tau) f^{\ast}(k_{\perp}\tau) = \frac{-i}{\tau},\\
&g(k_{\perp}\tau) \partial_{\tau} g^{\ast}(k_{\perp}\tau) - \partial_{\tau} g(k_{\perp}\tau) g^{\ast}(k_{\perp}\tau) = \frac{-i}{\tau}.\\
\end{split}
\end{equation}

Next we construct the mode expansions for  both  dynamical fields in terms of creation and annihilation operators
\begin{equation}
	\label{Eq:ModeExpansion}
\begin{split}
\tilde{\beta}^a(\tau, \mathbf{x}) &= \int \frac{d^2\mathbf{k}}{(2\pi)^2} \left(\hat{a}^a_{\mathbf{k}}f^{\ast}(k_{\perp}\tau) e^{i\mathbf{k}\cdot \mathbf{x}} + \hat{a}_{\mathbf{k}}^{a \dagger}f(k_{\perp}\tau) e^{-i\mathbf{k}\cdot \mathbf{x}} \right), \\
\beta_{\perp}^a(\tau, \mathbf{x}) &=\int \frac{d^2\mathbf{k}}{(2\pi)^2} \left(\hat{c}_{\mathbf{k}}^ag^{\ast}(k_{\perp} \tau) e^{i\mathbf{k}\cdot \mathbf{x}} + \hat{c}_{\mathbf{k}}^{a \dagger}g(k_{\perp}\tau) e^{-i\mathbf{k}\cdot \mathbf{x}} \right)\,.
\end{split}
\end{equation}
One can check that the normalization of the mode functions leads to the consistent commutation relations, that is
\begin{equation}
\begin{split}
&[\tilde{\beta}^a(\tau, \mathbf{x}), \tau\partial_{\tau} \tilde{\beta}^b(\tau, \mathbf{y})] = i\delta^{ab}\delta^{(2)}(\mathbf{x}-\mathbf{y}),\\
&[\beta_{\perp}^a(\tau, \mathbf{x}), \tau\partial_{\tau} \beta_{\perp}^b(\tau, \mathbf{y})] = i\delta^{ab}\delta^{(2)}(\mathbf{x}-\mathbf{y})\\
\end{split}
\end{equation}
for the fields and
\begin{equation}
\begin{split}
&\left[\hat{a}^a_{\mathbf{k}}, \hat{a}^{b\dagger}_{\mathbf{p}} \right] = (2\pi)^2 \delta^{(2)}(\mathbf{k}-\mathbf{p})\delta^{ab},\\
&\left[\hat{c}^a_{\mathbf{k}}, \hat{c}^{b\dagger}_{\mathbf{p}} \right] = (2\pi)^2 \delta^{(2)}(\mathbf{k}-\mathbf{p})\delta^{ab}\\
\end{split}
\end{equation}
for the creation and annihilation operators.
%The commutation relation of the creation and annihilation operators is what we need to have correct normalization when computing particle productions.
%\begin{equation}
%\begin{split}
%&\partial_{\tau} f^{\ast}(p_{\perp}\tau) \int d^2\mathbf{x} e^{i\mathbf{p}\cdot \mathbf{x}}\tilde{\beta}^a(\tau, \mathbf{x})  -f^{\ast}(p_{\perp}\tau) \int d^2\mathbf{x} e^{i\mathbf{p}\cdot \mathbf{x}} \partial_{\tau} \tilde{\beta}^a(\tau, \mathbf{x}) \\
%=& \partial_{\tau} f^{\ast}(p_{\perp}\tau) f(p_{\perp}\tau) \hat{a}_p^{a \dagger} - f^{\ast}(p_{\perp}\tau) \partial_{\tau} f(p_{\perp}\tau) \hat{a}_p^{a \dagger}\\
%=&-\frac{i}{\tau} \hat{a}_p^{a \dagger}
%\end{split}
%\end{equation}
The mode expansion~\eqref{Eq:ModeExpansion}  can be inverted in order to express  the creation operators in terms of the fields
\begin{equation}\label{eq:LSZ_reduction_formula_tau}
\begin{split}
\hat{a}_{\mathbf{p}}^{a \dagger}
=& -i\tau \sqrt{\frac{\pi}{4}}\left( H_1^{(2)}(p_{\perp}\tau) \overleftrightarrow{\partial_{\tau}} \tilde{\beta}^a(\tau, \mathbf{p})\right),\\
\hat{c}_{\mathbf{p}}^{a\dagger} =& - i\tau \sqrt{\frac{\pi}{4}}\left( H_0^{(2)}(p_{\perp}\tau) \overleftrightarrow{\partial_{\tau}} \beta^a_{\perp}(\tau, \mathbf{p})\right).\\
\end{split}
\end{equation}
Here  $f_1(x)\overleftrightarrow{\partial_x} f_2(x) = f_1(x) \partial_x f_2(x) - \partial_x f_1(x) f_2(x)$.
%These results are derived for \textit{free field theory}. However, for interacting fields, we make the similar generalization as the LSZ reduction formula and take Eq.~ \eqref{eq:LSZ_reduction_formula_tau} as the definitions for the time dependent creation operators.
Similar to Minkovski spacetime, we can start from the trivial identity for the creation operator for the asymptotic out state  $\tau\rightarrow \infty$:
\begin{equation}
\begin{split}
\hat{a}_{\mathbf{p}}^{a \dagger}(+\infty) =& \hat{a}_{\mathbf{p}}^{a \dagger}(\tau=0)+\int_0^{\infty} d\tau \partial_{\tau} \hat{a}_{\mathbf{p}}^{a \dagger}(\tau) \\
=& \hat{a}_{\mathbf{p}}^{a \dagger}(\tau=0)-i\sqrt{\frac{\pi}{4}}\int_0^{\infty} d\tau \frac{1}{\tau} H_1^{(2)}(p_{\perp}\tau) S_{\eta}^a(\tau, \mathbf{p}),
\end{split}
\end{equation}
\begin{equation}
\begin{split}
\hat{c}_{\mathbf{p}}^{a \dagger}(+\infty)= & \hat{c}_{\mathbf{p}}^{a \dagger}(\tau=0) +\int_0^{\infty} d\tau \, \partial_{\tau} \hat{c}_{\mathbf{p}}^{a \dagger}(\tau)\\
=&  \hat{c}_{\mathbf{p}}^{a \dagger}(\tau=0) -i\sqrt{\frac{\pi}{4}} \int_0^{\infty} d\tau \frac{1}{\tau} H_0^{(2)}(p_{\perp}\tau) S_{\perp}^a(\tau, \mathbf{p}).\\
\end{split}
\end{equation}
Here we used the fact that $H_1^{(2)}$ ($H_0^{(2)}$) satisfies the first (second) free field equation in Eqs.~\eqref{eq:LSZ_free_field_eqs}. Additionally we took into account that the fields $\tilde{\beta}(\tau, \mathbf{p})
$ and $\beta_{\perp}(\tau, \mathbf{p})$ are the (formal) solutions of  the non-linear Yang-Mills equations with the  source terms $S_{\eta}(\tau, \mathbf{p})$ and $S_{\perp}(\tau, \mathbf{p})$.
Here we see that the creation operators for the asymptotic out state have two distinct contributions. The first is due to the initial ``surface'' contribution $\tau=0$, while the second due to
the ``bulk'' evolution in the forward light cone.

We  can also compute $\hat{a}_p^{\dagger}(\tau=0)$ and $\hat{c}_p^{\dagger}(\tau=0)$:
\begin{equation}
\begin{split}
\hat{a}_{\mathbf{p}}^{\dagger}(\tau=0) = &-i\sqrt{\frac{\pi}{4} }\tau \left(H_1^{(2)}(p_{\perp}\tau)\partial_{\tau} \tilde{\beta}(\tau, \mathbf{p}) -\partial_{\tau}H_1^{(2)}(p_{\perp}\tau) \tilde{\beta}(\tau, \mathbf{p}) \right)\Big|_{\tau\rightarrow 0}\\
%=&-i\sqrt{ \frac{\pi}{4} }\tau \left( \frac{2i}{\pi p_{\perp}\tau }C_1 \frac{p_{\perp}}{2} - \frac{-2i}{\pi p_{\perp}\tau^2} C_1 \frac{p_{\perp}\tau}{2}\right)\\
=&C_1 \sqrt{\frac{1}{\pi}}, 
\end{split}
\end{equation}
\begin{equation}
\begin{split}
\hat{c}_{\mathbf{p}}^{\dagger}(\tau =0)=& -i\sqrt{\frac{\pi}{4}} \tau \left(H_0^{(2)}(p_{\perp}\tau) \partial_{\tau} \beta_{\perp}(\tau, \mathbf{p}) - \partial_{\tau} H_0^{(2)}(p_{\perp}\tau) \beta_{\perp}(\tau, \mathbf{p}) \right)\Big|_{\tau \rightarrow 0}\\
%=& +i\sqrt{\frac{\pi}{4}} \tau\left(\partial_{\tau} H_0^{(2)}(p_{\perp}\tau) \beta_{\perp}(\tau, \mathbf{p}) \right)\Big|_{\tau \rightarrow 0}\\
=&D_1\sqrt{\frac{1}{\pi}}. \\
\end{split}
\end{equation}
Where the constants $C_1$ and $D_1$ can be found using the initial conditions
$\beta(\tau=0, \mathbf{p}) =  \frac{p_{\perp}}{2} C_1$ and $\beta_{\perp}(\tau=0, \mathbf{p}) = D_1$.

Following the notations of  Ref.~\cite{McLerran:2016snu} we define the surface contributions $\mathfrak{S}_{\eta}$ and $\mathfrak{S}_{\perp}$ by
\begin{equation}
\begin{split}
&\mathfrak{S}_{\eta}(\mathbf{p}) = \hat{a}_{\mathbf{p}}^{\dagger}(\tau=0) = \frac{1}{\sqrt{\pi} p_{\perp}} 2 \beta(\tau=0, \mathbf{p}), \\
&\mathfrak{S}_{\perp}(\mathbf{p}) = \hat{c}_{\mathbf{p}}^{\dagger}(\tau=0) = \frac{1}{\sqrt{\pi}} \beta_{\perp}(\tau=0, \mathbf{p}).\\
\end{split}
\end{equation}
and  the bulk terms $\mathfrak{B}_{\eta}$ and $\mathfrak{B}_{\perp}$ by
\begin{equation}
\label{eq:bulks}
\begin{split}
&\mathfrak{B}_{\eta}(\mathbf{p}) =-i\sqrt{ \frac{\pi}{4}}\int_0^{\infty} d\tau \frac{1}{\tau} H_1^{(2)}(p_{\perp}\tau) S_{\eta}(\tau, \mathbf{p}),\\
&\mathfrak{B}_{\perp}(\mathbf{p})=- i\sqrt{\frac{\pi}{4}} \int_0^{\infty} d\tau \frac{1}{\tau} H_0^{(2)}(p_{\perp}\tau) S_{\perp}(\tau, \mathbf{p}).\\
\end{split}
\end{equation}
Thus the LSZ reduction formula can be written in the compact form
\begin{equation}
\begin{split}
&\hat{a}_{ \mathbf{p}}^{\dagger} = \mathfrak{S}_{\eta}(\mathbf{p}) + \mathfrak{B}_{\eta}(\mathbf{p}),\\
&\hat{c}_{ \mathbf{p}}^{\dagger} = \mathfrak{S}_{\perp}(\mathbf{p}) + \mathfrak{B}_{\perp}(\mathbf{p}).\\
\end{split}
\end{equation}
As the reader will see below, differentiating between the surface  (fully defined by the initial conditions at $\tau=0$) and the bulk terms (determined by the final state interactions) helps to pin-point the importance of the final state interactions. 

\subsection{Single inclusive gluon spectrum}

Using the creation and  annihilation we define the single inclusive gluon spectrum (for the fixed valence charge densities) as
\begin{equation}\label{eq:single_gluon_spectrum}
\begin{split}
	\frac{dN}{d^2\mathbf{k} dy} = &\frac{1}{(2\pi)^2} \Big(\hat{a}_{\mathbf{k}}^{\dagger}  \hat{a}_{\mathbf{k}} + \hat{c}_{\mathbf{k}}^{\dagger}  \hat{c}_{\mathbf{k}}
	\Big)\\
=&\frac{1}{(2\pi)^2} \sum_{\gamma=\eta,\perp}\Big(|\mathfrak{S}_{\gamma}(\mathbf{k})|^2 + \mathfrak{S}_{\gamma}(\mathbf{k}) \mathfrak{B}_{\gamma}^{\ast}(\mathbf{k}) + \mathfrak{S}^{\ast}_{\gamma}(\mathbf{k}) \mathfrak{B}_{\gamma}(\mathbf{k})+ |\mathfrak{B}_{\gamma}(\mathbf{k})|^2\Big)\,.\\
\end{split}
\end{equation}
Here $\gamma$ denotes two gluon polarizations $\gamma=\eta, \perp$; we sum over the polarizations as we are only interested in the polarization-blind single inclusive gluon production.

It is instructive  to analyze the  $\mathbf{k}\leftrightarrow -\mathbf{k}$ symmetry property  of each term in $dN/d^2\mathbf{k}dy$ as it does not require explicit expressions. This analysis will be helpful in isolating the odd harmonics in double gluon productions. 
As a first step, let's discuss the symmetries of the surface and bulk contributions.

The initial WW fields are real functions of $\mathbf{x}$; consequently the surface terms have the following symmetry in the momentum space
\begin{equation}
\begin{split}
&\mathfrak{S}_{\eta}^{a, \ast}(\mathbf{k}) = \mathfrak{S}_{\eta}^a(-\mathbf{k}),\\
&\mathfrak{S}_{\perp}^{a, \ast}(\mathbf{k}) = \mathfrak{S}_{\perp}^a(-\mathbf{k}).\\
\end{split}
\end{equation}

The bulk terms require a more detailed analysis. Superficially,  one can argue that since the source terms $S^a_{\eta}(\tau, \mathbf{x})$, $S^a_{\perp}(\tau, \mathbf{x})$ are real functions of $\mathbf{x}$ and consequently  $S^{a,\ast}_{\eta (\perp)}(\tau, \mathbf{k}) = S^{a}_{\eta (\perp)}(\tau, -\mathbf{k})$, we should expect that the entire bulk contribution has to be symmetric under
 $\mathbf{k}\leftrightarrow -\mathbf{k}$.  This argument however misses the fact that the Hankel functions in the definition~\eqref{eq:bulks} are complex valued. Taking this into account and explicitly separating the
 Hankel functions into the real and imaginary parts we can write
\begin{equation}
\begin{split}
\mathfrak{B}_{\eta}^a(\mathbf{k}) =&-i\sqrt{ \frac{\pi}{4}}\int_0^{\infty} \frac{d\tau}{\tau} J_1(k_{\perp}\tau) S_{\eta}^a(\tau, \mathbf{k}) -\sqrt{ \frac{\pi}{4}}\int_0^{\infty}  \frac{d\tau}{\tau} Y_1(k_{\perp}\tau) S_{\eta}^a(\tau, \mathbf{k}) \\
=&\widetilde{\mathfrak{B}}_{\eta}^a(\mathbf{k}) + \overline{\mathfrak{B}}_{\eta}^a(\mathbf{k}).\\
\end{split}
\end{equation}
Note that in the configuration space, the first term is purely imaginary while  the second term is  purely  real.
There is an overall phase difference between two term of $\pi/2$, see a related discussion in Ref.~\cite{Kovchegov:2018jun}.
The newly introduced functions satisfy
\begin{equation}
\begin{split}
&\widetilde{\mathfrak{B}}_{\eta}^{a,\ast} (\mathbf{k}) = -\widetilde{\mathfrak{B}}_{\eta}^a(-\mathbf{k}),\\
&\overline{\mathfrak{B}}_{\eta}^{a, \ast}(\mathbf{k}) = \overline{\mathfrak{B}}_{\eta}^a(-\mathbf{k}).
\end{split}
\end{equation}
Similar separation can be defined for the transverse component of the bulk term
\begin{equation}
\begin{split}
\mathfrak{B}_{\perp}^a(\mathbf{k}) =&-i\sqrt{ \frac{\pi}{4}}\int_0^{\infty} \frac{d\tau}{\tau} J_0(k_{\perp}\tau) S_{\perp}^a(\tau, \mathbf{k}) -\sqrt{ \frac{\pi}{4}}\int_0^{\infty}  \frac{d\tau}{\tau} Y_0(k_{\perp}\tau) S_{\perp}^a(\tau, \mathbf{k}) \\
=&\widetilde{\mathfrak{B}}_{\perp}^a(\mathbf{k}) + \overline{\mathfrak{B}}_{\perp}^a(\mathbf{k})\\
\end{split}
\end{equation}
with the properties
\begin{equation}
\begin{split}
&\widetilde{\mathfrak{B}}_{\perp}^{a,\ast} (\mathbf{k}) = -\widetilde{\mathfrak{B}}_{\perp}^a(-\mathbf{k}),\\
&\overline{\mathfrak{B}}_{\perp}^{a, \ast}(\mathbf{k}) = \overline{\mathfrak{B}}_{\perp}^a(-\mathbf{k}).
\end{split}
\end{equation}

We are now in position to discuss the symmetry properties of each term in Eq.~\eqref{eq:single_gluon_spectrum}.  The pure surface term is symmetric with respect to $\mathbf{k}\leftrightarrow -\mathbf{k}$.
\begin{equation}
|\mathfrak{S}_{\gamma}(\mathbf{k})|^2 = \mathfrak{S}^{a,\ast}_{\gamma}(\mathbf{k})\mathfrak{S}^{a}_{\gamma}(\mathbf{k}) = \mathfrak{S}^{a}_{\gamma}(-\mathbf{k})\mathfrak{S}^{a}_{\gamma}(\mathbf{k}).
\end{equation}
 The interference terms between the surface and bulk contributions have two contributions with different properties under  $\mathbf{k}\leftrightarrow -\mathbf{k}$
\begin{equation}
\begin{split}
&\mathfrak{S}_{\gamma}(\mathbf{k}) \mathfrak{B}_{\gamma}^{\ast}(\mathbf{k}) + \mathfrak{S}^{\ast}_{\gamma}(\mathbf{k}) \mathfrak{B}_{\gamma}(\mathbf{k})\\
=&\left(-\mathfrak{S}^a_{\gamma}(\mathbf{k})\widetilde{\mathfrak{B}}_{\gamma}^{a}(-\mathbf{k})+\mathfrak{S}^{a}_{\gamma}(-\mathbf{k}) \widetilde{\mathfrak{B}}^a_{\gamma}(\mathbf{k}) \right)+\left(\mathfrak{S}^a_{\gamma}(\mathbf{k})\overline{\mathfrak{B}}^a_{\gamma}(-\mathbf{k})) + \mathfrak{S}^{a}_{\gamma}(-\mathbf{k}) \overline{\mathfrak{B}}^a_{\gamma}(\mathbf{k}))\right).\\
\end{split}
\end{equation}
We grouped the terms according to the symmetry:  the first (second)  piece involving $\widetilde{\mathfrak{B}}_{\gamma}^a$  ($\overline{\mathfrak{B}}_{\gamma}^a$)  is antisymmetric (symmetric) with respect to $\mathbf{k}\leftrightarrow -\mathbf{k}$. As was identified in Ref.~\cite{McLerran:2016snu}, the antisymmetric part is ultimately responsible for generating a non-vanishing contribution to odd harmonics at the lowest non-trivial order.
The pure bulk term reads
\begin{equation}
\begin{split}
&|\mathfrak{B}_{\gamma}(\mathbf{k})|^2 = \mathfrak{B}^{a,\ast}_{\gamma}(\mathbf{p})\mathfrak{B}^a_{\gamma}(\mathbf{k}) \\
=&\left(-\widetilde{\mathfrak{B}}^{a}_{\gamma}(-\mathbf{k})\widetilde{\mathfrak{B}}^{a}_{\gamma}(\mathbf{k}) +  \overline{\mathfrak{B}}^{a}_{\gamma}(-\mathbf{k})\overline{\mathfrak{B}}^{a}_{\gamma}(\mathbf{k})\right) + \left(-\widetilde{\mathfrak{B}}^{a}_{\gamma}(-\mathbf{k})\overline{\mathfrak{B}}^{a}_{\gamma}(\mathbf{k}) +\widetilde{\mathfrak{B}}^{a}_{\gamma}(\mathbf{k}) \overline{\mathfrak{B}}^{a}_{\gamma}(-\mathbf{k}) \right).
\end{split}
\end{equation}
The contribution in the first bracket is symmetric  while in the second bracket is  antisymmetric under $\mathbf{k} \leftrightarrow  -\mathbf{k}$.  To summarize this discussion,
the bulk terms are complex-values  and their imaginary parts contribute to odd harmonics either through the bulk surface interference  or the pure bulk terms. Note that overall $	\frac{dN}{d^2\mathbf{k} dy} $ is real.

In practice, our calculations will be done for power series expansion for  the single gluon distribution.
We have
\begin{equation}
\begin{split}
&\mathfrak{S}_{\gamma} = g\mathfrak{S}_{\gamma}^{(1)} + g^3\mathfrak{S}_{\gamma}^{(3)}+ g^5\mathfrak{S}_{\gamma}^{(5)}+ \ldots\\
&\mathfrak{B}_{\gamma} =  g^3\mathfrak{B}_{\gamma}^{(3)}+ g^5\mathfrak{B}_{\gamma}^{(5)}+ \ldots\\
\end{split}
\end{equation}
This power series expansions were terminate at the fifth order, as it is sufficient
to extract the first saturation contribution to single inclusive gluon production, which is up to order $g^6$:
\begin{equation}
\begin{split}
\frac{dN}{d^2\mathbf{k} dy} =&\frac{1}{(2\pi)^2}\sum_{\gamma = \eta, \perp}\Bigg( g^2 \left[|\mathfrak{S}_{\gamma}^{(1)}(\mathbf{k})|^2 \right]
\\ & +g^4\Big[\mathfrak{S}_{\gamma}^{(1)\ast}(\mathbf{k})\left( \mathfrak{S}_{\gamma}^{(3)}(\mathbf{k})+\mathfrak{B}_{\gamma}^{(3)}(\mathbf{k})\right)
+\mathfrak{S}_{\gamma}^{(1)}(\mathbf{k}) \left(\mathfrak{S}_{\gamma}^{(3)\ast}(\mathbf{k}) +  \mathfrak{B}_{\gamma}^{(3)\ast}(\mathbf{k})\right)\Big]\\
&+g^6\Big[ |\mathfrak{S}_{\gamma}^{(3)}(\mathbf{k})|^2 +|\mathfrak{B}_{\gamma}^{(3)}(\mathbf{k})|^2+\mathfrak{S}_{\gamma}^{(1)\ast}(\mathbf{k})\left( \mathfrak{S}_{\gamma}^{(5)}(\mathbf{k})+\mathfrak{B}_{\gamma}^{(5)}(\mathbf{k})\right)\\
&+\mathfrak{S}_{\gamma}^{(1)}(\mathbf{k}) \left(\mathfrak{S}_{\gamma}^{(5)\ast}(\mathbf{k}) +  \mathfrak{B}_{\gamma}^{(5)\ast}(\mathbf{k})\right)+\mathfrak{S}_{\gamma}^{(3)}(\mathbf{k}) \mathfrak{B}_{\gamma}^{(3)\ast}(\mathbf{k})+\mathfrak{S}_{\gamma}^{(3)\ast}(\mathbf{k}) \mathfrak{B}_{\gamma}^{(3)}(\mathbf{k})\Big]\Bigg)\\
\equiv&\frac{1}{(2\pi)^2} \Big(g^2 n^{(2)}(\mathbf{k}) + g^4n^{(4)}(\mathbf{k}) + g^6 n^{(6)}(\mathbf{k}) \Big).
\end{split}
\end{equation}
In the following we calculate each order $n^{(i)}(\mathbf{k})$ separately. From our previous analysis, it is obvious that, for a fixed configuration of the valence charges,  the leading order $n^{(2)}(\mathbf{k})$ is even under $\mathbf{k}\leftrightarrow -\mathbf{k}$ while all higher order terms have both even and odd contributions. Note that $g^4$ terms is cubic in $\rho_P$ and thus, at least in the Gaussian MV model,  does not contribute to the configuration/event averaged single inclusive gluon production.

\section{The Gluon Production Amplitude}
\label{Sect:CalcAmp}
We now proceed with assembling all the pieces together in order to find the functional $\frac{dN}{d^2\mathbf{k} dy} [\rho_P, \rho_T]$. First we explicitly write down the surface and the bulk terms.

\subsection{The surface terms}
\label{Sect:Surf}
The surface terms are defined by the initial condition on the light cone $\tau=0$; thus to find them one does not require solving the classical Yang-Mills equations in the forward light cone $\tau>0$.
Starting from the definitions
\begin{equation}
\begin{split}
&\mathfrak{S}_{\eta}(\mathbf{k}) = \frac{1}{\sqrt{\pi} k_{\perp}} 2\beta(\tau=0, \mathbf{k}) ,\\
&\mathfrak{S}_{\perp}(\mathbf{k}) = \frac{1}{\sqrt{\pi}}\beta_{\perp}(\tau=0,\mathbf{k}) = \frac{1}{\sqrt{\pi} k_{\perp}} i\epsilon_{ih}\mathbf{k}_h \beta_i(\tau=0,\mathbf{k})\\
\end{split}
\end{equation}
we obtain that  the order-$g$ surface terms are simply
\begin{equation}\label{eq:surface_terms_g1}
\begin{split}
&\mathfrak{S}_{\eta}^{(1)}(\mathbf{k}) = \frac{1}{\sqrt{\pi}k_{\perp}} b_{\eta}(\mathbf{k}), \\
&\mathfrak{S}_{\perp}^{(1)}(\mathbf{k}) = \frac{1}{\sqrt{\pi} k_{\perp}} b_{\perp}(\mathbf{k}).\\
\end{split}
\end{equation}
The order-$g^3$ and order-$g^5$ contribution to the surface terms can be also extracted straightforwardly
\begin{equation}\label{eq:surface_eta_g3}
\mathfrak{S}_{\eta}^{(3)} = \frac{-i\mathbf{k}^i }{\sqrt{\pi}k_{\perp}}\zeta_{(3)}^i(\mathbf{k})+\frac{ig}{\sqrt{\pi}k_{\perp}} \int \frac{d^2\mathbf{p}}{(2\pi)}\left[ b_{\eta}(\mathbf{p}), b_{\|}(\mathbf{k}-\mathbf{p})\right],
\end{equation}
\begin{equation}\label{eq:surface_perp_g3}
\begin{split}
\mathfrak{S}_{\perp}^{(3)} =&\frac{i\epsilon^{ih}\mathbf{k}^h }{\sqrt{\pi}k_{\perp}} \zeta_{(3)}^i(\mathbf{k}) +\frac{ig}{\sqrt{\pi}k_{\perp}} \int\frac{d^2\mathbf{p}}{(2\pi)^2} \frac{\mathbf{k}\cdot\mathbf{p}}{p_{\perp}^2} [b_{\perp}(\mathbf{p}), b_{\|}(\mathbf{k}-\mathbf{p})]\\
&\qquad -\frac{1}{2} \frac{ig}{\sqrt{\pi}k_{\perp}} \int\frac{d^2\mathbf{p}}{(2\pi)^2} (\mathbf{k}\times \mathbf{p}) [b_{\|}(\mathbf{p}), b_{\|}(\mathbf{k}-\mathbf{p})]\\
\end{split}
\end{equation}
and
\begin{equation}\label{eq:surface_perp_g5}
\begin{split}
\mathfrak{S}^{(5)}_{\perp}(\mathbf{k})
=&\frac{i\epsilon_{lh}\mathbf{k}_h}{\sqrt{\pi}k_{\perp}} \zeta^l_{(5)}(\mathbf{k})+ \frac{\epsilon^{lh}\mathbf{k}^h}{\sqrt{\pi}k_{\perp}}\int \frac{d^2\mathbf{p}}{(2\pi)^2} [ b_{\|}(\mathbf{k}-\mathbf{p}), \zeta^l_{(3)}(\mathbf{p})]\\
&+\frac{1}{\sqrt{\pi}}\int \frac{d^2\mathbf{p}}{(2\pi)^2}\frac{\mathbf{k}\cdot(\mathbf{k}-\mathbf{p})}{k_{\perp} |\mathbf{k}-\mathbf{p}|^2}\frac{\mathbf{p}^l}{p_{\perp}^2}  \left[\zeta^l_{(3)}(\mathbf{p}), b_{\perp}(\mathbf{k}-\mathbf{p})\right]\\
&-\frac{1}{\sqrt{\pi}}\int \frac{d^2\mathbf{p}}{(2\pi)^2}\frac{d^2\mathbf{q}}{(2\pi)^2}\frac{\mathbf{k}\cdot(\mathbf{k}-\mathbf{p})}{k_{\perp} |\mathbf{k}-\mathbf{p}|^2}\frac{\mathbf{p}\times \mathbf{q}}{p_{\perp}^2q_{\perp}^2}[[b_{\|}(\mathbf{p}-\mathbf{q}), b_{\perp}(\mathbf{q})], b_{\perp}(\mathbf{k}-\mathbf{p})]\\
&-\frac{1}{\sqrt{\pi}}\int \frac{d^2\mathbf{p}}{(2\pi)^2}\frac{d^2\mathbf{q}}{(2\pi)^2}\frac{\mathbf{k}\cdot(\mathbf{k}-\mathbf{p})}{k_{\perp} |\mathbf{k}-\mathbf{p}|^2}\frac{\mathbf{p}\cdot\mathbf{q}}{2p_{\perp}^2}[[b_{\|}(\mathbf{p}-\mathbf{q}), b_{\|}(\mathbf{q})], b_{\perp}(\mathbf{k}-\mathbf{p})]\\
&-\frac{1}{2\sqrt{\pi}} \int \frac{d^2\mathbf{p}}{(2\pi)^2} \frac{d^2\mathbf{q}}{(2\pi)^2}\frac{\mathbf{k}\cdot\mathbf{q}}{k_{\perp}q_{\perp}^2} [ b_{\|}(\mathbf{k}-\mathbf{p}), [b_{\|}(\mathbf{p}-\mathbf{q}),b_{\perp}(\mathbf{q})]]\\
&-\frac{1}{3\sqrt{\pi}} \int \frac{d^2\mathbf{p}}{(2\pi)^2} \frac{d^2\mathbf{q}}{(2\pi)^2}\frac{\mathbf{q}\times \mathbf{k}}{k_{\perp}} [ b_{\|}(\mathbf{k}-\mathbf{p}), [b_{\|}(\mathbf{p}-\mathbf{q}),b_{\|}(\mathbf{q})]],\\
\end{split}
\end{equation}
\begin{equation}\label{eq:surface_eta_g5}
\begin{split}
\mathfrak{S}_{\eta}^{(5)}(\mathbf{k})
=& -\frac{i\mathbf{k}}{\sqrt{\pi} k_{\perp}} \zeta_{(5)}^{i}(\mathbf{k}) - i \int \frac{d^2\mathbf{p}}{(2\pi)^2} [ b_{\|}(\mathbf{k}-\mathbf{p}), -i\mathbf{p}^i \zeta_{(3)}^i(\mathbf{p})]\\
& +\frac{1}{\sqrt{\pi}k_{\perp}} \int \frac{d^2\mathbf{p}}{(2\pi)^2}\frac{\mathbf{p}^l}{p_{\perp}^2}[\zeta_{(3)}^l(\mathbf{p}), b_{\eta}(\mathbf{k}-\mathbf{p})]\\
&-\frac{1}{\sqrt{\pi}k_{\perp}}\int \frac{d^2\mathbf{p}}{(2\pi)^2}\frac{d^2\mathbf{q}}{(2\pi)^2} \frac{\mathbf{p}\times\mathbf{q}}{p_{\perp}^2q_{\perp}^2}[[b_{\|}(\mathbf{p}-\mathbf{q}), b_{\perp}(\mathbf{q})],b_{\eta}(\mathbf{k}-\mathbf{p})]\\
&-\frac{1}{\sqrt{\pi}k_{\perp}}\int \frac{d^2\mathbf{p}}{(2\pi)^2}\frac{d^2\mathbf{q}}{(2\pi)^2}  \frac{\mathbf{p}\cdot\mathbf{q}}{2p_{\perp}^2} [[b_{\|}(\mathbf{p}-\mathbf{q}), b_{\|}(\mathbf{q})], b_{\eta}(\mathbf{k}-\mathbf{p})]\\
&-\frac{1}{2\sqrt{\pi}k_{\perp}} \int \frac{d^2\mathbf{p}}{(2\pi)^2}\frac{d^2\mathbf{q}}{(2\pi)^2} [b_{\|}(\mathbf{k}-\mathbf{p}), [b_{\|}(\mathbf{p}-\mathbf{q}), b_{\eta}(\mathbf{q})]].\\
\end{split}
\end{equation}

\subsection{The bulk terms}
\label{Sect:Bulk}
In this section we evaluate the bulk terms $\mathfrak{B}_{\eta,\perp}^{(3)}$  and $\mathfrak{B}_{\eta,\perp}^{(5)}$.

\subsubsection*{The bulk terms $\mathfrak{B}_{\eta}^{(3)}$ and $\mathfrak{B}_{\perp}^{(3)}$ }
The order-$g^3$ bulk terms are defined by
\begin{equation}
\begin{split}
&\mathfrak{B}_{\eta}^{(3)}(\mathbf{k}) =-i\sqrt{ \frac{\pi}{4}}\int_0^{\infty} d\tau \frac{1}{\tau} H_1^{(2)}(k_{\perp}\tau) S_{\eta}^{(3)}(\tau, \mathbf{k}),\\
&\mathfrak{B}_{\perp}^{(3)}(\mathbf{k}) =-i\sqrt{ \frac{\pi}{4}}\int_0^{\infty} d\tau \frac{1}{\tau} H_0^{(2)}(k_{\perp}\tau) S_{\perp}^{(3)}(\tau, \mathbf{k}).\\
\end{split}
\end{equation}
Here the corresponding source terms are
\begin{equation}
\label{Eq:Seta3}
\begin{split}
S_{\eta}^{(3)}(\tau, \mathbf{k}) = &-i\tau^3 \int d^2\mathbf{x} e^{i\mathbf{k}\cdot \mathbf{x}} \left([\partial_i\beta_i^{(1)}, \beta^{(1)}]+2[\beta_i^{(1)}, \partial_i \beta^{(1)}] \right)\\
=& i\tau^2  \int \frac{d^2\mathbf{p}}{(2\pi)} \Bigg( \frac{2\mathbf{k}\times \mathbf{p}}{p_{\perp}^2|\mathbf{k}-\mathbf{p}|}J_0(p_{\perp}\tau)  J_1(|\mathbf{k}-\mathbf{p}|\tau)\Big[b_{\perp}(\mathbf{p}), b_{\eta}(\mathbf{k}-\mathbf{p})\Big] \Bigg)\\
\end{split}
\end{equation}
and
\begin{equation}
\label{Eq:Sperp3}
\begin{split}
&S_{\perp}^{(3)}(\tau,\mathbf{k}) = \frac{-i\epsilon^{ji} \mathbf{k}_j}{k_{\perp}} S_i^{(3)}\\
=&-i\tau^2 \int \frac{d^2\mathbf{p}}{(2\pi)^2}\Bigg(\frac{ \mathbf{p}\times \mathbf{k}}{k_{\perp}p_{\perp}|\mathbf{k}-\mathbf{p}|}\Big[b_{\eta}(\mathbf{p}), b_{\eta}( \mathbf{k}-\mathbf{p})\Big] J_1(p_{\perp}\tau) J_1(|\mathbf{k}-\mathbf{p}|\tau)\\
&+\frac{(\mathbf{k}\times \mathbf{p})(\mathbf{k}\cdot \mathbf{p} -p^2_{\perp}-k_{\perp}^2)}{k_{\perp}p_{\perp}^2|\mathbf{k}-\mathbf{p}|^2} \Big[b_{\perp}(\mathbf{p}), b_{\perp}(\mathbf{k}-\mathbf{p})\Big] J_0(p_{\perp}\tau)J_0(|\mathbf{k}-\mathbf{p}|\tau) \Bigg)\,.\\
\end{split}
\end{equation}
From Eq.~\eqref{Eq:Seta3} and Eq.~\eqref{Eq:Sperp3}
we obtain the order-$g^3$ bulk terms
\begin{equation}
\label{Eq:Bulk3}
\begin{split}
\mathfrak{B}_{\eta}^{(3)}(\mathbf{k})=&\frac{1}{\sqrt{\pi} k_{\perp}} \int \frac{d^2\mathbf{p}}{(2\pi)^2}  \frac{\mathbf{k}\times \mathbf{p}}{p_{\perp}^2 |\mathbf{k}-\mathbf{p}|^2} \left( \frac{\mathbf{k}\cdot(\mathbf{k}-\mathbf{p})}{|\mathbf{k}\times \mathbf{p}|} + i\right)\left [b_{\perp}(\mathbf{p}), b_{\eta}(\mathbf{k}-\mathbf{p})\right],\\
\mathfrak{B}_{\perp}^{(3)}(\mathbf{k}) = &-\frac{1}{2\sqrt{\pi} k_{\perp}} \int \frac{d^2\mathbf{p}}{(2\pi)^2} \frac{\mathbf{p}\times \mathbf{k}}{p_{\perp}^2|\mathbf{k}-\mathbf{p}|^2}\left(\frac{\mathbf{p}\cdot(\mathbf{p}-\mathbf{k})}{|\mathbf{p}\times \mathbf{k}|} - i\right) \left[b_{\eta}(\mathbf{p}), b_{\eta}(\mathbf{k}-\mathbf{p})\right]\\
&+ \frac{(\mathbf{p}\cdot(\mathbf{k}-\mathbf{p}) -k_{\perp}^2)}{p_{\perp}^2|\mathbf{k}-\mathbf{p}|^2} \frac{\mathbf{k}\times \mathbf{p}}{|\mathbf{p}\times \mathbf{k}|} \left[b_{\perp}(\mathbf{p}), b_{\perp}(\mathbf{k}-\mathbf{p})\right]\,.\\
\end{split}
\end{equation}
In computing the bulk terms, we need to calculate the proper time integral from $\tau=0$ to $\tau=\infty$. The integrands involve products of Bessel functions. 
The integral formulas used are summarized in Appendix \ref{ap:integral_formulas_three_bessels}.  Results in eq. \eqref{Eq:Bulk3} have already been obtained in ref. \cite{McLerran:2016snu}. We now move on to compute the new order-$g^5$ bulk terms.

\subsubsection*{The bulk term $\mathfrak{B}_{\eta}^{(5)}(\mathbf{k})$}
The first task is to compute explicitly the source term  at order-$g^5$
\begin{equation}\label{eq:S_eta_g5}
\begin{split}
S^{(5)}_{\eta}(\tau,\mathbf{k}) = &-g\tau^3 \int\frac{d^2\mathbf{q}}{(2\pi)^2} (\mathbf{k}+\mathbf{q})_i\left[\beta_i^{(1)}(\tau, \mathbf{k}-\mathbf{q}), \beta^{(3)}(\tau, \mathbf{q})\right]\\
&-g\tau^3 \int\frac{d^2\mathbf{q}}{(2\pi)^2}  (2\mathbf{k}-\mathbf{q})_i\left[\beta_i^{(3)}(\tau, \mathbf{q}), \beta^{(1)}(\tau, \mathbf{k}-\mathbf{q})\right]\\
&-g^2\tau^3\int\frac{d^2\mathbf{q}}{(2\pi)^2} \int\frac{d^2\mathbf{p}}{(2\pi)^2} \left[\beta_i^{(1)}(\tau, \mathbf{k}-\mathbf{q}), \left[\beta_i^{(1)}(\tau, \mathbf{p}), \beta^{(1)}(\tau, \mathbf{q}-\mathbf{p})\right]\right],
\end{split}
\end{equation}
 We substitute the solutions $\beta^{(1)}(\tau, \mathbf{k})$, $\beta_i^{(1)}(\tau, \mathbf{k})$, $\beta^{(3)}(\tau, \mathbf{k})$, $\beta_i^{(3)}(\tau, \mathbf{k})$ given in Eqs.~\eqref{eq:sol_order_g}, \eqref{eq:sol_order_g3_first}, \eqref{eq:sol_order_g3_second}, and \eqref{eq:sol_order_g3_third}
 in order to  obtain the explicit  expression for $S_{\eta}^{(5)}(\tau,\mathbf{k})$,
\begin{equation}
\begin{split}
S^{(5)}_{\eta}(\tau,\mathbf{k})
 =&-i\tau^2\int\frac{d^2\mathbf{q}}{(2\pi)^2}\frac{2\mathbf{q}\times \mathbf{k}}{q_{\perp}|\mathbf{k}-\mathbf{q}|^2} \left[ 2\beta^{(3)}(\tau=0,\mathbf{q}) , b_{\perp}(\mathbf{k}-\mathbf{q})\right]J_1(q_{\perp}\tau) J_0(|\mathbf{k}-\mathbf{q}|\tau)\\
&+\tau^2\int\frac{d^2\mathbf{q}}{(2\pi)^2}\int\frac{d^2\mathbf{p}}{(2\pi)^2}\frac{2\mathbf{q}\times \mathbf{k}}{|\mathbf{k}-\mathbf{q}|^2}\frac{\mathbf{q}\times \mathbf{p}}{p_{\perp}^2|\mathbf{q}-\mathbf{p}|^2}\Big[[b_{\perp}(\mathbf{p}), b_{\eta}(\mathbf{q}-\mathbf{p})], b_{\perp}(\mathbf{k}-\mathbf{q})\Big] \\
&\qquad\times\int_{-\pi}^{\pi}\frac{d\phi}{2\pi}\frac{\mathbf{q}\cdot(\mathbf{q}-2\mathbf{p})+w_{\perp}^2}{q_{\perp}^2-w_{\perp}^2}\left(\frac{J_1(w_{\perp}\tau)}{w_{\perp}}-\frac{J_1(q_{\perp}\tau)}{q_{\perp}}\right)J_0(|\mathbf{k}-\mathbf{q}|\tau) \\
&+ i\tau^2 \int\frac{d^2\mathbf{q}}{(2\pi)^2} \frac{2\mathbf{k}\times \mathbf{q}}{q_{\perp}|\mathbf{k}-\mathbf{q}|} \Big[\beta_{\perp}^{(3)}(\tau=0, \mathbf{q}), b_{\eta}(\mathbf{k}-\mathbf{q})\Big]J_0(q_{\perp}\tau)J_1(|\mathbf{k}-\mathbf{q}|\tau)\\
&+\tau^2 \int\frac{d^2\mathbf{q}}{(2\pi)^2}  \int \frac{d^2\mathbf{p}}{(2\pi)^2}\frac{2\mathbf{k}\times \mathbf{q}}{q^2_{\perp}|\mathbf{k}-\mathbf{q}|}\frac{ \mathbf{q}\times \mathbf{p}}{2p^2_{\perp}|\mathbf{q}-\mathbf{p}|^2}\int_{-\pi}^{\pi}\frac{d\phi}{2\pi}  \frac{1}{q_{\perp}^2-w_{\perp}^2}\Big((-p_{\perp}^2-|\mathbf{q}-\mathbf{p}|^2+w_{\perp}^2)\\
&\qquad\Big[\big[b_{\eta}(\mathbf{p}), b_{\eta}( \mathbf{q}-\mathbf{p})\big],  b_{\eta}(\mathbf{k}-\mathbf{q})\Big]+2(\mathbf{p}\cdot\mathbf{q}-p_{\perp}^2-q_{\perp}^2) \Big[\big[b_{\perp}(\mathbf{p}), b_{\perp}(\mathbf{q}-\mathbf{p})\big],  b_{\eta}(\mathbf{k}-\mathbf{q})\Big] \Big)\\
&\qquad\qquad\times (J_0(w_{\perp}\tau)-J_0(q_{\perp}\tau))J_1(|\mathbf{k}-\mathbf{q}|\tau)\\
&+\tau^2 \int\frac{d^2\mathbf{q}}{(2\pi)^2} \int\frac{d^2\mathbf{p}}{(2\pi)^2}\frac{(2\mathbf{k}-\mathbf{q})\cdot \mathbf{q}}{q_{\perp}^2|\mathbf{k}-\mathbf{q}|}   \frac{\mathbf{q}\cdot(\mathbf{q}-2\mathbf{p})}{4p_{\perp}^2 |\mathbf{q}-\mathbf{p}|^2} \int_{-\pi}^{\pi} \frac{d\phi}{2\pi}\frac{1}{w_{\perp}^2}\Big( (-q_{\perp}^2+w_{\perp}^2+2\mathbf{p}\cdot(\mathbf{q}-\mathbf{p}))\\
&\qquad \Big[\left[b_{\eta}(\mathbf{p}), b_{\eta}(\mathbf{q}-\mathbf{p})\right], b_{\eta}(\mathbf{k}-\mathbf{q})\Big]+2\mathbf{p}\cdot(\mathbf{q}-\mathbf{p}) \Big[\left[b_{\perp}(\mathbf{p}), b_{\perp}(\mathbf{q}-\mathbf{p})\right], b_{\eta}(\mathbf{k}-\mathbf{q})\Big]   \Big)\\
&\qquad\qquad\times (1- J_0(w_{\perp}\tau))J_1(|\mathbf{k}-\mathbf{q}|\tau)\\
&+\tau^2\int\frac{d^2\mathbf{q}}{(2\pi)^2} \int\frac{d^2\mathbf{p}}{(2\pi)^2} \frac{(\mathbf{k}-\mathbf{q})\cdot \mathbf{p}}{p_{\perp}^2|\mathbf{q}-\mathbf{p}||\mathbf{k}-\mathbf{q}|^2} \Big[b_{\perp}(\mathbf{k}-\mathbf{q}), [b_{\perp}(\mathbf{p}), b_{\eta}(\mathbf{q}-\mathbf{p})]\Big]\\
&\qquad\qquad \times J_1(|\mathbf{q}-\mathbf{p}|\tau)J_0(p_{\perp}\tau) J_0(|\mathbf{k}-\mathbf{q}|\tau).\\
\end{split}
\end{equation}
Using the above result,  we can evaluate the bulk term $\mathfrak{B}_{\eta}^{(5)}$ by 
\begin{equation}
\begin{split}
\mathfrak{B}_{\eta}^{(5)}(\mathbf{k}) =-i\sqrt{ \frac{\pi}{4}}\int_0^{\infty} d\tau \frac{1}{\tau} H_1^{(2)}(k_{\perp}\tau) S_{\eta}^{(5)}(\tau, \mathbf{k}).
\end{split}
\end{equation}
Again, it requires integrations over the proper time from $\tau=0$ to $\tau=\infty$. All the integrals involved have closed form results. The integral formulas used are given in the Appendix~\ref{App:BInt}.  We only present the final result here
\begin{equation}
\begin{split}
\mathfrak{B}^{(5)}_{\eta}(\mathbf{k}) =&\frac{1}{\sqrt{\pi}k_{\perp}}\int\frac{d^2\mathbf{q}}{(2\pi)^2} \left[ 2\beta^{(3)}(\tau=0,\mathbf{q}) , b_{\perp}(\mathbf{k}-\mathbf{q})\right]\frac{\mathbf{k}\times \mathbf{q}}{q_{\perp}^2|\mathbf{k}-\mathbf{q}|^2}\left(\frac{\mathbf{k}\cdot\mathbf{q}}{|\mathbf{k}\times \mathbf{q}| }+i\right)\\
&+ \frac{1}{\sqrt{\pi}k_{\perp}}\int\frac{d^2\mathbf{q}}{(2\pi)^2}  \Big[q_{\perp}\beta_{\perp}^{(3)}(\tau=0, \mathbf{q}), b_{\eta}(\mathbf{k}-\mathbf{q})\Big]\frac{\mathbf{k}\times \mathbf{q}}{q_{\perp}^2|\mathbf{k}-\mathbf{q}|^2} \left(\frac{\mathbf{k}\cdot(\mathbf{k}-\mathbf{q})}{|\mathbf{k}\times \mathbf{q}|} + i \right)\\
&+\frac{i}{\sqrt{\pi}k_{\perp}}\int\frac{d^2\mathbf{q}}{(2\pi)^2}\frac{d^2\mathbf{p}}{(2\pi)^2}\Big[[b_{\perp}(\mathbf{p}), b_{\eta}(\mathbf{q}-\mathbf{p})], b_{\perp}(\mathbf{k}-\mathbf{q})\Big]\frac{\mathcal{I}_1 (\mathbf{p}, \mathbf{q}, \mathbf{k})}{p_{\perp}^2|\mathbf{q}-\mathbf{p}|^2|\mathbf{k}-\mathbf{q}|^2} \\
&+\frac{i}{\sqrt{\pi}k_{\perp}} \int\frac{d^2\mathbf{q}}{(2\pi)^2} \frac{d^2\mathbf{p}}{(2\pi)^2}\Big[\big[b_{\eta}(\mathbf{p}), b_{\eta}( \mathbf{q}-\mathbf{p})\big],  b_{\eta}(\mathbf{k}-\mathbf{q})\Big]\frac{\mathcal{I}_2(\mathbf{p}, \mathbf{q}, \mathbf{k})}{p^2_{\perp}|\mathbf{q}-\mathbf{p}|^2|\mathbf{k}-\mathbf{q}|^2}\\
&+\frac{i}{\sqrt{\pi} k_{\perp}}\int\frac{d^2\mathbf{q}}{(2\pi)^2}  \frac{d^2\mathbf{p}}{(2\pi)^2}\Big[\big[b_{\perp}(\mathbf{p}), b_{\perp}(\mathbf{q}-\mathbf{p})\big],  b_{\eta}(\mathbf{k}-\mathbf{q})\Big]\frac{\mathcal{I}_3(\mathbf{p},\mathbf{q}, \mathbf{k})}{p^2_{\perp}|\mathbf{q}-\mathbf{p}|^2|\mathbf{k}-\mathbf{q}|^2}\,.\\
\end{split}
\end{equation}
Here $\mathcal{I}_{1,2,3}(\mathbf{p}, \mathbf{q}, \mathbf{k})$ are functions of the momentum only and do not depend on the target Wilson line or the projectile color charge density. The introduction of these three functions is because the angular integrations cannot be carried out analytically. We therefore denote the various parts containing the angular integrals by these auxilliary functions:
\begin{equation}
\begin{split}
&\mathcal{I}_1 (\mathbf{p}, \mathbf{q}, \mathbf{k})\\
= &\pi k_{\perp}(\mathbf{k}\times \mathbf{q})(\mathbf{q}\times\mathbf{p})\int_{-\pi}^{\pi}\frac{d\phi}{2\pi}\frac{|\mathbf{q}-\mathbf{p}|^2-p_{\perp}^2+w_{\perp}^2}{q_{\perp}^2-w_{\perp}^2}\\
&\qquad \times \Big(\frac{1}{ w_{\perp}}L_{011}(|\mathbf{k}-\mathbf{q}|, w_{\perp},k_{\perp})- \frac{1}{q_{\perp}}L_{011}(|\mathbf{k}-\mathbf{q}|, q_{\perp}, k_{\perp})\Big) \\
&+\frac{1}{4}\pi k_{\perp}(\mathbf{k}-\mathbf{q})\cdot\mathbf{p} \int_{-\pi}^{\pi} \frac{d\phi}{2\pi} \frac{|\mathbf{q}-\mathbf{p}|^2-p_{\perp}^2 + w_{\perp}^2}{w_{\perp}}  L_{011}(|\mathbf{k}-\mathbf{q}|, w_{\perp},k_{\perp}), \\
\end{split}
\end{equation}
\begin{equation}
\begin{split}
&\mathcal{I}_2(\mathbf{p}, \mathbf{q}, \mathbf{k}) \\
=&\frac{1}{2q_{\perp}^2}(\mathbf{k}\times \mathbf{q})(\mathbf{q}\times \mathbf{p})\pi |\mathbf{k}-\mathbf{q}|k_{\perp}\\
&\quad \times\int_{-\pi}^{\pi}\frac{d\phi}{2\pi} \left(1+\frac{2\mathbf{p}\cdot(\mathbf{p}-\mathbf{q})}{q_{\perp}^2-w_{\perp}^2}\right) \Big(L_{011}(w_{\perp}, |\mathbf{k}-\mathbf{q}|, k_{\perp})-  L_{011}(q_{\perp}, |\mathbf{k}-\mathbf{q}|, k_{\perp})\Big)\\
&+\frac{1}{8q_{\perp}^2}(k_{\perp}^2-|\mathbf{k}-\mathbf{q}|^2)(|\mathbf{q}-\mathbf{p}|^2-p_{\perp}^2)\pi |\mathbf{k}-\mathbf{q}|k_{\perp}\\
&\qquad \times \int_{-\pi}^{\pi} \frac{d\phi}{2\pi}\left(1 -\frac{|\mathbf{q}-\mathbf{p}|^2+p_{\perp}^2}{w_{\perp}^2}\right)\Big(L_{011}(w_{\perp}, |\mathbf{k}-\mathbf{q}|,k_{\perp}) -L_{011}(0,|\mathbf{k}-\mathbf{q}|, k_{\perp}) \Big), \\
\end{split}
\end{equation}
\begin{equation}
\begin{split}
&\mathcal{I}_3(\mathbf{p}, \mathbf{q}, \mathbf{k}) \\
=& \frac{1}{q_{\perp}^2} (\mathbf{k}\times \mathbf{q})(\mathbf{q}\times \mathbf{p}) (p_{\perp}^2+q_{\perp}^2-\mathbf{p}\cdot\mathbf{q}) \pi|\mathbf{k}-\mathbf{q}|k_{\perp}\\
&\qquad \times \int_{-\pi}^{\pi}\frac{d\phi}{2\pi}  \frac{1}{q_{\perp}^2-w_{\perp}^2}\Big(L_{011}(w_{\perp}, |\mathbf{k}-\mathbf{q}|, k_{\perp})- L_{011}(q_{\perp}, |\mathbf{k}-\mathbf{q}|, k_{\perp})\Big)\\
&+\frac{1}{4q_{\perp}^2} \left(k_{\perp}^2-|\mathbf{k}-\mathbf{q}|^2\right)\left(|\mathbf{q}-\mathbf{p}|^2-p_{\perp}^2\right)\mathbf{p}\cdot(\mathbf{q}-\mathbf{p})\pi |\mathbf{k}-\mathbf{q}| k_{\perp} \\
&\qquad\times\int_{-\pi}^{\pi} \frac{d\phi}{2\pi}\frac{1}{w_{\perp}^2} \Big(L_{011}(w_{\perp}, |\mathbf{k}-\mathbf{q}|, k_{\perp})-L_{011}(0, |\mathbf{k}-\mathbf{q}|, k_{\perp}) \Big), \\
\end{split}
\end{equation}
where $w_{\perp} =\sqrt{p_{\perp}^2+|\mathbf{q}-\mathbf{p}|^2-2p_{\perp}|\mathbf{q}-\mathbf{p}|\cos\phi}$. The definition of the function $L_{011}(a,b,c)$ can be found in the Appendix \ref{ap:integral_formulas_three_bessels}. It is expressed in terms of simple elementary functions although containing Heaviside step functions depending on the relative magnitude of $a, b, c$.  Superficially it  appear that  functions $\mathcal{I}_{1,2,3}(\mathbf{p}, \mathbf{q}, \mathbf{k})$ might have multiple singularities due to vanishing denominators in the integrands. We demonstrate in Appendix~\ref{ap:Is} that all singularities are removable.

\subsubsection*{The bulk term $\mathfrak{B}_{\perp}^{(5)}$}
For the transverse polarization, the corresponding order-$g^5$ source term has the expression
\begin{equation}\label{eq:S_perp_g5}
\begin{split}
S^{(5)}_{\perp}(\tau, \mathbf{k}) =&\frac{i\epsilon^{il}\mathbf{k}_l}{k_{\perp}} \tau^2 \int \frac{d^2\mathbf{q}}{(2\pi)^2}  (\mathbf{k}-2\mathbf{q})_i\left( [\tilde{\beta}^{(3)}(\tau, \mathbf{q}), \tilde{\beta}^{(1)}(\tau, \mathbf{k}-\mathbf{q})]+[\beta_j^{(3)}(\tau, \mathbf{q}), \beta_j^{(1)}(\tau, \mathbf{k}-\mathbf{q})] \right)\\
&\qquad+\left(-(2\mathbf{k}-\mathbf{q})_j [\beta_j^{(3)}(\tau, \mathbf{q}), \beta_i^{(1)}(\tau, \mathbf{k}-\mathbf{q})]+(\mathbf{k}+\mathbf{q})_j[\beta_i^{(3)}(\tau, \mathbf{q}), \beta_j^{(1)}(\tau, \mathbf{k}-\mathbf{q})] \right)\\
&\qquad+ \int \frac{d^2\mathbf{q}}{(2\pi)^2}  \frac{d^2\mathbf{p}}{(2\pi)^2} \Big( [\tilde{\beta}^{(1)}(\tau, \mathbf{k}-\mathbf{q}), [\beta_i^{(1)}(\tau, \mathbf{p}), \tilde{\beta}^{(1)}(\tau, \mathbf{q}-\mathbf{p})]] \\
&\qquad\qquad+ [\beta_j^{(1)}(\tau, \mathbf{k}-\mathbf{q}), [\beta_i^{(1)}(\tau, \mathbf{p}), \beta_j^{(1)}(\tau, \mathbf{q}-\mathbf{p})]]\Big)\,.
\end{split}
\end{equation}
Again, using the solutions $\beta^{(1)}(\tau, \mathbf{k})$, $\beta_i^{(1)}(\tau, \mathbf{k})$, $\beta^{(3)}(\tau, \mathbf{k})$, $\beta_i^{(3)}(\tau, \mathbf{k})$ given in Eqs.~\eqref{eq:sol_order_g}, \eqref{eq:sol_order_g3_first} \eqref{eq:sol_order_g3_second}, and \eqref{eq:sol_order_g3_third}, one obtains its explicit expression
\begin{equation}
\begin{split}
&S_{\perp}^{(5)}(\tau,\mathbf{k})\\
=&i\tau^2\frac{1}{k_{\perp}}\int \frac{d^2\mathbf{q}}{(2\pi)^2}  \frac{2(\mathbf{k}\times \mathbf{q})}{q_{\perp} |\mathbf{k}-\mathbf{q}|}\Big[2\beta^{(3)}(\tau=0,\mathbf{q}), b_{\eta}(\mathbf{k}-\mathbf{q})\Big]J_1(q_{\perp}\tau) J_1(|\mathbf{k}-\mathbf{q}|\tau)\\
&-\tau^2\frac{1}{k_{\perp}}\int \frac{d^2\mathbf{q}}{(2\pi)^2} \int \frac{d^2\mathbf{p}}{(2\pi)^2} \frac{2(\mathbf{k}\times \mathbf{q})(\mathbf{q}\times \mathbf{p})}{p_{\perp}^2|\mathbf{q}-\mathbf{p}|^2|\mathbf{k}-\mathbf{q}|}\Big[[b_{\perp}(\mathbf{p}), b_{\eta}(\mathbf{q}-\mathbf{p})], b_{\eta}(\mathbf{k}-\mathbf{q})\Big] \\
&\qquad\qquad \times\int_{-\pi}^{\pi}\frac{d\phi}{2\pi}\frac{\mathbf{q}\cdot(\mathbf{q}-2\mathbf{p})+w_{\perp}^2}{q_{\perp}^2-w_{\perp}^2}\left(\frac{J_1(w_{\perp}\tau)}{w_{\perp}}-\frac{J_1(q_{\perp}\tau)}{q_{\perp}}\right)J_1(|\mathbf{k}-\mathbf{q}|\tau)\\
&+i\tau^2\frac{1}{k_{\perp}}\int \frac{d^2\mathbf{q}}{(2\pi)^2}\frac{(k_{\perp}^2+q_{\perp}^2-\mathbf{k}\cdot\mathbf{q})(2\mathbf{k}\times \mathbf{q})}{q_{\perp} |\mathbf{k}-\mathbf{q}|^2} [\beta_{\perp}^{(3)}(\tau=0, \mathbf{q}),b_{\perp}(\mathbf{k}-\mathbf{q})]J_0(q_{\perp}\tau)J_0(|\mathbf{k}-\mathbf{q}|\tau)\\
& +\tau^2\frac{1}{k_{\perp}}\int \frac{d^2\mathbf{q}}{(2\pi)^2}\int \frac{d^2\mathbf{p}}{(2\pi)^2}\frac{(k_{\perp}^2+q_{\perp}^2-\mathbf{k}\cdot\mathbf{q})(\mathbf{k}\times \mathbf{q})( \mathbf{q}\times \mathbf{p})}{q^2_{\perp} |\mathbf{k}-\mathbf{q}|^2p^2_{\perp}|\mathbf{q}-\mathbf{p}|^2} \int_{-\pi}^{\pi}\frac{d\phi}{2\pi}  \frac{1}{q_{\perp}^2-w_{\perp}^2}\Big((w_{\perp}^2-p_{\perp}^2-|\mathbf{q}-\mathbf{p}|^2)\\
&\qquad\qquad\times\Big[[b_{\eta}(\mathbf{p}), b_{\eta}( \mathbf{q}-\mathbf{p})], b_{\perp}(\mathbf{k}-\mathbf{q})\Big]+2(\mathbf{p}\cdot\mathbf{q}-p_{\perp}^2-q_{\perp}^2) \Big[[b_{\perp}(\mathbf{p}), b_{\perp}(\mathbf{q}-\mathbf{p})], b_{\perp}(\mathbf{k}-\mathbf{q})\Big] \Big)\\
&\qquad \qquad\times (J_0(w_{\perp}\tau)-J_0(q_{\perp}\tau))J_0(|\mathbf{k}-\mathbf{q}|\tau)\\
&+\tau^2\frac{1}{k_{\perp}}\int \frac{d^2\mathbf{q}}{(2\pi)^2}\int\frac{d^2\mathbf{p}}{(2\pi)^2} \frac{(2\mathbf{k}-\mathbf{q})\cdot \mathbf{q}\,\mathbf{k}\cdot(\mathbf{k}-\mathbf{q}) \mathbf{q}\cdot(\mathbf{q}-2\mathbf{p})}{4p_{\perp}^2 |\mathbf{q}-\mathbf{p}|^2|\mathbf{k}-\mathbf{q}|^2q_{\perp}^2} \int_{-\pi}^{\pi} \frac{d\phi}{2\pi}\frac{1}{w_{\perp}^2}\Big((w_{\perp}^2-q_{\perp}^2+2\mathbf{p}\cdot(\mathbf{q}-\mathbf{p}))\\
& \qquad\qquad\times \left[[b_{\eta}(\mathbf{p}), b_{\eta}(\mathbf{q}-\mathbf{p})],b_{\perp}(\mathbf{k}-\mathbf{q})\right] + 2\mathbf{p}\cdot(\mathbf{q}-\mathbf{p})\Big[b_{\perp}(\mathbf{p}), b_{\perp}(\mathbf{q}-\mathbf{p})],b_{\perp}(\mathbf{k}-\mathbf{q})\Big]    \Big)\\
&\qquad\qquad \times (1- J_0(w_{\perp}\tau))J_0(|\mathbf{k}-\mathbf{q}|\tau)
\\
&+\tau^2\frac{1}{k_{\perp}}  \int \frac{d^2\mathbf{q}}{(2\pi)^2}  \frac{d^2\mathbf{p}}{(2\pi)^2}  \frac{\mathbf{p}\cdot \mathbf{k}}{|\mathbf{q}-\mathbf{p}||\mathbf{k}-\mathbf{q}| p_{\perp}^2} \Big[ b_{\eta}(\mathbf{k}-\mathbf{q}),[b_{\perp}(\mathbf{p}), b_{\eta}(\mathbf{q}-\mathbf{p})]\Big] \\
&\qquad\quad \times J_1(|\mathbf{q}-\mathbf{p}|\tau)J_0(p_{\perp}\tau)J_1(|\mathbf{k}-\mathbf{q}|\tau)\\
&-\tau^2\frac{1}{k_{\perp}}  \int \frac{d^2\mathbf{q}}{(2\pi)^2}  \frac{d^2\mathbf{p}}{(2\pi)^2} \frac{(\mathbf{k}-\mathbf{q})\cdot(\mathbf{q}-\mathbf{p}) \mathbf{p}\cdot \mathbf{k}}{|\mathbf{q}-\mathbf{p}|^2p_{\perp}^2 |\mathbf{k}-\mathbf{q}|^2} \Big[b_{\perp}(\mathbf{k}-\mathbf{q}), [b_{\perp}(\mathbf{p}), b_{\perp}(\mathbf{q}-\mathbf{p})]\Big]\\
&\qquad \quad\times J_0(|\mathbf{q}-\mathbf{p}|\tau)J_0(p_{\perp}\tau)J_0(|\mathbf{k}-\mathbf{q}|\tau).\\
\end{split}
\end{equation}
The source term is then used to compute the order-$g^5$ bulk term, contributing to single gluon production amplitude, by 
\begin{equation}
\mathfrak{B}_{\perp}^{(5)}(\mathbf{p})=- i\sqrt{\frac{\pi}{4}} \int_0^{\infty} d\tau \frac{1}{\tau} H_0^{(2)}(p_{\perp}\tau) S^{(5)}_{\perp}(\tau, \mathbf{p}).
\end{equation}
We skip the details of doing the proper time integral and present the final result here
\begin{equation}
\begin{split}
\mathfrak{B}_{\perp}^{(5)}(\mathbf{k})
=&\frac{1}{\sqrt{\pi}k_{\perp}}\int \frac{d^2\mathbf{q}}{(2\pi)^2}  \Big[2\beta^{(3)}(\tau=0,\mathbf{q}), b_{\eta}(\mathbf{k}-\mathbf{q})\Big]\frac{\mathbf{k}\times \mathbf{q}}{q_{\perp}^2 |\mathbf{k}-\mathbf{q}|^2}\left(\frac{\mathbf{q}\cdot(\mathbf{q}-\mathbf{k})}{|\mathbf{q}\times \mathbf{k}|} -i\right)\\
&+\frac{1}{\sqrt{\pi}k_{\perp}}\int \frac{d^2\mathbf{q}}{(2\pi)^2}\left[q_{\perp}\beta_{\perp}^{(3)}(\tau=0, \mathbf{q}),b_{\perp}(\mathbf{k}-\mathbf{q})\right]\frac{(\mathbf{k}\times \mathbf{q})(k_{\perp}^2+q_{\perp}^2-\mathbf{k}\cdot\mathbf{q})}{q^2_{\perp} |\mathbf{k}-\mathbf{q}|^2|\mathbf{k}\times \mathbf{q}|} \\
&+\frac{i}{\sqrt{\pi}k_{\perp}}\int \frac{d^2\mathbf{q}}{(2\pi)^2}\frac{d^2\mathbf{p}}{(2\pi)^2}\Big[\left[b_{\perp}(\mathbf{p}), b_{\eta}(\mathbf{q}-\mathbf{p})\right], b_{\eta}(\mathbf{k}-\mathbf{q})\Big] \frac{\mathcal{G}_1(\mathbf{p}, \mathbf{q}, \mathbf{k})}{p_{\perp}^2|\mathbf{q}-\mathbf{p}|^2|\mathbf{k}-\mathbf{q}|^2} \\
&+i\frac{1}{\sqrt{\pi}k_{\perp}}\int \frac{d^2\mathbf{q}}{(2\pi)^2}\frac{d^2\mathbf{p}}{(2\pi)^2}\Big[[b_{\eta}(\mathbf{p}), b_{\eta}(\mathbf{q}-\mathbf{p})],b_{\perp}(\mathbf{k}-\mathbf{q})\Big] \frac{\mathcal{G}_2(\mathbf{p}, \mathbf{q}, \mathbf{k})}{p_{\perp}^2 |\mathbf{q}-\mathbf{p}|^2|\mathbf{k}-\mathbf{q}|^2} \\
&+i\frac{1}{\sqrt{\pi}k_{\perp}}\int \frac{d^2\mathbf{q}}{(2\pi)^2}\frac{d^2\mathbf{p}}{(2\pi)^2}\Big[[b_{\perp}(\mathbf{p}), b_{\perp}(\mathbf{q}-\mathbf{p})],b_{\perp}(\mathbf{k}-\mathbf{q})\Big]    \frac{\mathcal{G}_3(\mathbf{p}, \mathbf{q}, \mathbf{k})}{p_{\perp}^2 |\mathbf{q}-\mathbf{p}|^2|\mathbf{k}-\mathbf{q}|^2}.\\
\end{split}
\end{equation}
%Evaluating this expression required computing  integrals involving a product of three and four Bessel functions, see Appendix~\ref{App:BInt}.
%This expression was organized according to the color structure.  To simplify the notation
Similar to the analysis for $\mathfrak{B}_{\eta}^{(5)}(\mathbf{k})$ in the previous section,  we introduced  
another three  auxilliary functions $\mathcal{G}_{1,2,3}(\mathbf{p}, \mathbf{q}, \mathbf{k})$ defined as 
\begin{equation}
\begin{split}
&\mathcal{G}_1(\mathbf{p}, \mathbf{q}, \mathbf{k})\\
=& \pi |\mathbf{k}-\mathbf{q}|(\mathbf{k}\times \mathbf{q})(\mathbf{q}\times \mathbf{p})\int_{-\pi}^{\pi}\frac{d\phi}{2\pi}\frac{\mathbf{q}\cdot(\mathbf{q}-2\mathbf{p})+w_{\perp}^2}{q_{\perp}^2-w_{\perp}^2}\left(\frac{1}{w_{\perp}}L_{110}(w_{\perp}, |\mathbf{k}-\mathbf{q}|, k_{\perp})-\frac{1}{q_{\perp}}L_{110}(q_{\perp}, |\mathbf{k}-\mathbf{q}|, k_{\perp})\right)\\
&+\pi |\mathbf{k}-\mathbf{q}|(\mathbf{p}\cdot \mathbf{k})\int_{-\pi}^{\pi}\frac{d \phi}{2\pi}\frac{|\mathbf{q}-\mathbf{p}|^2-p_{\perp}^2 + w_{\perp}^2}{4w_{\perp}}L_{110}(w_{\perp}, |\mathbf{k}-\mathbf{q}|, k_{\perp}),\\
%=& (\mathbf{k}\times \mathbf{q})(\mathbf{q}\times \mathbf{p})\int_{-\pi}^{\pi}\frac{d\phi}{2\pi}\frac{\mathbf{q}\cdot(\mathbf{q}-2\mathbf{p})+w_{\perp}^2}{q_{\perp}^2-w_{\perp}^2}\Big(\frac{1}{w^2_{\perp}}\Big((w_{\perp}^2-2\mathbf{k}\cdot \mathbf{q} + q_{\perp}^2)\tilde{L}_{000}(w_{\perp}, |\mathbf{k}-\mathbf{q}|, k_{\perp}) -i\Big)\\
%&\qquad \qquad-\frac{1}{q^2_{\perp}}\left((2\mathbf{q}\cdot(\mathbf{q}-\mathbf{k})) \tilde{L}_{000}(q_{\perp}, |\mathbf{k}-\mathbf{q}|, k_{\perp}) -i\right)\Big)\\
%&+(\mathbf{p}\cdot \mathbf{k})\int_{-\pi}^{\pi}\frac{d \phi}{2\pi}\frac{|\mathbf{q}-\mathbf{p}|^2-p_{\perp}^2 + w_{\perp}^2}{4w^2_{\perp}}\Big((w_{\perp}^2+|\mathbf{k}-\mathbf{q}|^2 - k_{\perp}^2)\tilde{L}_{000}(w_{\perp}, |\mathbf{k}-\mathbf{q}|, k_{\perp}) -i\Big)\,.\\
\end{split}
\end{equation}
%Here we used (see Appendix~\ref{ap:Ls})
%\begin{equation}
%L_{110}(a, b, c) = \frac{1}{\pi a b} \left( (a^2+b^2-c^2)\tilde{L}_{000}(a,b,c) -i \right)
%\end{equation}
%with $\tilde{L}_{000}(a,b,c) = \frac{\pi}{2} L_{000}(a,b,c)$. 
\begin{equation}
\begin{split}
\mathcal{G}_2(\mathbf{p}, \mathbf{q}, \mathbf{k}) =&\frac{1}{q_{\perp}^2}(k_{\perp}^2+q_{\perp}^2-\mathbf{k}\cdot\mathbf{q})(\mathbf{k}\times \mathbf{q})( \mathbf{q}\times \mathbf{p})\frac{\pi}{2}\int_{-\pi}^{\pi}\frac{d\phi}{2\pi}  \frac{1}{q_{\perp}^2-w_{\perp}^2}(-w_{\perp}^2+p_{\perp}^2+|\mathbf{q}-\mathbf{p}|^2)\\
&\times\Big(L_{000}(w_{\perp}, |\mathbf{k}-\mathbf{q}|, k_{\perp}) -L_{000}(q_{\perp}, |\mathbf{k}-\mathbf{q}|, k_{\perp})\Big) \\
&+\frac{1}{4q_{\perp}^2}(2\mathbf{k}-\mathbf{q})\cdot \mathbf{q}\,\mathbf{k}\cdot(\mathbf{k}-\mathbf{q}) \mathbf{q}\cdot(\mathbf{q}-2\mathbf{p})\frac{\pi}{2}\int_{-\pi}^{\pi} \frac{d\phi}{2\pi}\frac{1}{w_{\perp}^2}(w_{\perp}^2-q_{\perp}^2+2\mathbf{p}\cdot(\mathbf{q}-\mathbf{p}))\\
&\times \Big(L_{000}(w_{\perp}, |\mathbf{k}-\mathbf{q}|,k_{\perp}-L_{000}(0, |\mathbf{k}-\mathbf{q}|, k_{\perp})\Big)\,, 
\\
\end{split}
\end{equation}
\begin{equation}
\begin{split}
\mathcal{G}_3(\mathbf{p}, \mathbf{q}, \mathbf{k})=&\frac{\pi }{q_{\perp}^2}(k_{\perp}^2+q_{\perp}^2-\mathbf{k}\cdot\mathbf{q})(\mathbf{k}\times \mathbf{q})( \mathbf{q}\times \mathbf{p})(-\mathbf{p}\cdot\mathbf{q}+p_{\perp}^2+q_{\perp}^2)\\
&\qquad\qquad \times \int_{-\pi}^{\pi}\frac{d\phi}{2\pi}  \frac{1}{q_{\perp}^2-w_{\perp}^2} \Big(L_{000}(w_{\perp}, |\mathbf{k}-\mathbf{q}|, k_{\perp}) - L_{000}(q_{\perp}, |\mathbf{k}-\mathbf{q}|, k_{\perp})\Big) \\
&+ \frac{\pi}{4q_{\perp}^2}(2\mathbf{k}-\mathbf{q})\cdot \mathbf{q}\,\mathbf{k}\cdot(\mathbf{k}-\mathbf{q}) \mathbf{q}\cdot(\mathbf{q}-2\mathbf{p})\mathbf{p}\cdot(\mathbf{q}-\mathbf{p}) \\
&\qquad\qquad \times \int_{-\pi}^{\pi} \frac{d\phi}{2\pi}\frac{1}{w_{\perp}^2} \Big( L_{000}(w_{\perp}, |\mathbf{k}-\mathbf{q}|,k_{\perp}-L_{000}(0, |\mathbf{k}-\mathbf{q}|, k_{\perp})\Big) 
\\
&-\frac{1}{2} (\mathbf{k}\times\mathbf{q})(\mathbf{q}\times\mathbf{p})\frac{\pi}{2}\int_{-\pi}^{\pi} \frac{d\phi}{2\pi}  L_{000}(w_{\perp}, |\mathbf{k}-\mathbf{q}|, k_{\perp}).\\
\end{split}
\end{equation}
The explicit expression of functions $L_{000}(a,b,c)$ and $L_{110}(a,b,c)$ are given  in Appendix \ref{ap:integral_formulas_three_bessels}. The analytic structure of these functions are studied in Appendix~\ref{ap:Is}. Althought it is impossible to obtain  closed form expressions for these auxilliary functions, after taking care of the removable singularities, they are easily evaluated numerically for given momenta $\mathbf{p}, \mathbf{q}, \mathbf{k}$. \\

%%%%%%%%%%%%%%%%%%%%%%%%%%%%%%%%%%%%%%%%%%%%%%%%%%%%%%%%%%%
%%%%%%%%%%%%%%%%%%%%%%%%%%%%%%%%%%%%%%%%%%%%%%%%%%%%%%%%%%%
%%%%%%%%%%%%%%%%%%%%%%%%%%%%%%%%%%%%%%%%%%%%%%%%%%%%%%%%%%%
%%%%%%%%%%%%%%%%%%%%%%%%%%%%%%%%%%%%%%%%%%%%%%%%%%%%%%%%%%%
\section{Calculating the Single Inclusive Gluon Production}
\label{Sect:CalcSI}
We are now ready to proceed with evaluating the single gluon production up to order  $g^{6}$.  We will eventually express the gluon production in terms of the proton color charge density $\rho_P^a(\mathbf{k})$ and the target Wilson line $U^{ab}(\mathbf{k})$. 
The following explicit expressions will be repeatedly used. We list them for reference.
\begin{equation}\label{eqs:betaperp_rhoU}
\begin{split}
&b^c_{\eta}(\mathbf{k})
=\int\frac{d^2\mathbf{p}}{(2\pi)^2} \frac{(\mathbf{k}-\mathbf{p})\cdot \mathbf{p}}{p_{\perp}^2} \rho_P^a(\mathbf{p}) U^{ac}(\mathbf{k}-\mathbf{p}),\\
& b^c_{\perp}(\mathbf{k})  
= \int \frac{d^2\mathbf{p}}{(2\pi)^2} \frac{\mathbf{k}\times \mathbf{p}}{ p_{\perp}^2}\rho_P^a(\mathbf{p}) U^{ac}(\mathbf{k}-\mathbf{p})\,,\\
&b_{\|}^{c}(\mathbf{k}) = \int \frac{d^2\mathbf{p}}{(2\pi)^2} \frac{-\mathbf{k}\cdot\mathbf{p}}{k_{\perp}^2p_{\perp}^2} \rho_P^a(\mathbf{p})U^{ac}(\mathbf{k}-\mathbf{p})\,.\\
\end{split}
\end{equation}
Further more, the eikonally color rotated order-$g^3$ and order-$g^5$ Weiszacker-Williams fields are 
\begin{equation}
\begin{split}
&\zeta_{(3)}^{c,l}(\mathbf{q}) = \int \frac{d^2\mathbf{p}}{(2\pi)^2} \alpha_{P, (3)}^{a,l}(\mathbf{p})U^{ac}(\mathbf{q}-\mathbf{p}),\\
&\zeta_{(5)}^{c,l}(\mathbf{k}) = \int \frac{d^2\mathbf{q}}{(2\pi)^2} \alpha_{P, (5)}^{a,l}(\mathbf{q})U^{ac}(\mathbf{k}-\mathbf{q})\\
\end{split}
\end{equation}
with 
\begin{equation}
\begin{split}
\alpha_{P,(3)}^{a,l}(\mathbf{p}) =&\frac{i}{2}\left(\delta^{lj}-\frac{\mathbf{p}^l\mathbf{p}^j}{p_{\perp}^2} \right)\int \frac{d^2\mathbf{p}_1}{(2\pi)^2}\frac{-1}{|\mathbf{p}-\mathbf{p}_1|^2}\frac{i\mathbf{p}_1^j}{p_{1}^2} \Big[\rho_P(\mathbf{p}-\mathbf{p}_1), \rho_P(\mathbf{p}_1)\Big]\\
=&\frac{1}{2}\left(\delta^{lj}-\frac{\mathbf{p}^l\mathbf{p}^j}{p_{\perp}^2} \right)\int \frac{d^2\mathbf{p}_1}{(2\pi)^2}\frac{\mathbf{p}_1^j}{|\mathbf{p}-\mathbf{p}_1|^2p_{1}^2} \Big[\rho_P(\mathbf{p}-\mathbf{p}_1), \rho_P(\mathbf{p}_1)\Big], \\
\end{split}
\end{equation}
\begin{equation}
\begin{split}
\alpha_{P,(5)}^{a,l}(\mathbf{q})=&\left(\delta^{ij}-\frac{\mathbf{q}^i\mathbf{q}^j}{q_{\perp}^2} \right)\Big(\frac{1}{2} \int \frac{d^2\mathbf{p}}{(2\pi)^2} \frac{d^2\mathbf{p}_1}{(2\pi)^2}\frac{1}{p_{\perp}^2p_1^2}\frac{i(\mathbf{q}-\mathbf{p})^j}{|\mathbf{q}-\mathbf{p}|^2} [[\rho_P(\mathbf{p}_1), \rho_P(\mathbf{p}-\mathbf{p}_1)], \rho_P(\mathbf{q}-\mathbf{p})]\\
&+\frac{1}{6}\int \frac{d^2\mathbf{p}}{(2\pi)^2} \frac{d^2\mathbf{p}_1}{(2\pi)^2} \frac{i(\mathbf{p}-\mathbf{p}_1)^j}{|\mathbf{p}-\mathbf{p}_1|^2} \frac{1}{p_1^2|\mathbf{q}-\mathbf{p}|^2}[[\rho_P(\mathbf{p}_1), \rho_P(\mathbf{p}-\mathbf{p}_1)], \rho_P(\mathbf{q}-\mathbf{p})]\Big) .
\end{split}
\end{equation}
%%%%%%%%%%%%%%%%%%%%%%%%%%%%%%%%%%%%%%%%%%
%%%%%%%%%%%%%%%%%%%%%%%%%%%%%%%%%%%%%%%%%%
%%%%%%%%%%%%%%%%%%%%%%%%%%%%%%%%%%%%%%%%%%
%%%%%%%%%%%%%%%%%%%%%%%%%%%%%%%%%%%%%%%%%%
%%%%%%%%%%%%%%%%%%%%%%%%%%%%%%%%%%%%%%%%%%
%%%%%%%%%%%%%%%%%%%%%%%%%%%%%%%%%%%%%%%%%%

\subsection{Leading order  gluon production}
\label{Sect:LOSI}

The leading order gluon production is defined by
\begin{equation}
n^{(2)}(\mathbf{k})= |\mathfrak{S}_{\eta}^{(1)}(\mathbf{k})|^2+|\mathfrak{S}_{\perp}^{(1)}(\mathbf{k})|^2\,.
\end{equation}
Using the explicit forms for
\begin{equation}\label{eq:surface_one_eta}
\mathfrak{S}_{\eta}^{(1)}(\mathbf{k})  = \frac{1}{\sqrt{\pi}k_{\perp}} b_{\eta}(\mathbf{k}) =\frac{1}{\sqrt{\pi}k_{\perp}}  \int\frac{d^2\mathbf{p}}{(2\pi)^2} \frac{(\mathbf{k}-\mathbf{p})\cdot \mathbf{p}}{p_{\perp}^2} \rho^a(\mathbf{p}) U^{ac}(\mathbf{k}-\mathbf{p}) T^c,
\end{equation}
\begin{equation}\label{eq:surface_one_perp}
\mathfrak{S}_{\perp}^{(1)}(\mathbf{k})  =\frac{1}{\sqrt{\pi}k_{\perp}} b_{\perp}(\mathbf{k}) =\frac{1}{\sqrt{\pi}k_{\perp}} \int \frac{d^2\mathbf{p}}{(2\pi)^2} \frac{\mathbf{k}\times \mathbf{p}}{ p_{\perp}^2}\rho^a(\mathbf{p}) U^{ac}(\mathbf{k}-\mathbf{p}) T^c
\end{equation}
we obtain that at the leading order
\begin{equation}
\begin{split}
&\frac{dN_{\rm LO}}{d^2\mathbf{k} dy} =  \frac{1}{(2\pi)^2} g^2 n^{(2)}(\mathbf{k})\\
=&\frac{g^2}{(2\pi)^3} \frac{1}{\pi k_{\perp}^2} (\delta^{ij}\delta^{mn}+\epsilon^{ij}\epsilon^{mn}) \int\frac{d^2\mathbf{p}}{(2\pi)^2} \frac{d^2\mathbf{q}}{(2\pi)^2} \frac{\mathbf{p}_i (\mathbf{k}-\mathbf{p})_j}{\mathbf{p}^2} \frac{\mathbf{q}_m (\mathbf{k}-\mathbf{q})_n}{\mathbf{q}^2}\\
&\times \rho^a(\mathbf{p}) \left[U(\mathbf{k}-\mathbf{p}) U^{\dagger}(\mathbf{k}-\mathbf{q})\right]_{ad} \rho^d(-\mathbf{q})\,.
\end{split}
\end{equation}
This result reproduces the dilute-dense  single gluon spectrum calculated in Refs.~\cite{Dumitru:2001ux,Blaizot:2004wu,Blaizot:2008yb}.
Note that this expression is given in the functional form with explicit dependence on $\rho_P$ and $\rho_T$.

\begin{figure}[t]
\centering % \begin{center}/\end{center} takes some additional vertical space
\includegraphics[width = 0.8\textwidth]{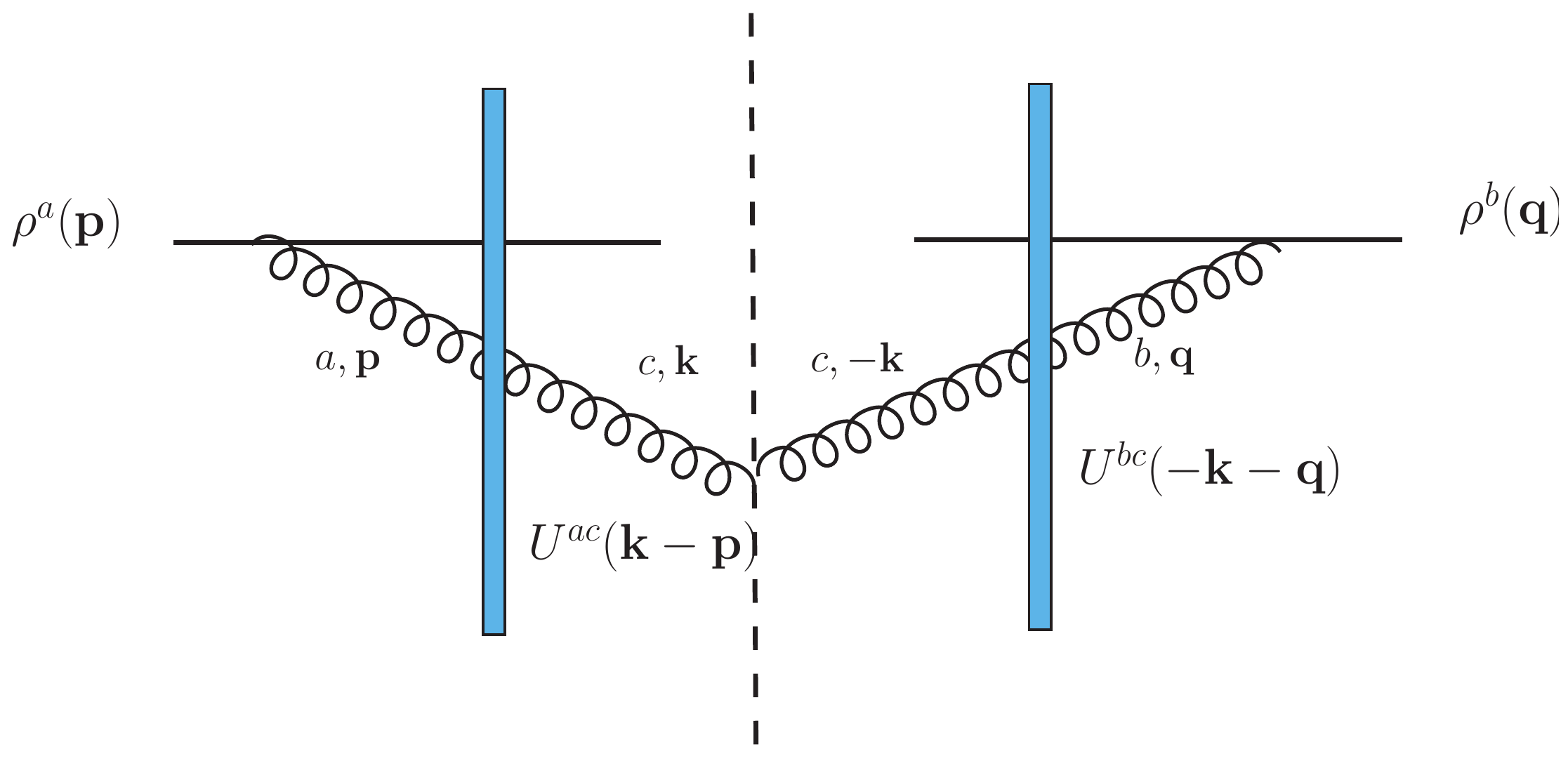}
\caption{Color structure for $|M_{(1)}|^2$. Two gluons lines cross the target nucleus indicating the two Wilson lines. The dashed line separates the amplitude on the left and the complex conjugate amplitude on the right. }
\label{fig:M1_squared}
\end{figure}

For comparison with high order corrections, we use the \textit{Lipatov vertex} \cite{Fadin:1975cb,Kuraev:1977fs, Kovchegov:2012mbw}
\begin{equation}
L^i(\mathbf{p},\mathbf{k}) = \left(\frac{\mathbf{p}^i}{p_{\perp}^2} - \frac{\mathbf{k}^i}{k_{\perp}^2}\right).
\end{equation}
Then the leading order gluon production can be expressed in a more informative form 
\begin{equation}
n^{(2)}(\mathbf{k})= - \frac{1}{\pi} \int\frac{d^2\mathbf{p}}{(2\pi)^2} \frac{d^2\mathbf{q}}{(2\pi)^2} L^i(\mathbf{p}, \mathbf{k}) L^i(\mathbf{q}, -\mathbf{k}) \rho^a(\mathbf{p})\rho^b(\mathbf{q})U^{ac}(\mathbf{k}-\mathbf{p})U^{bc}(-\mathbf{k}-\mathbf{q}). 
\end{equation}
The color structure can be shown by Fig.~\ref{fig:M1_squared}. The coefficient function is just the Lipatov vertex squared.\\
%In order to compare with Ref.~\cite{Kovchegov:1998bi}, we also perform the average over the proton color charge density using the MV model. In the momentum space, the MV models leads to the following correlator:
% \begin{equation}
%\langle \rho^a(\mathbf{p}) \rho^d(\mathbf{q})\rangle = \delta^{ad}(2\pi)^2\delta^{(2)}(\mathbf{p}+\mathbf{q}) \mu_P^2,
%\end{equation}
%which helps us to simplify the leading order single gluon spectrum to
%\begin{equation}
%\begin{split}
%&\Big\langle \frac{dN}{d^2\mathbf{k}}(\rho_P, \rho_T)\Big\rangle_{\rho_P}=\frac{g^2\mu_P^2}{(2\pi)^2} \frac{1}{\pi k_{\perp}^2}  \int\frac{d^2\mathbf{p}}{(2\pi)^2} \frac{|\mathbf{k}-\mathbf{p}|^2}{p_{\perp}^2}\mathrm{Tr} \left[U(\mathbf{k}-\mathbf{p}) U^{T}(-\mathbf{k}+\mathbf{p})\right].
%\end{split}
%\end{equation}

%%%%%%%%%%%%%%%%%%%%%%%%%%%%%%%%%%%%%%%%%%%%%%%%%
%%%%%%%%%%%%%%%%%%%%%%%%%%%%%%%%%%%%%%%%%%%%%%%%%%%
%%%%%%%%%%%%%%%%%%%%%%%%%%%%%%%%%%%%%%%%%%%%%%%%%
%%%%%%%%%%%%%%%%%%%%%%%%%%%%%%%%%%%%%%%%%%%%%%%%%%%
\subsection{Next-to-leading order gluon production}
\label{Sect:NLOSI}
The order-$g^4$ contribution to the single gluon production is defined by
\begin{equation}
n^{(4)}(\mathbf{k})=\sum_{\gamma = \eta, \perp}\Big[\mathfrak{S}_{\gamma}^{(1)\ast}(\mathbf{k})\left( \mathfrak{S}_{\gamma}^{(3)}(\mathbf{k})+\mathfrak{B}_{\gamma}^{(3)}(\mathbf{k})\right)+\mathfrak{S}_{\gamma}^{(1)}(\mathbf{k}) \left(\mathfrak{S}_{\gamma}^{(3)\ast}(\mathbf{k}) +  \mathfrak{B}_{\gamma}^{(3)\ast}(\mathbf{k})\right)\Big]\,.
\end{equation}
The surface terms $\mathfrak{S}_{\perp}^{(1)}(\mathbf{k})$, $\mathfrak{S}_{\eta}^{(1)}(\mathbf{k})$, $\mathfrak{S}_{\perp}^{(3)}(\mathbf{k})$ and $\mathfrak{S}_{\eta}^{(3)}(\mathbf{k})$   have been calculated in Eqs.~\eqref{eq:surface_terms_g1}, \eqref{eq:surface_eta_g3}, and \eqref{eq:surface_perp_g3}. The  bulk terms at order-$g^3$ are given in Eqs.~\eqref{Eq:Bulk3}.

Collecting the results for the surface and bulk terms at order-$g^3$, we have
\begin{equation}
\begin{split}
&\mathfrak{S}_{\eta}^{(3)}(\mathbf{k}) + \mathfrak{B}_{\eta}^{(3)}(\mathbf{k}) \\
= & \frac{-i\mathbf{k}^i }{\sqrt{\pi}k_{\perp}}\zeta_{(3)}^i(\mathbf{k})+\frac{ig}{\sqrt{\pi}k_{\perp}} \int \frac{d^2\mathbf{p}}{(2\pi)}\left[ b_{\eta}(\mathbf{p}), b_{\|}(\mathbf{k}-\mathbf{p})\right]\\
&+\frac{1}{\sqrt{\pi} k_{\perp}} \int \frac{d^2\mathbf{p}}{(2\pi)^2}  \frac{\mathbf{k}\times \mathbf{p}}{p_{\perp}^2 |\mathbf{k}-\mathbf{p}|^2} \left( \frac{\mathbf{k}\cdot(\mathbf{k}-\mathbf{p})}{|\mathbf{k}\times \mathbf{p}|} + i\right) [b_{\perp}(\mathbf{p}), b_{\eta}(\mathbf{k}-\mathbf{p})], \\
\end{split}
\end{equation}
and
\begin{equation}
\begin{split}
&\mathfrak{S}_{\perp}^{(3)}(\mathbf{k}) + \mathfrak{B}_{\perp}^{(3)}(\mathbf{k}) \\
=&\frac{i\epsilon^{ih}\mathbf{k}^h }{\sqrt{\pi}k_{\perp}} \zeta_{(3)}^i(\mathbf{k}) +\frac{ig}{\sqrt{\pi}k_{\perp}} \int\frac{d^2\mathbf{p}}{(2\pi)^2} \frac{\mathbf{k}\cdot\mathbf{p}}{p_{\perp}^2} [b_{\perp}(\mathbf{p}), b_{\|}(\mathbf{k}-\mathbf{p})] \\
&-\frac{1}{2} \frac{ig}{\sqrt{\pi}k_{\perp}} \int\frac{d^2\mathbf{p}}{(2\pi)^2} (\mathbf{k}\times \mathbf{p}) [b_{\|}(\mathbf{p}), b_{\|}(\mathbf{k}-\mathbf{p})]\\
&-\frac{1}{2\sqrt{\pi} k_{\perp}} \int \frac{d^2\mathbf{p}}{(2\pi)^2} \frac{\mathbf{p}\times \mathbf{k}}{p_{\perp}^2|\mathbf{k}-\mathbf{p}|^2}\left(\frac{\mathbf{p}\cdot(\mathbf{p}-\mathbf{k})}{|\mathbf{p}\times \mathbf{k}|} - i\right) [b_{\eta}(\mathbf{p}), b_{\eta}(\mathbf{k}-\mathbf{p})]\\
&+ \frac{(\mathbf{p}\cdot(\mathbf{k}-\mathbf{p}) -k_{\perp}^2)}{p_{\perp}^2|\mathbf{k}-\mathbf{p}|^2} \frac{\mathbf{k}\times \mathbf{p}}{|\mathbf{p}\times \mathbf{k}|} [b_{\perp}(\mathbf{p}), b_{\perp}(\mathbf{k}-\mathbf{p})]\,.
\end{split}
\end{equation}

\begin{figure}
\centering
\begin{subfigure}[t]{0.75\textwidth}
   \includegraphics[width=1\linewidth]{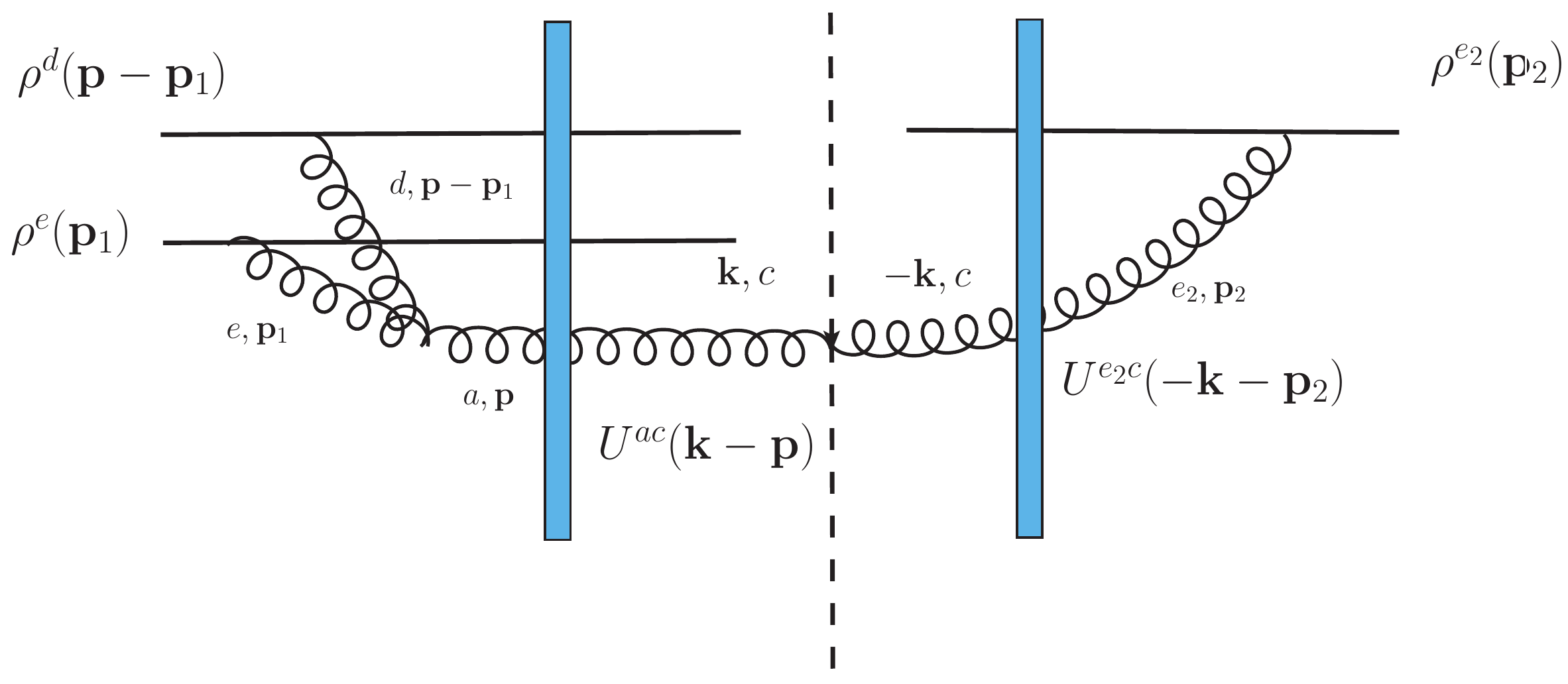}
   \caption{}
   \label{fig:M1M3_2wl} 
\end{subfigure}
\begin{subfigure}[t]{0.75\textwidth}
   \includegraphics[width=1\linewidth]{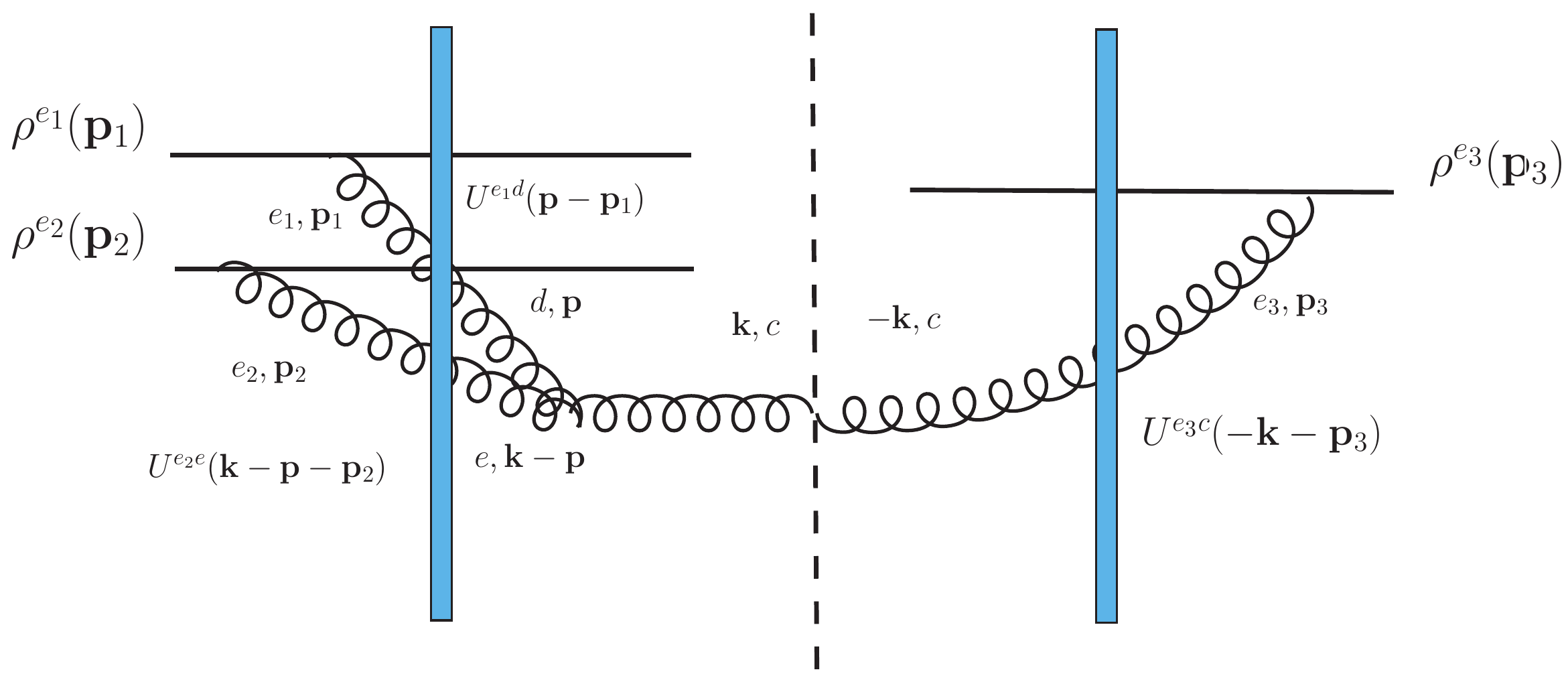}
   \caption{}
   \label{fig:M1M3_3wl}
\end{subfigure}
\caption{Color structure for terms in the $M^{\ast}_{(1)}M_{(3)}$ that contain two  and three Wilson lines: (a) two gluon lines cross the target nucleus and (b) three gluon lines cross the target nucleus. The dashed line separates the amplitude on the left and the complex conjugate amplitude on the right.  }
\label{fig:M1M3}
\end{figure}

From these surface terms and bulk terms, after some tediou algebra, substituting eqs \eqref{eqs:betaperp_rhoU}, one obtains the next-to-leading order single inclusive gluon production
\begin{equation}\label{eq:NLO_single_gluon_production}
\begin{split}
n^{(4)}(\mathbf{k}) 
=&\frac{1}{\pi }\int_{\mathbf{p}, \mathbf{p}_1, \mathbf{p}_2}\mathcal{E}_1(\mathbf{p}, \mathbf{p}_1, \mathbf{p}_2, \mathbf{k})  f^{ade} \rho_{P}^d(\mathbf{p}-\mathbf{p}_1) \rho_P^e(\mathbf{p}_1)\rho_P^{e_2}(\mathbf{p}_2)U^{ac}(\mathbf{k}-\mathbf{p})U^{e_2 c}(-\mathbf{k}-\mathbf{p}_2)\\
&+\frac{1}{\pi } \int_{\mathbf{p}, \mathbf{p}_1, \mathbf{p}_2, \mathbf{p}_3}\mathcal{E}_2(\mathbf{p}, \mathbf{p}_1, \mathbf{p}_2, \mathbf{p}_3, \mathbf{k}) f^{cde} \rho_P^{e_1}(\mathbf{p}_1)  \rho_P^{e_2}(\mathbf{p}_2)\rho_P^{e_3}(\mathbf{p}_3)\\
&\qquad \times U^{e_1 d}(\mathbf{p}-\mathbf{p}_1)  U^{e_2 e}(\mathbf{k}-\mathbf{p}-\mathbf{p}_2) U^{e_3 c}(-\mathbf{k}-\mathbf{p}_3)\\
&+c.c.
\end{split}
\end{equation}
Here $c.c.$ represents the complex conjugate terms and we have introduced the shorthand notation $\int_{\mathbf{p}} = \int \frac{d^2\mathbf{p}}{(2\pi)^2}$. The color structure (functional dependence on $\rho_P$ and $U$) are correctly captured by the two diagrams in Fig.~\ref{fig:M1M3}. The kinematic  function $\mathcal{E}_1$ has the expression
\begin{equation}\label{eq:coefficientfunc_E1}
\mathcal{E}_1(\mathbf{p}, \mathbf{p}_1, \mathbf{p}_2, \mathbf{k}) 
=\frac{1}{2}\frac{(\mathbf{p}_1\times \mathbf{p})}{|\mathbf{p}-\mathbf{p}_1|^2 p_1^2}\left(\frac{(\mathbf{p}\times \mathbf{p}_2)}{p_{\perp}^2 p_2^2} +\frac{ (\mathbf{p}\times \mathbf{k}) }{p_{\perp}^2 k^2_{\perp}}\right).
\end{equation}
%where we have used the identity
%$(\mathbf{k}\cdot\mathbf{p})(\mathbf{k}\times \mathbf{p}_2) - (\mathbf{k}\times \mathbf{p})(\mathbf{k}\cdot\mathbf{p}_2) = (\mathbf{p}\times \mathbf{p}_2)k_{\perp}^2$.
One immediately sees that the complex conjugate amplitude is represented by the Lipatov vertex
\begin{equation}
L^j(\mathbf{p}_2, -\mathbf{k}) = \left(\frac{\mathbf{p}_2^j}{p_2^2} + \frac{\mathbf{k}^j}{k_{\perp}^2}\right). 
\end{equation}
We thus introduce an effective vertex at order-$g^3$, 
\begin{equation}\label{eq:effective_vertex_g3_first}
\Gamma^j_{1} (\mathbf{p}, \mathbf{p}_1)  = \frac{1}{2}\frac{(\mathbf{p}_1\times \mathbf{p})}{|\mathbf{p}-\mathbf{p}_1|^2 p_1^2} \frac{\epsilon^{ij} \mathbf{p}^i}{p_{\perp}^2}.
\end{equation}
The coefficient function in eq. \eqref{eq:coefficientfunc_E1} follows simply as the product of the two effective vertices. 
\begin{equation}
\mathcal{E}_1(\mathbf{p}, \mathbf{p}_1, \mathbf{p}_2, \mathbf{k})  = \Gamma^j_{1} (\mathbf{p}_1, \mathbf{p})L^j(\mathbf{p}_2, -\mathbf{k}) 
\end{equation}
See Fig.~\ref{fig:M1M3_2wl}  for the momentum assignments and the color structure.

The second kinematic function in eq. \eqref{eq:NLO_single_gluon_production} is
\begin{equation}
\begin{split}
&\mathcal{E}_2(\mathbf{p}, \mathbf{p}_1, \mathbf{p}_2, \mathbf{p}_3, \mathbf{k})  \\
=&-\frac{(\mathbf{p}-\mathbf{p}_1)\cdot\mathbf{p}_1}{p_1^2} \frac{(\mathbf{k}-\mathbf{p})\cdot\mathbf{p}_2}{|\mathbf{k}-\mathbf{p}|^2 p_2^2} \frac{(\mathbf{k}+\mathbf{p}_3)\cdot\mathbf{p}_3}{ k^2_{\perp}p_3^2}\\
&-\frac{\mathbf{p} \times \mathbf{p}_1}{p_1^2}\frac{(\mathbf{k}-\mathbf{p}-\mathbf{p}_2)\cdot\mathbf{p}_2}{p_2^2} \frac{(\mathbf{k}\times \mathbf{p})}{p_{\perp}^2 |\mathbf{k}-\mathbf{p}|^2}\left(i \frac{\mathbf{k}\cdot(\mathbf{k}-\mathbf{p})}{|\mathbf{k}\times \mathbf{p}|} -1\right)\frac{(\mathbf{k}+\mathbf{p}_3)\cdot\mathbf{p}_3}{ k^2_{\perp}p_3^2}\\
&-\left[\frac{(\mathbf{k}\cdot\mathbf{p})(\mathbf{p} \times \mathbf{p}_1)}{p_{\perp}^2p_1^2}  +\frac{1}{2}\frac{(\mathbf{k}\times \mathbf{p})(\mathbf{p}\cdot\mathbf{p}_1)}{p_{\perp}^2p_1^2}  \right]\frac{(\mathbf{k}-\mathbf{p})\cdot\mathbf{p}_2}{|\mathbf{k}-\mathbf{p}|^2p_2^2}\frac{\mathbf{k}\times \mathbf{p}_3}{ k^2_{\perp}p_3^2}\\
&-\frac{1}{2}\frac{(\mathbf{p}-\mathbf{p}_1)\cdot\mathbf{p}_1}{p_1^2}\frac{(\mathbf{k}-\mathbf{p}-\mathbf{p}_2)\cdot\mathbf{p}_2}{p_2^2}\frac{(\mathbf{k}\times\mathbf{p}) }{p_{\perp}^2|\mathbf{k}-\mathbf{p}|^2}\left(i\frac{\mathbf{p}\cdot(\mathbf{p}-\mathbf{k})}{|\mathbf{p}\times \mathbf{k}|} +1\right)\frac{\mathbf{k}\times \mathbf{p}_3}{ k^2_{\perp}p_3^2} \\
&+ \frac{1}{2}\frac{\mathbf{p}\times \mathbf{p}_1}{p_1^2}\frac{(\mathbf{k}-\mathbf{p})\times \mathbf{p}_2}{p_2^2}\frac{(\mathbf{p}\cdot(\mathbf{k}-\mathbf{p}) -k_{\perp}^2)}{p_{\perp}^2|\mathbf{k}-\mathbf{p}|^2}  i\frac{\mathbf{k}\times \mathbf{p}}{|\mathbf{k}\times \mathbf{p}|}\frac{\mathbf{k}\times \mathbf{p}_3}{ k^2_{\perp}p_3^2} 
\end{split}
\end{equation}
Again, isolating the Lipatov vertex in the complex conjugate amplitude 
\begin{equation}
L^j(\mathbf{p}_3, -\mathbf{k}) = \left(\frac{\mathbf{p}_3^j}{p_3^2} + \frac{\mathbf{k}^j}{k_{\perp}^2}\right),
\end{equation}
we introduce another effective vertex at order-$g^3$,
\begin{equation}\label{eq:effective_vertex_g3_second}
\begin{split}
&\Gamma^j_2(\mathbf{p}, \mathbf{p}_1, \mathbf{p}_2, \mathbf{k})  \\
=&-\Bigg\{\frac{(\mathbf{p}-\mathbf{p}_1)\cdot\mathbf{p}_1}{p_1^2} \frac{(\mathbf{k}-\mathbf{p})\cdot\mathbf{p}_2}{|\mathbf{k}-\mathbf{p}|^2 p_2^2} +\frac{\mathbf{p} \times \mathbf{p}_1}{p_1^2}\frac{(\mathbf{k}-\mathbf{p}-\mathbf{p}_2)\cdot\mathbf{p}_2}{|\mathbf{k}-\mathbf{p}|^2p_2^2} \frac{(\mathbf{k}\times \mathbf{p})}{p_{\perp}^2 }\left(i \frac{\mathbf{k}\cdot(\mathbf{k}-\mathbf{p})}{|\mathbf{k}\times \mathbf{p}|} -1\right)\Bigg\}\frac{\mathbf{k}^j}{k_{\perp}^2}\\
&-\Bigg\{\left[\frac{(\mathbf{p} \times \mathbf{p}_1)}{p_1^2}  \frac{(\mathbf{k}\cdot\mathbf{p})}{p_{\perp}^2}+\frac{1}{2}\frac{(\mathbf{p}\cdot\mathbf{p}_1)}{p_1^2}  \frac{(\mathbf{k}\times \mathbf{p})}{p_{\perp}^2}\right]\frac{(\mathbf{k}-\mathbf{p})\cdot\mathbf{p}_2}{|\mathbf{k}-\mathbf{p}|^2p_2^2}\\
&\qquad+\frac{1}{2}\frac{(\mathbf{p}-\mathbf{p}_1)\cdot\mathbf{p}_1}{p_1^2}\frac{(\mathbf{k}-\mathbf{p}-\mathbf{p}_2)\cdot\mathbf{p}_2}{p_2^2|\mathbf{k}-\mathbf{p}|^2} \frac{(\mathbf{k}\times\mathbf{p}) }{p_{\perp}^2}\left(i\frac{\mathbf{p}\cdot(\mathbf{p}-\mathbf{k})}{|\mathbf{p}\times \mathbf{k}|} +1\right)\\
&\qquad + \frac{i}{2}\frac{\mathbf{p}\times \mathbf{p}_1}{p_1^2}\frac{(\mathbf{k}-\mathbf{p})\times \mathbf{p}_2}{p_2^2|\mathbf{k}-\mathbf{p}|^2}\frac{\mathbf{k}\times \mathbf{p}}{p_{\perp}^2}  \frac{k_{\perp}^2+p_{\perp}^2-\mathbf{p}\cdot\mathbf{k}}{|\mathbf{k}\times \mathbf{p}|}\Bigg\}\frac{\epsilon^{lj}\mathbf{k}^l}{k_{\perp}^2}\,.
\end{split}
\end{equation}
As a consequence, the coefficient function $\mathcal{E}_2$ can be expressed as a product of the two effective vertices
\begin{equation}
\mathcal{E}_2(\mathbf{p}, \mathbf{p}_1, \mathbf{p}_2, \mathbf{p}_3, \mathbf{k}) =\Gamma^j_2(\mathbf{p}, \mathbf{p}_1, \mathbf{p}_2, \mathbf{k}) L^j(\mathbf{p}_3, -\mathbf{k})   \,.
\end{equation}
See Fig.~\ref{fig:M1M3_3wl} for the assignments of the  various momenta.

Note that $\mathcal{E}_2$ has both real  and imaginary parts. The imaginary part is always accompanied by the sign function $\mathrm{sgn}(\mathbf{k}\times \mathbf{p}) = (\mathbf{k}\times \mathbf{p})/|\mathbf{k}\times \mathbf{p}|$.  It is precisely this imaginary part that will be responsible for  the asymmetry between $n^{(4)}(\mathbf{k})$ and $n^{(4)}(-\mathbf{k})$.  In $n^{(4)}(\mathbf{k})$, the odd part under $\mathbf{k}\leftrightarrow -\mathbf{k}$ has already been extracted in Ref.~\cite{McLerran:2016snu}.  The results were contrasted with  those obtained from a diagrammatic approach of Ref.~\cite{Chirilli:2015tea} in Ref.~\cite{Kovchegov:2018jun}. The even part under $\mathbf{k}\leftrightarrow -\mathbf{k}$ is computed here for the first time.
Also note that since $n^{(4)}(\mathbf{k})$ is cubic in $\rho_{P}$, it does not result in  any nontrivial contribution to the single inclusive gluon production after averaging over the projectile color charge densities in the McLerran-Venugopalan model. However, it will be important for double inclusive gluon productions (as $n^{(4)}(\mathbf{k}) n^{(4)}(\mathbf{p})$   is proportional to $\rho_{P}^{6}$, which does not vanish in the McLerran-Venugopalan model).

Considering the color structures for $n^{(4)}(\mathbf{k})$ in Eq.~\eqref{eq:NLO_single_gluon_production},  the terms can be  grouped into two categories according to the number of target Wilson lines involved:  two or three Wilson lines.  From Fig.~\ref{fig:order_g_production} and Fig.~\ref{fig:order_g3_production}, there seem to be a lot of possible combinations of diagrams when multiplying the order-$g^3$ amplitude and the order-$g$ complex conjugate amplitude. However, as far as color  and transverse momentum flows are concerned, all possible graphs can be properly represented by the two diagrams in Fig.~\ref{fig:M1M3}. 
With the help of the effective vertices $L^j(\mathbf{p}_3, \mathbf{k})$, $\Gamma_1^j(\mathbf{p}, \mathbf{p}_1)$, $\Gamma_2^j(\mathbf{p},  \mathbf{p}_1, \mathbf{p}_2, \mathbf{k}) $ , one only needs to directly compute the two diagrams shown in fig. \ref{fig:M1M3} to obtain the correct next to leading order correction to the single gluon production.

Finally, as a consistence check, when the target Wilson line $U(\mathbf{x})=1$,  meaning the absence of the nucleus, there will be no gluon productions. Using $U^{ab}(\mathbf{p}) \rightarrow \delta^{ab}\delta^{(2)}(\mathbf{p})$, it is easy to check that $n^{(4)}(\mathbf{k})$ expressed in Eq.~\eqref{eq:NLO_single_gluon_production} vanishes. 

%%%%%%%%%%%%%%%%%%%%%%%%%%%%%%%%%%%%%%%%%%%%%%%%%%%%%
%%%%%%%%%%%%%%%%%%%%%%%%%%%%%%%%%%%%%%%%%%%%%%%%%%%%
%%%%%%%%%%%%%%%%%%%%%%%%%%%%%%%%%%%%%%%%%%%%%%%%%%%%%
%%%%%%%%%%%%%%%%%%%%%%%%%%%%%%%%%%%%%%%%%%%%%%%%%%%%
\subsection{First saturation correction to single gluon production}
\label{Sect:NNLOSI}
In this section, we calculate the order-$g^6$ contribution to the single inclusive gluon production, the so-called first saturation correction. This contribution is quartic in $\rho_{P}$ and will give the first non-vanishing correction to the configuration-averaged single inclusive gluon production in the McLerran-Venugopalan model. 

At order-$g^6$, the single gluon production is defined  by
\begin{equation}
\label{Eq:n6}
\begin{split}
n^{(6)}(\mathbf{k})=&\sum_{\gamma=\eta, \perp}\Big[ (\mathfrak{S}_{\gamma}^{(3)}(\mathbf{k})+ \mathfrak{B}_{\gamma}^{(3)}(\mathbf{k}))(\mathfrak{S}_{\gamma}^{(3)\ast}(\mathbf{k})+ \mathfrak{B}_{\gamma}^{(3)\ast}(\mathbf{k}))\\
&+\mathfrak{S}_{\gamma}^{(1)\ast}(\mathbf{k})\left( \mathfrak{S}_{\gamma}^{(5)}(\mathbf{k})+\mathfrak{B}_{\gamma}^{(5)}(\mathbf{k})\right)+\mathfrak{S}_{\gamma}^{(1)}(\mathbf{k}) \left(\mathfrak{S}_{\gamma}^{(5)\ast}(\mathbf{k}) +  \mathfrak{B}_{\gamma}^{(5)\ast}(\mathbf{k})\right)\Big].\\
\end{split}
\end{equation}
We  have obtained all the surface and bulk terms needed to compute the final expression for $n^{(6)}(\mathbf{k})$ (see Sects.~\ref{Sect:Surf} and \ref{Sect:Bulk}). 
The explicit calculation is very long and tedious (it needs substitution of eqs \eqref{eqs:betaperp_rhoU}). We thus only report the final result here. All the terms are grouped into three categories according to the number of target Wilson lines they contain: two, three and four Wilson lines.  

\begin{figure}
\centering
\begin{subfigure}[t]{0.75\textwidth}
   \includegraphics[width=1\linewidth]{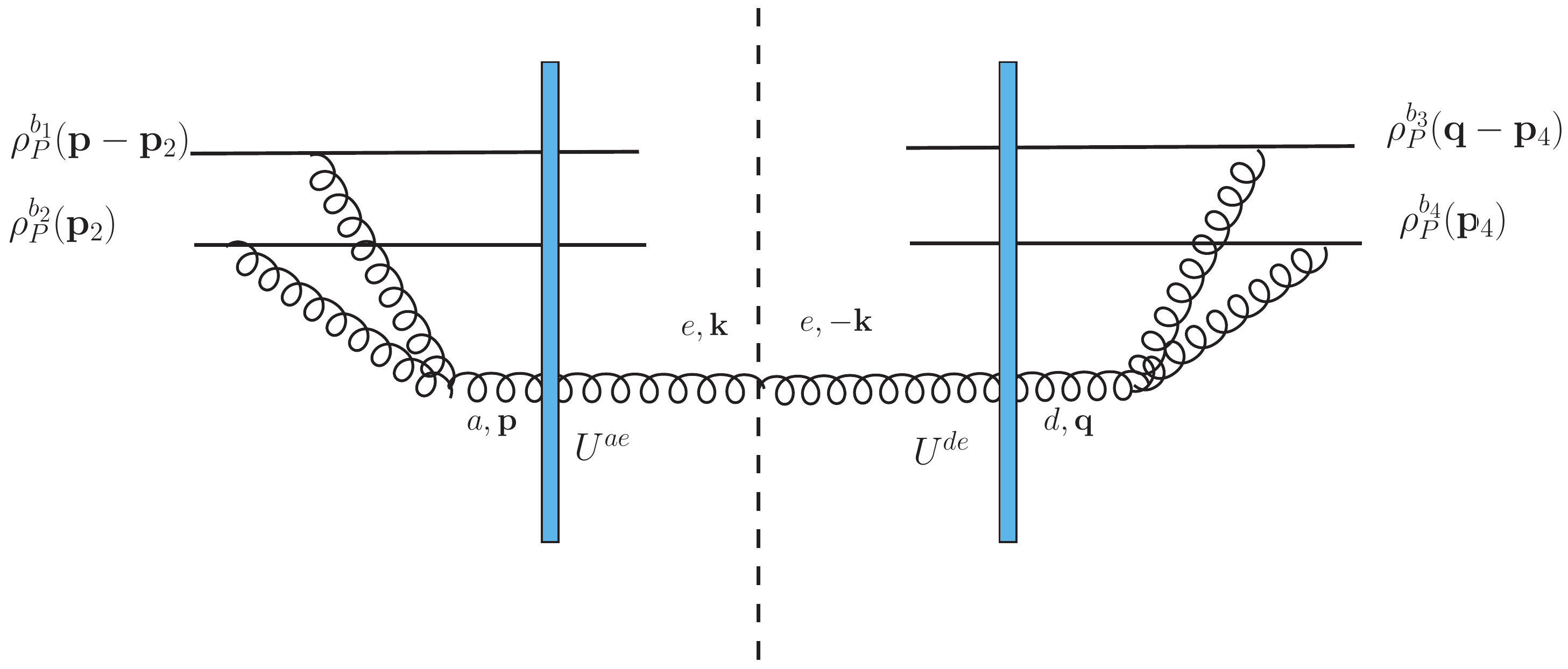}
   \caption{}
   \label{fig:M3_squared_2wl} 
\end{subfigure}
\begin{subfigure}[t]{0.75\textwidth}
   \includegraphics[width=1\linewidth]{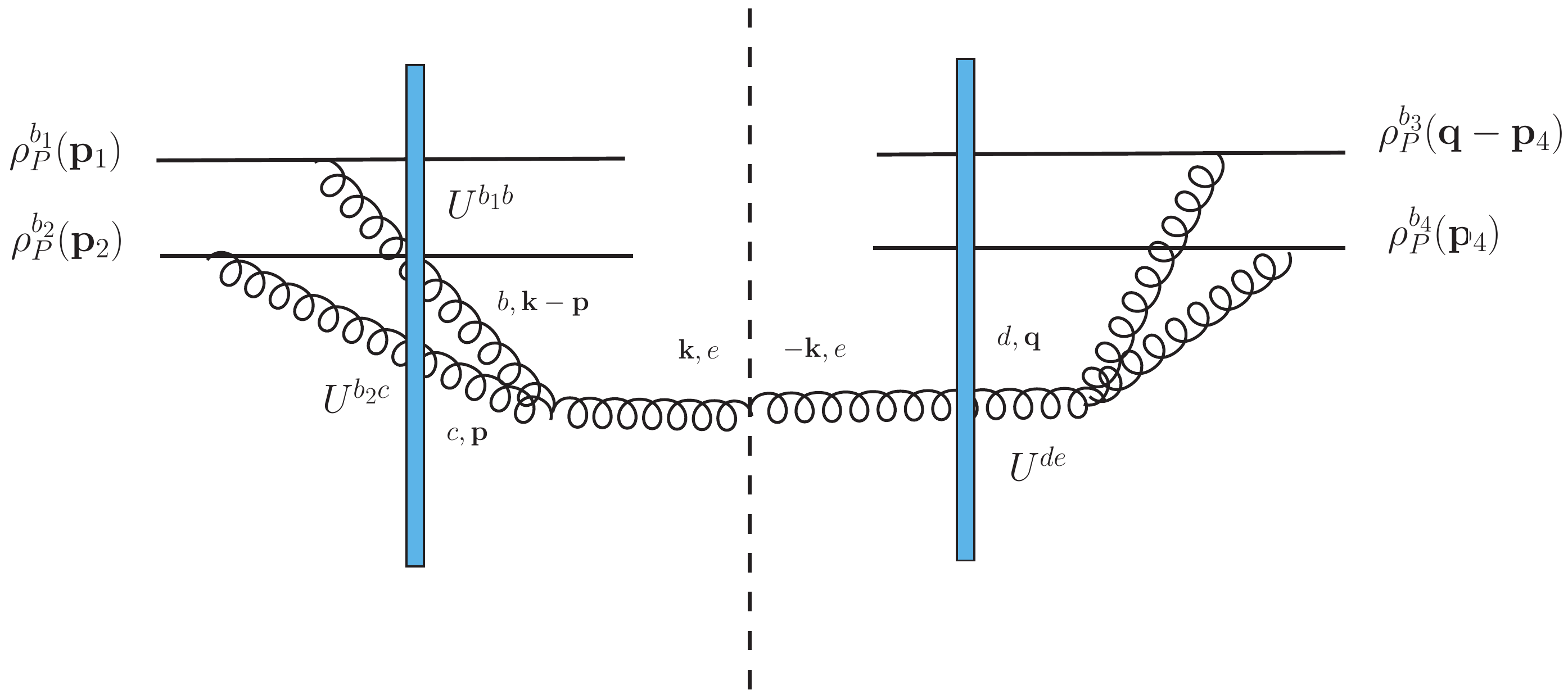}
   \caption{}
   \label{fig:M3_squared_3wl}
\end{subfigure}
\begin{subfigure}[t]{0.75\textwidth}
   \includegraphics[width=1\linewidth]{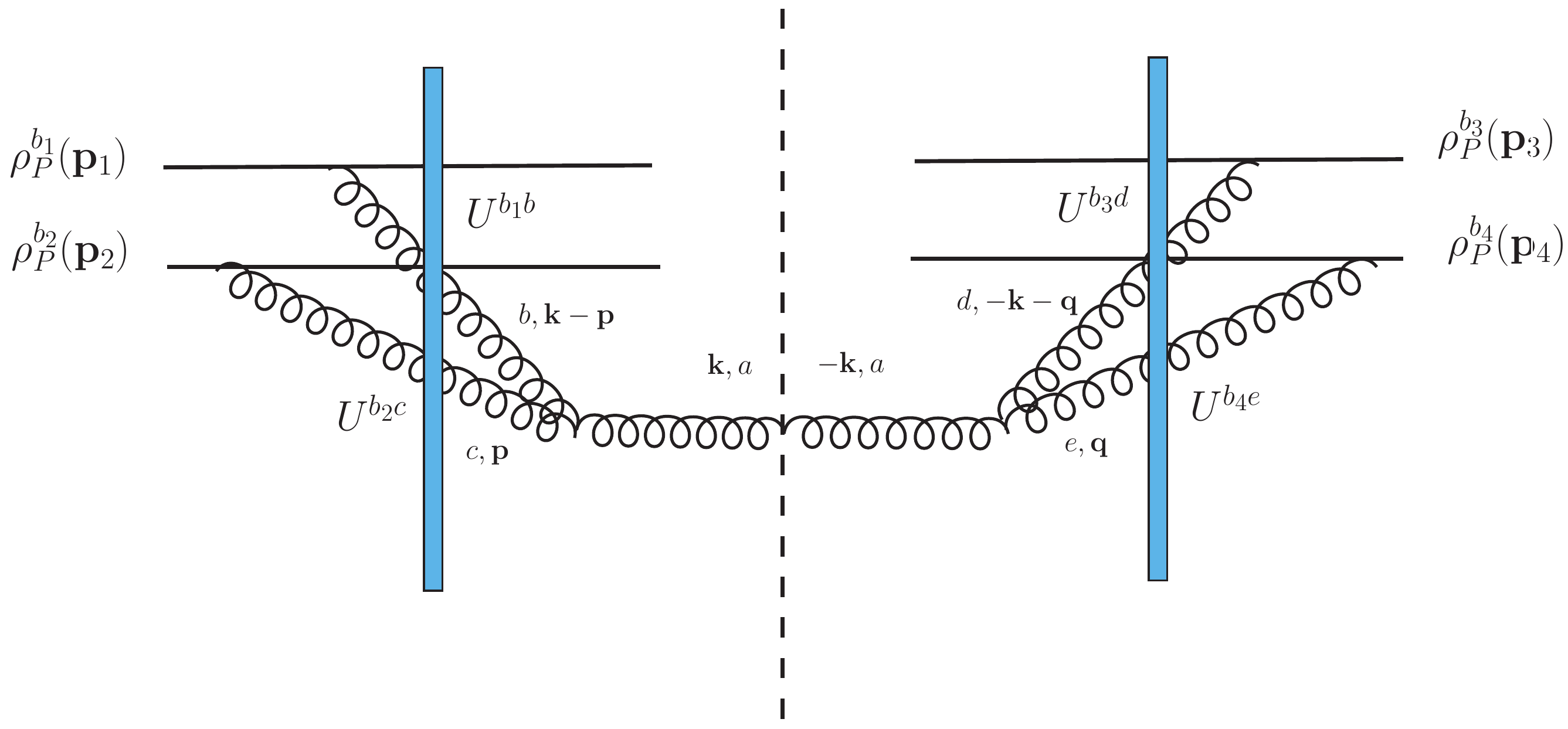}
   \caption{}
   \label{fig:M3_squared_4wl}
\end{subfigure}
\caption{Color structure for terms in the $|M_{(3)}|^2$ that contain two,  three and four Wilson lines. The dashed line separates the amplitude and the complex conjugate amplitude. In (a) two gluon lines crossing the target nucleus, (b) three gluon lines crossing the target nucleus and (c) four gluon lines crossing the target nucleus.}
\label{fig:M3_squared}
\end{figure}

We start with the order-$g^3$ amplitude squared
\begin{equation}\label{eq:M3_squared_result}
\begin{split}
&\sum_{\gamma = \perp, \eta}\Big(\mathfrak{S}_{\gamma}^{(3)}(\mathbf{k}) + \mathfrak{B}_{\gamma}^{(3)}(\mathbf{k})\Big)\Big(\mathfrak{S}_{\gamma}^{(3)\ast}(\mathbf{k}) + \mathfrak{B}_{\gamma}^{(3)\ast}(\mathbf{k})\Big) \\
=&\frac{1}{\pi}\int_{\mathbf{p}, \mathbf{q}, \mathbf{p}_2, \mathbf{p}_4}\mathcal{H}_1(\mathbf{p}, \mathbf{q}, \mathbf{p}_2, \mathbf{p}_4) f^{ab_1b_2}f^{db_3b_4}\rho_P^{b_1}(\mathbf{p}-\mathbf{p}_2)\rho_P^{b_2}(\mathbf{p}_2) \rho_P^{b_3}(\mathbf{q}-\mathbf{p}_4)\rho_P^{b_4}(\mathbf{p}_4) \\
&\qquad\qquad \times U^{ae}(\mathbf{k}-\mathbf{p})U^{de}(-\mathbf{k}-\mathbf{q})\\
&+\frac{1}{\pi}\int_{\mathbf{p}, \mathbf{q}, \mathbf{p}_1, \mathbf{p}_2, \mathbf{p}_4} \mathcal{H}_2(\mathbf{p},\mathbf{q}, \mathbf{p}_1, \mathbf{p}_2, \mathbf{p}_4, \mathbf{k}) f^{db_3b_4}f^{ebc}\rho^{b_1}_P(\mathbf{p}_1)\rho^{b_2}_P(\mathbf{p}_2)\rho^{b_3}_P(\mathbf{q}-\mathbf{p}_4)\rho^{b_4}_P(\mathbf{p}_4)\\
&\qquad\qquad \times U^{b_1b}(\mathbf{k}-\mathbf{p}-\mathbf{p}_1)U^{b_2c}(\mathbf{p}-\mathbf{p}_2)U^{de}(-\mathbf{k}-\mathbf{q})\\
&+\frac{1}{\pi}\int_{\mathbf{p}, \mathbf{q}, \mathbf{p}_1, \mathbf{p}_2, \mathbf{p}_3, \mathbf{p}_4}\mathcal{H}_3(\mathbf{p},\mathbf{q}, \mathbf{p}_1, \mathbf{p}_2, \mathbf{p}_3, \mathbf{p}_4, \mathbf{k})  f^{abc}f^{ade} \rho_P^{b_1}(\mathbf{p}_1)\rho_P^{b_2}(\mathbf{p}_2)\rho_P^{b_3}(\mathbf{p}_3)  \rho_P^{b_4}(\mathbf{p}_4)\\
&\qquad\qquad \times U^{b_1 b}(\mathbf{k}-\mathbf{p}-\mathbf{p}_1)U^{b_2 c}(\mathbf{p}-\mathbf{p}_2) U^{b_3d}(-\mathbf{k}-\mathbf{q}-\mathbf{p}_3) U^{b_4 e}(\mathbf{q}-\mathbf{p}_4)\\
&+{\rm c.c.}\,.
\end{split}
\end{equation}
The three scalar functions  $\mathcal{H}_1(\mathbf{p}, \mathbf{q}, \mathbf{p}_2, \mathbf{p}_4)$, $\mathcal{H}_2(\mathbf{p},\mathbf{q}, \mathbf{p}_2, \mathbf{p}_3, \mathbf{p}_4, \mathbf{k})$, $\mathcal{H}_3(\mathbf{p},\mathbf{q}, \mathbf{p}_1, \mathbf{p}_2, \mathbf{p}_3, \mathbf{p}_4, \mathbf{k})$ only depend on transverse momentum. We present their explicit expressions in the following.
\begin{equation}
\mathcal{H}_1(\mathbf{p}, \mathbf{q}, \mathbf{p}_2, \mathbf{p}_4)=-\frac{1}{4} \frac{(\mathbf{p}_2\times\mathbf{p})}{|\mathbf{p}-\mathbf{p}_2|^2p_2^2}\frac{(\mathbf{p}\cdot\mathbf{q})}{p_{\perp}^2q_{\perp}^2} \frac{(\mathbf{p}_4\times \mathbf{q})}{|\mathbf{q}-\mathbf{p}_4|^2p_4^2}. 
\end{equation}
We have used the identity
$
(\mathbf{p}_2\times\mathbf{q})p_{\perp}^2 - (\mathbf{p}\times\mathbf{q})(\mathbf{p}\cdot\mathbf{p}_2) = (\mathbf{p}\cdot\mathbf{q})(\mathbf{p}_2\times \mathbf{p})$. It is obvious that $\mathcal{H}_1$ is a real function of  transverse momentum. Up to a minus sign, $\mathcal{H}_1$ is nothing but the product of the two effective vertex introduced in  Eq.~ \eqref{eq:effective_vertex_g3_first}
\begin{equation}
\mathcal{H}_1(\mathbf{p}, \mathbf{q}, \mathbf{p}_2, \mathbf{p}_4) = -\Gamma^{j}_1(\mathbf{p}_2, \mathbf{p}) \Gamma^{j}_1(\mathbf{p}_4, \mathbf{q})\,.
\end{equation}
See Fig.~\ref{fig:M3_squared_2wl} for the momentum assignments.\\

There are six terms in $\mathcal{H}_2$,
\begin{equation}
\begin{split}
&\mathcal{H}_2(\mathbf{p}, \mathbf{q}, \mathbf{p}_1, \mathbf{p}_2, \mathbf{p}_4, \mathbf{k})\\
=&\Bigg\{\frac{(\mathbf{k}-\mathbf{p})\cdot\mathbf{p}_1}{p_1^2|\mathbf{k}-\mathbf{p}|^2} \frac{(\mathbf{p}-\mathbf{p}_2)\cdot\mathbf{p}_2}{p_2^2}+\frac{(\mathbf{k}-\mathbf{p}-\mathbf{p}_1)\cdot\mathbf{p}_1}{p_1^2|\mathbf{k}-\mathbf{p}|^2} \frac{\mathbf{p}\times \mathbf{p}_2}{p_2^2}  \frac{\mathbf{k}\times\mathbf{p}}{p_{\perp}^2}\left(i\frac{\mathbf{k}\cdot(\mathbf{k}-\mathbf{p})}{|\mathbf{k}\times \mathbf{p}|} -1\right)\Bigg\}\\
&\quad \times \frac{1}{2 }\frac{\mathbf{k}\times \mathbf{q}}{k_{\perp}^2q_{\perp}^2}\frac{\mathbf{p}_4\times \mathbf{q}}{|\mathbf{q}-\mathbf{p}_4|^2p_4^2} +\Bigg\{-\frac{(\mathbf{k}-\mathbf{p})\cdot\mathbf{p}_1}{p_1^2|\mathbf{k}-\mathbf{p}|^2} \left[\frac{\mathbf{p}\times \mathbf{p}_2}{p_2^2} \frac{\mathbf{k}\cdot\mathbf{p}}{p_{\perp}^2}+\frac{1}{2 } \frac{\mathbf{p}\cdot \mathbf{p}_2}{p_2^2}\frac{\mathbf{k}\times\mathbf{p}}{p_{\perp}^2} \right]\\
&\quad +\frac{1}{2 }\frac{(\mathbf{k}-\mathbf{p}-\mathbf{p}_1)\cdot\mathbf{p}_1}{p_1^2|\mathbf{k}-\mathbf{p}|^2} \frac{(\mathbf{p}-\mathbf{p}_2)\cdot \mathbf{p}_2}{p_2^2}  \frac{\mathbf{k}\times\mathbf{p}}{p_{\perp}^2}\left(i\frac{\mathbf{p}\cdot(\mathbf{k}-\mathbf{p})}{|\mathbf{k}\times \mathbf{p}|} -1\right)\\
&\quad -\frac{i}{2 }\frac{(\mathbf{k}-\mathbf{p})\times\mathbf{p}_1}{p_1^2|\mathbf{k}-\mathbf{p}|^2} \frac{\mathbf{p}\times\mathbf{p}_2}{p_2^2}  \frac{\mathbf{k}\times\mathbf{p}}{p_{\perp}^2}\frac{k_{\perp}^2+p_{\perp}^2 - \mathbf{k}\cdot\mathbf{p}}{|\mathbf{k}\times \mathbf{p}|}\Bigg\}\frac{1}{2}\frac{\mathbf{k}\cdot\mathbf{q}}{k_{\perp}^2q_{\perp}^2} \frac{\mathbf{p}_4\times \mathbf{q}}{|\mathbf{q}-\mathbf{p}_4|^2p_4^2}.
\end{split}
\end{equation}
Using the effective vertices introduced in Eqs.~\eqref{eq:effective_vertex_g3_first} and \eqref{eq:effective_vertex_g3_second}, $\mathcal{H}_2$ is precisely the product of these two vertices
\begin{equation}
\mathcal{H}_2(\mathbf{p}, \mathbf{q}, \mathbf{p}_1, \mathbf{p}_2, \mathbf{p}_4, \mathbf{k}) = \Gamma_2^j(\mathbf{p}_1,\mathbf{p}_2, \mathbf{p},\mathbf{k})\Gamma_1^j(\mathbf{p}_4, \mathbf{q})\,.
\end{equation}
See Fig.~\ref{fig:M3_squared_3wl} for the various momentum definitions.
$\mathcal{H}_2$ contains both real and imaginary parts. The imaginary part involves the single sign function $\mathrm{sgn}(\mathbf{k}\times \mathbf{q}) = \mathbf{k}\times \mathbf{q}/ |\mathbf{k}\times \mathbf{q}|$ and will be responsible for the asymmetry between $n^{(6)}(\mathbf{k})$ and $ n^{(6)}(-\mathbf{k})$. \\

There are thirteen terms in $\mathcal{H}_3$, 
\begin{equation}
\begin{split}
&\mathcal{H}_3(\mathbf{p},\mathbf{q}, \mathbf{p}_1, \mathbf{p}_2, \mathbf{p}_3, \mathbf{p}_4, \mathbf{k})\\
=&-\frac{1}{ 2k_{\perp}^2}\frac{(\mathbf{k}-\mathbf{p})\cdot\mathbf{p}_1}{|\mathbf{k}-\mathbf{p}|^2p_1^2}  \frac{(\mathbf{p}-\mathbf{p}_2)\cdot\mathbf{p}_2}{p_2^2} \frac{(\mathbf{k}+\mathbf{q})\cdot\mathbf{p}_3}{|\mathbf{k}+\mathbf{q}|^2 p_3^2} \frac{(\mathbf{q}-\mathbf{p}_4)\cdot\mathbf{p}_4}{p_4^2}\\
&+\frac{1}{2 k_{\perp}^2}\frac{(\mathbf{k}-\mathbf{p})\cdot\mathbf{p}_1}{|\mathbf{k}-\mathbf{p}|^2p_1^2}  \frac{\mathbf{p}\times\mathbf{p}_2}{p_2^2} \frac{(\mathbf{k}+\mathbf{q})\cdot\mathbf{p}_3}{|\mathbf{k}+\mathbf{q}|^2 p_3^2} \frac{\mathbf{q}\times\mathbf{p}_4}{p_4^2}\\
&-\frac{1}{2 k_{\perp}^2}\frac{(\mathbf{k}-\mathbf{p})\cdot\mathbf{p}_1}{|\mathbf{k}-\mathbf{p}|^2p_1^2}  \frac{\mathbf{p}\times\mathbf{p}_2}{p_2^2}\left[\frac{\mathbf{k}\cdot\mathbf{p}}{ p_{\perp}^2} -\frac{1}{4}\frac{(\mathbf{p}\times\mathbf{k})}{ p_{\perp}^2}\right] \frac{(\mathbf{k}+\mathbf{q})\cdot\mathbf{p}_3}{|\mathbf{k}+\mathbf{q}|^2 p_3^2} \frac{\mathbf{q}\cdot\mathbf{p}_4}{p_4^2}\frac{(\mathbf{q}\times\mathbf{k})}{q_{\perp}^2} \\
&-\frac{i}{ k^2_{\perp}} \frac{(\mathbf{k}-\mathbf{p})\cdot\mathbf{p}_1}{|\mathbf{k}-\mathbf{p}|^2p_1^2}  \frac{(\mathbf{p}-\mathbf{p}_2)\cdot\mathbf{p}_2}{p_2^2} \frac{(\mathbf{k}+\mathbf{q}+\mathbf{p}_3)\cdot\mathbf{p}_3}{|\mathbf{k}+\mathbf{q}|^2 p_3^2} \frac{\mathbf{q}\times\mathbf{p}_4}{p_4^2} \frac{\mathbf{k}\times \mathbf{q}}{q_{\perp}^2 } \left( \frac{\mathbf{k}\cdot(\mathbf{k}+\mathbf{q})}{|\mathbf{k}\times \mathbf{q}|} - i\right)\\
&-\frac{i}{2k^2_{\perp}}\frac{(\mathbf{k}-\mathbf{p})\cdot\mathbf{p}_1}{|\mathbf{k}-\mathbf{p}|^2p_1^2}  \left[\frac{\mathbf{p}\times\mathbf{p}_2}{p_2^2}\frac{\mathbf{k}\cdot\mathbf{p}}{ p_{\perp}^2} -\frac{1}{2}\frac{\mathbf{p}\cdot\mathbf{p}_2}{p_2^2}\frac{(\mathbf{p}\times\mathbf{k})}{p_{\perp}^2} \right]\\
&\qquad \times\Bigg[\frac{(\mathbf{k}+\mathbf{q}+\mathbf{p}_3)\cdot\mathbf{p}_3}{ |\mathbf{k}+\mathbf{q}|^2p_3^2} \frac{(\mathbf{q}-\mathbf{p}_4)\cdot\mathbf{p}_4}{p_4^2} \frac{\mathbf{k}\times \mathbf{q}}{q_{\perp}^2}\left(\frac{\mathbf{q}\cdot(\mathbf{q}+\mathbf{k})}{|\mathbf{q}\times \mathbf{k}|} +i\right)\\
&\qquad\quad+ \frac{(\mathbf{k}+\mathbf{q})\times \mathbf{p}_3}{|\mathbf{k}+\mathbf{q}|^2 p_3^2} \frac{\mathbf{q}\times\mathbf{p}_4}{p_4^2} \frac{ \mathbf{k}\times \mathbf{q}}{q_{\perp}^2} \frac{(k_{\perp}^2 + q_{\perp}^2+\mathbf{k}\cdot\mathbf{q})}{|\mathbf{k}\times \mathbf{q}|}\Bigg]\\
&-\frac{1}{2k^2_{\perp}}\frac{(\mathbf{k}-\mathbf{p}-\mathbf{p}_1)\cdot\mathbf{p}_1}{|\mathbf{k}-\mathbf{p}|^2p_1^2}  \frac{\mathbf{p}\times\mathbf{p}_2}{p_2^2}\frac{\mathbf{k}\times \mathbf{p}}{p_{\perp}^2 } \frac{(\mathbf{k}+\mathbf{q}+\mathbf{p}_3)\cdot\mathbf{p}_3}{|\mathbf{k}+\mathbf{q}|^2 p_3^2} \frac{\mathbf{q}\times\mathbf{p}_4}{p_4^2}  \frac{\mathbf{k}\times \mathbf{q}}{q_{\perp}^2 } \\
&\qquad\qquad \times \left( \frac{\mathbf{k}\cdot(\mathbf{k}-\mathbf{p})}{|\mathbf{k}\times \mathbf{p}|} + i\right) \left( \frac{\mathbf{k}\cdot(\mathbf{k}+\mathbf{q})}{|\mathbf{k}\times \mathbf{q}|} - i\right)\\
&-\frac{1}{4 k^2_{\perp}}\frac{(\mathbf{k}-\mathbf{p})\times\mathbf{p}_1}{|\mathbf{k}-\mathbf{p}|^2p_1^2}  \frac{\mathbf{p}\times\mathbf{p}_2}{p_2^2} \frac{ \mathbf{p}\times \mathbf{k}}{p_{\perp}^2}\frac{(\mathbf{k}+\mathbf{q}+\mathbf{p}_3)\cdot\mathbf{p}_3}{|\mathbf{k}+\mathbf{q}|^2 p_3^2} \frac{(\mathbf{q}-\mathbf{p}_4)\cdot\mathbf{p}_4}{p_4^2}\frac{\mathbf{k}\times \mathbf{q}}{q_{\perp}^2}\\
&\qquad\times  \frac{(k_{\perp}^2 + p_{\perp}^2-\mathbf{k}\cdot\mathbf{p})}{|\mathbf{p}\times \mathbf{k}|}\left(\frac{\mathbf{q}\cdot(\mathbf{q}+\mathbf{k})}{|\mathbf{q}\times \mathbf{k}|} +i\right)\\
&-\frac{1}{8 k^2_{\perp}} \frac{(\mathbf{k}-\mathbf{p}-\mathbf{p}_1)\cdot\mathbf{p}_1}{|\mathbf{k}-\mathbf{p}|^2p_1^2}  \frac{(\mathbf{p}-\mathbf{p}_2)\cdot\mathbf{p}_2}{p_2^2}\frac{\mathbf{p}\times \mathbf{k}}{p_{\perp}^2} \frac{(\mathbf{k}+\mathbf{q})\times\mathbf{p}_3}{|\mathbf{k}+\mathbf{q}|^2 p_3^2} \frac{\mathbf{q}\times\mathbf{p}_4}{p_4^2}\frac{\mathbf{k}\times\mathbf{q} }{q_{\perp}^2}\\
& \qquad\qquad \times \frac{(k_{\perp}^2 + q_{\perp}^2+\mathbf{k}\cdot\mathbf{q})}{|\mathbf{q}\times \mathbf{k}|}\left(\frac{\mathbf{p}\cdot(\mathbf{p}-\mathbf{k})}{|\mathbf{p}\times \mathbf{k}|} - i\right)\\
&-\frac{1}{8 k^2_{\perp}}\frac{(\mathbf{k}-\mathbf{p})\times\mathbf{p}_1}{|\mathbf{k}-\mathbf{p}|^2p_1^2}  \frac{(\mathbf{p}\times\mathbf{p}_2)}{p_2^2} \frac{\mathbf{p}\times \mathbf{k} }{p_{\perp}^2} \frac{(\mathbf{k}+\mathbf{q})\times\mathbf{p}_3}{|\mathbf{k}+\mathbf{q}|^2 p_3^2} \frac{(\mathbf{q}\times\mathbf{p}_4)}{p_4^2}\frac{ \mathbf{k}\times \mathbf{q}}{q_{\perp}^2} \\
&\qquad \qquad \times  \frac{(k_{\perp}^2 + q_{\perp}^2+\mathbf{k}\cdot\mathbf{q})}{|\mathbf{q}\times \mathbf{k}|}\frac{(k_{\perp}^2 + p_{\perp}^2-\mathbf{k}\cdot\mathbf{p})}{|\mathbf{p}\times \mathbf{k}|}.\\
\end{split}
\end{equation}

It is an easy exercise to show that $\mathcal{H}_3$ can be expressed in terms of the effective vertex we introduced in Eq.~\eqref{eq:effective_vertex_g3_second}
\begin{equation}
\begin{split}
&\mathcal{H}_3(\mathbf{p},\mathbf{q}, \mathbf{p}_1, \mathbf{p}_2, \mathbf{p}_3, \mathbf{p}_4, \mathbf{k}) + \mathcal{H}_3^{\ast}(\mathbf{p},\mathbf{q}, \mathbf{p}_1, \mathbf{p}_2, \mathbf{p}_3, \mathbf{p}_4, \mathbf{k})\\
 = &-\Gamma_2^j(\mathbf{p}_1, \mathbf{p}_2, \mathbf{p}, \mathbf{k}) \Gamma_2^{j\ast}(\mathbf{p}_3, \mathbf{p}_4,\mathbf{q}, -\mathbf{k}).\\
 \end{split}
\end{equation}
See Fig.~\ref{fig:M3_squared_4wl} for the various momentum assignments in the amplitude and the complex conjugate amplitude.

There are two sign functions present in the expression of $\mathcal{H}_3$, $\mathrm{sgn}(\mathbf{k}\times \mathbf{p})$ and $\mathrm{sgn}(\mathbf{k}\times \mathbf{q})$.   Imginary parts of $\mathcal{H}_3$ contain only one of the sign functions but not two at the same time. On the other hand, some terms in the real part contain product of both  sign functions.  It is clear that a single sign function is responsible for breaking the symmetry between $\mathbf{k}\leftrightarrow -\mathbf{k}$.

As for the color structure  of the order-$g^3$ amplitude squared Eq.~\eqref{eq:M3_squared_result}, although there are many diagrams when multiplying the amplitude with the conjugate amplitude, according to the number of Wilson lines, they can be representd by the three diagrams in Fig.~\ref{fig:M3_squared}. Depending on how many gluon lines crossing the target nucleus in the diagram, they corresponds to two-, three- and four-Wilson line terms. With the help of the effective vertices Eqs.~\eqref{eq:effective_vertex_g3_first} and \eqref{eq:effective_vertex_g3_second}, the precise expression can be computed directly from these three diagrams.\\

%\begin{figure}[t]
%\centering % \begin{center}/\end{center} takes some additional vertical space
%\includegraphics[width = 0.8\textwidth]{M3_squared_twoWL.pdf}
%\caption{Color structure for terms in the $|M_{(3)}|^2$ that contain two Wilson lines. The dashed line separates the amplitude and the complex conjugate amplitude. There are two gluon lines crossing the target nucleus.}
%\label{fig:M3_squared_2wl}
%\end{figure}

We move on to present the final result for the crossing terms between the order-$g$ and order-$g^5$ gluon production amplitudes. Again, the terms are grouped into three categories according to the number of target Wilson lines.
\begin{equation}\label{eq:M1M5_final_result}
\begin{split}
&\sum_{\gamma = \perp, \eta}\mathfrak{S}_{\gamma}^{(1)\ast} (\mathbf{k})\Big(\mathfrak{S}_{\gamma}^{(5)}(\mathbf{k}) + \mathfrak{B}_{\gamma}^{(5)}(\mathbf{k})\Big) +\mathfrak{S}_{\gamma}^{(1)} (\mathbf{k})\Big(\mathfrak{S}_{\gamma}^{(5)\ast}(\mathbf{k}) + \mathfrak{B}_{\gamma}^{(5)\ast}(\mathbf{k})\Big) \\
=&\frac{1}{\pi}\int_{\mathbf{p}, \mathbf{q}, \mathbf{p}_2, \mathbf{p}_4} \mathcal{F}_1(\mathbf{p}, \mathbf{q}, \mathbf{p}_2, \mathbf{p}_4, \mathbf{k}) f^{dab_3}f^{ab_1b_2} \rho_P^{b_1}(\mathbf{p}-\mathbf{p}_2)\rho_P^{b_2}(\mathbf{p}_2) \rho_P^{b_3}(\mathbf{q}-\mathbf{p})\rho_P^{b_4}(\mathbf{p}_4)\\
&\qquad\qquad \times U^{de}(\mathbf{k}-\mathbf{q})U^{b_4e}(-\mathbf{k}-\mathbf{p}_4)\\
&+\frac{1}{\pi}\int_{\mathbf{p}, \mathbf{q}, \mathbf{p}_2, \mathbf{p}_3, \mathbf{p}_4} \mathcal{F}_2(\mathbf{p}, \mathbf{q}, \mathbf{p}_2, \mathbf{p}_3, \mathbf{p}_4, \mathbf{k}) f^{bb_1b_2} f^{ade} \rho_P^{b_1}(\mathbf{p}-\mathbf{p_2})\rho_P^{b_2}(\mathbf{p}_2)\rho_P^{b_3}(\mathbf{p}_3)\rho_P^{b_4}(\mathbf{p}_4)\\
&\qquad\qquad \times U^{ba}(\mathbf{q}-\mathbf{p}) U^{b_3 d}(\mathbf{k}-\mathbf{q}-\mathbf{p}_3)U^{b_4 e}(-\mathbf{k}-\mathbf{p}_4)\\
&+\frac{1}{\pi}\int_{\mathbf{p}, \mathbf{q}, \mathbf{p}_1, \mathbf{p}_2, \mathbf{p}_3, \mathbf{p}_4} \mathcal{F}_3(\mathbf{p}, \mathbf{q}, \mathbf{p}_1, \mathbf{p}_2, \mathbf{p}_3, \mathbf{p}_4, \mathbf{k})f^{abc}f^{ade}\rho^{b_1}_P(\mathbf{p}_1)\rho^{b_2}_P(\mathbf{p}_2)\rho^{b_3}_P(\mathbf{p}_3)\rho^{b_4}_P(\mathbf{p}_4)\\
&\qquad\qquad\times U^{b_1b}(\mathbf{p}-\mathbf{p}_1)U^{b_2c}(\mathbf{q}-\mathbf{p}-\mathbf{p}_2) U^{b_3 d}(\mathbf{k}-\mathbf{q}-\mathbf{p}_3) U^{b_4e}(-\mathbf{k}-\mathbf{p}_4)\\
&+{\rm c.c.}
\end{split}
\end{equation}

\begin{figure}
\centering
\begin{subfigure}[t]{0.75\textwidth}
   \includegraphics[width=1\linewidth]{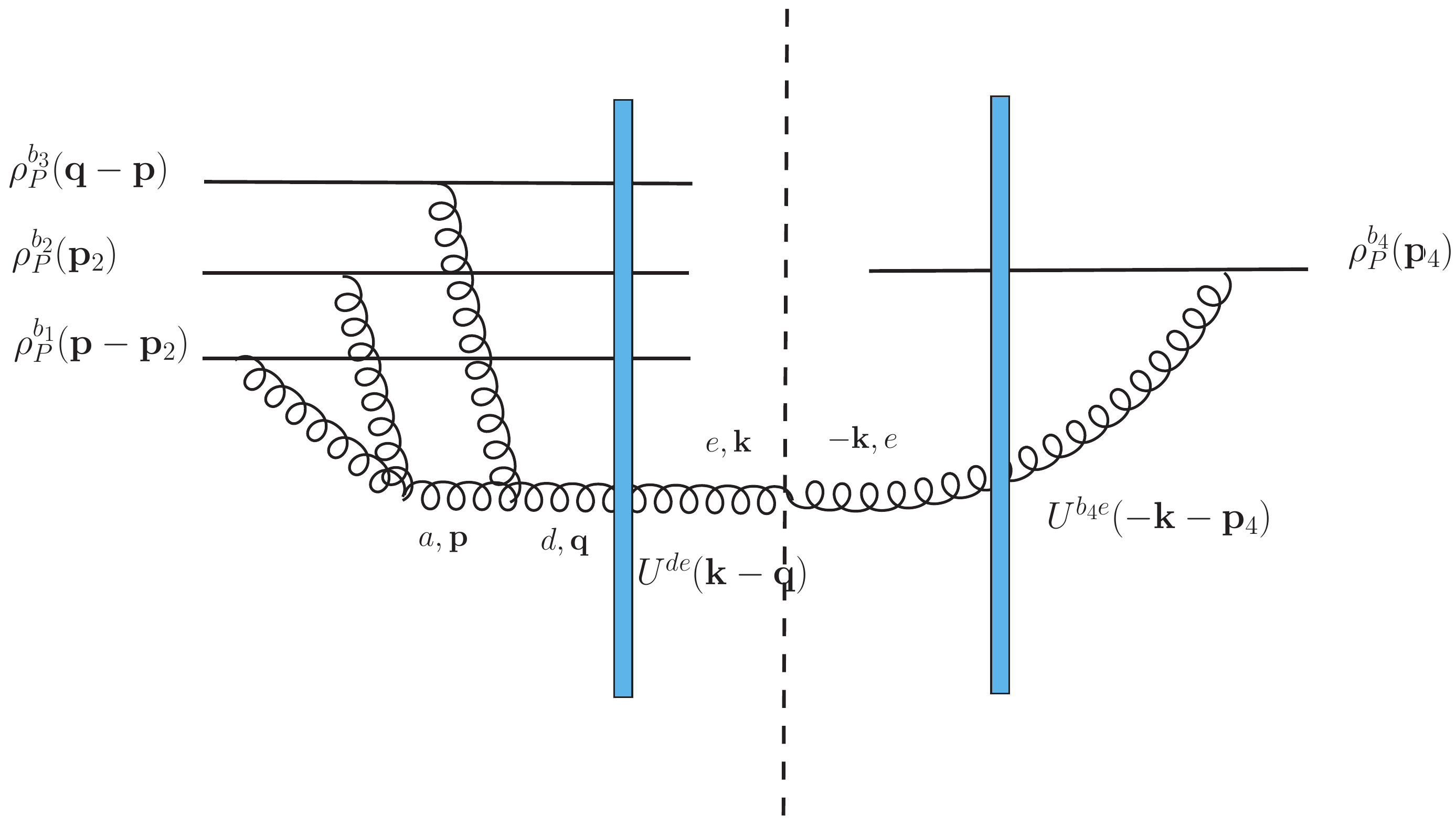}
   \caption{}
   \label{fig:M1M5_squared_2wl} 
\end{subfigure}
\begin{subfigure}[t]{0.75\textwidth}
   \includegraphics[width=1\linewidth]{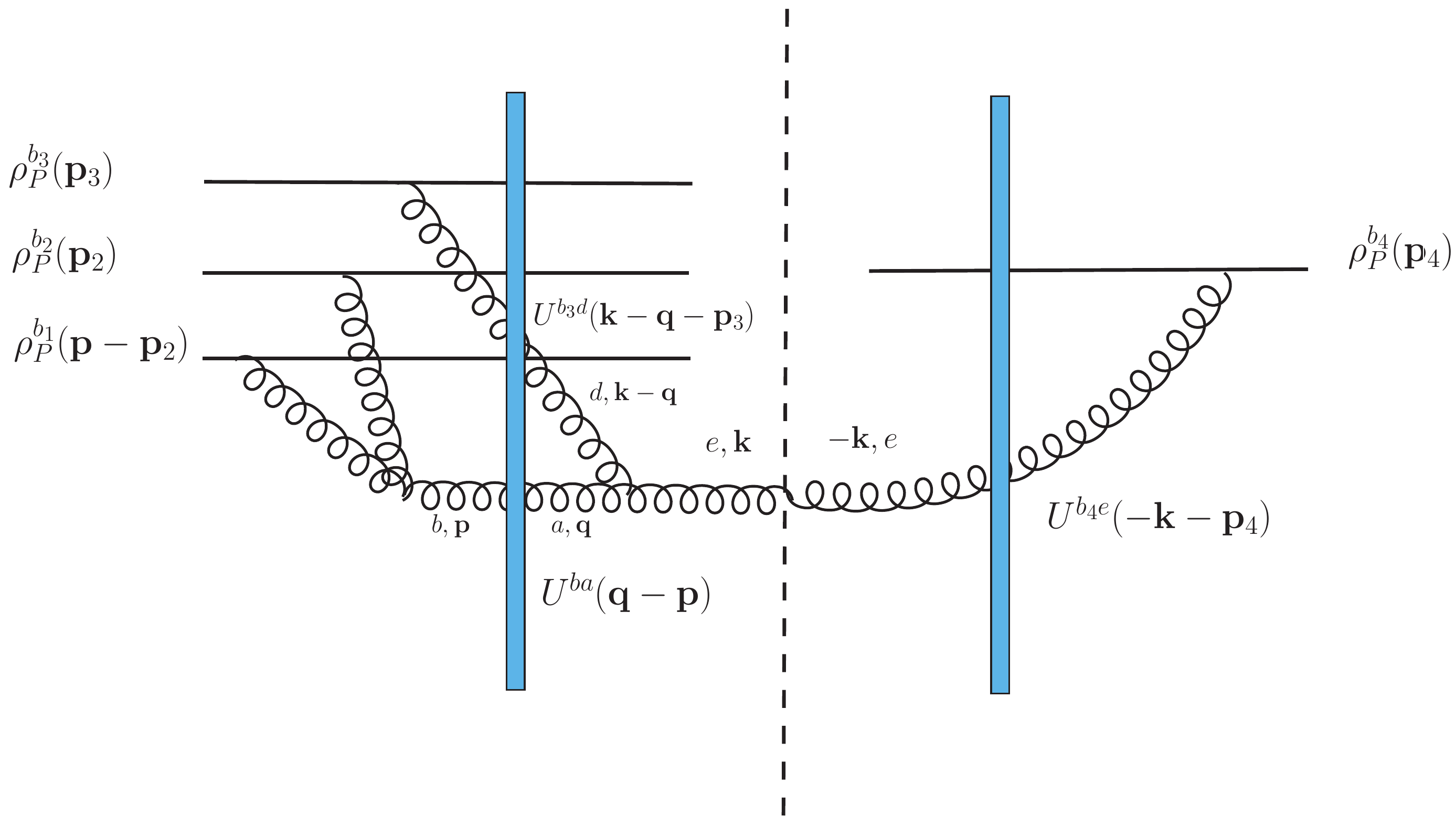}
   \caption{}
   \label{fig:M1M5_squared_3wl}
\end{subfigure}
\begin{subfigure}[t]{0.75\textwidth}
   \includegraphics[width=1\linewidth]{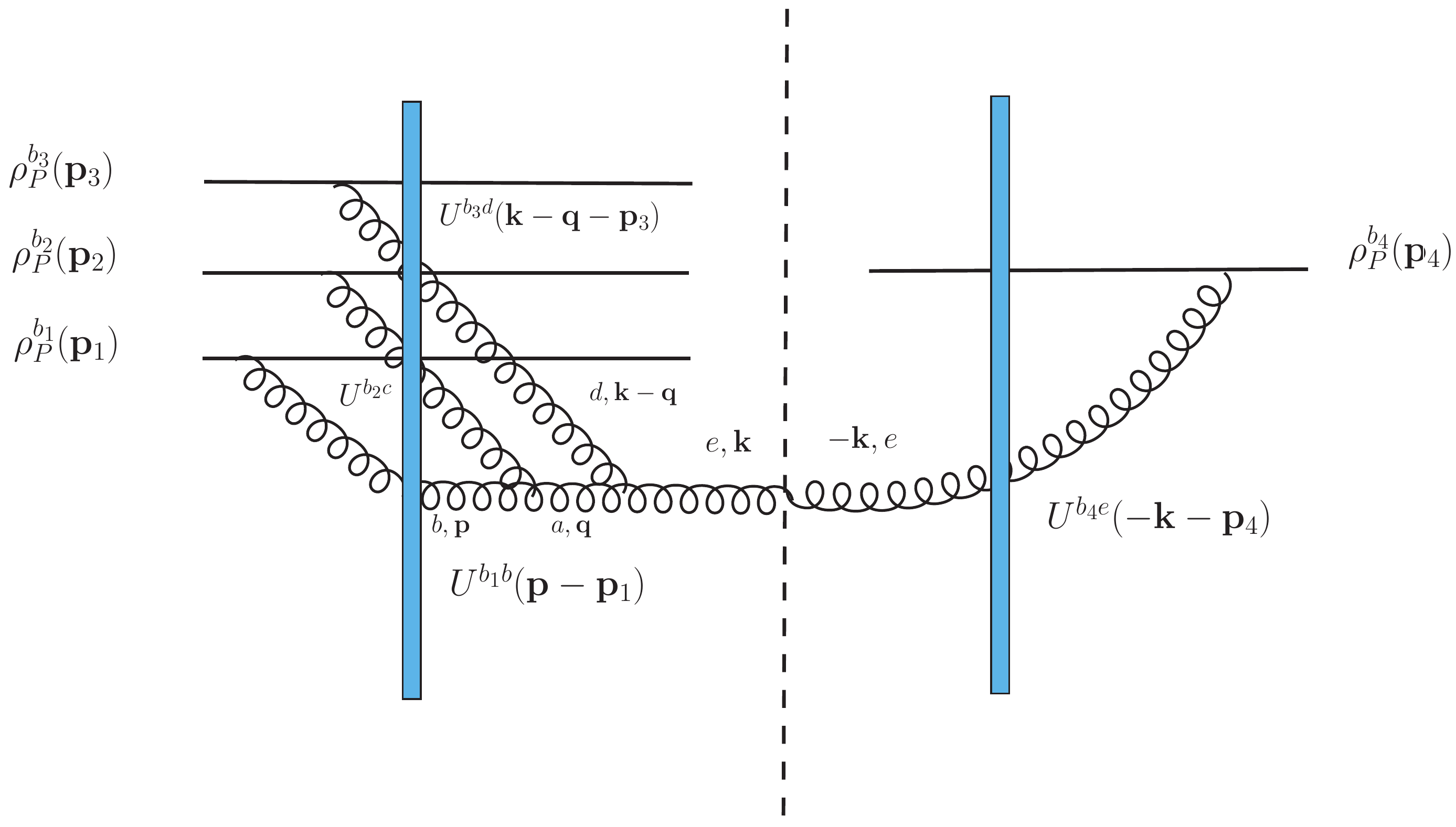}
   \caption{}
   \label{fig:M1M5_squared_4wl}
\end{subfigure}
\caption{Color structures for the terms in $M_{(1)}^{\ast}M_{(5)}$ that contain two,  three and four Wilson lines. The dashed line separates the amplitude and the complex conjugate amplitude. In (a) two gluon lines crossing the target nucleus, (b) three gluon lines crossing the target nucleus and (c) four gluon lines crossing the target nucleus.}
\label{fig:M1M5}
\end{figure}

The three kinematic functions $\mathcal{F}_1, \mathcal{F}_2, \mathcal{F}_3$ only depend on transverse momentum. Their explicit expressions are presented in the following. For $\mathcal{F}_1$,
\begin{equation}
\mathcal{F}_1(\mathbf{p}, \mathbf{q}, \mathbf{p}_2, \mathbf{p}_4, \mathbf{k}) = \frac{1}{2q_{\perp}^2|\mathbf{p}-\mathbf{p}_2|^2|\mathbf{q}-\mathbf{p}|^2} \left(\frac{-(\mathbf{p}\times \mathbf{q})}{p_{\perp}^2} + \frac{\mathbf{p}_2\times \mathbf{q}}{3p_2^2} \right) \left(\frac{(\mathbf{k} \times \mathbf{q})}{k_{\perp}^2}+ \frac{\mathbf{p}_4 \times \mathbf{q}}{p_4^2}\right).
\end{equation}
Subtracting the Lipatov vertex in the complex conjugate amplitude
\begin{equation}
L^j(\mathbf{p}_4, -\mathbf{k}) = \left(\frac{\mathbf{p}_4^j}{p_4^2}+\frac{\mathbf{k}^j}{k_{\perp}^2}\right),
\end{equation}
we introduce an effective vertex at order-$g^5$, 
\begin{equation}
\Upsilon^j_{1}(\mathbf{p}, \mathbf{q}, \mathbf{p}_2) = \frac{1}{2|\mathbf{p}-\mathbf{p}_2|^2|\mathbf{q}-\mathbf{p}|^2} \left(\frac{\mathbf{p}\times \mathbf{q}}{p_{\perp}^2} - \frac{\mathbf{p}_2\times \mathbf{q}}{3p_2^2} \right)\frac{\epsilon^{ij}\mathbf{q}^i}{q_{\perp}^2}.
\end{equation}
Then the coefficient function can be expressed as a  product of two effective vertices
\begin{equation}
\mathcal{F}_1(\mathbf{p}, \mathbf{q}, \mathbf{p}_2, \mathbf{p}_4, \mathbf{k}) =\Upsilon^j_{1}(\mathbf{p}, \mathbf{q}, \mathbf{p}_2)  L^j(\mathbf{p}_4, -\mathbf{k}) \,.
\end{equation}
See Fig.~\ref{fig:M1M5_squared_2wl} for the momentum assignments. It is interesting to note that the effective vertex $\Upsilon_1^j(\mathbf{p}, \mathbf{q},\mathbf{q}_2)$ is independent of the momentum $\mathbf{k}$. 
It is also apparent that $\mathcal{F}_1$ is a real function. \\

There are eight terms in $\mathcal{F}_2$. The final result can also be expressed as a product of two effective vertices
\begin{equation}
\mathcal{F}_2(\mathbf{p},\mathbf{q},\mathbf{p}_2, \mathbf{p}_3, \mathbf{p}_4, \mathbf{k}) = \Upsilon^j_2(\mathbf{p},\mathbf{q},\mathbf{p}_2, \mathbf{p}_3, \mathbf{k}) L^j(\mathbf{p}_4, -\mathbf{k})
\end{equation}
We decompose the order-$g^5$ effective vertex $\Upsilon^j_2$ into two parts
\begin{equation}
\Upsilon^j_2(\mathbf{p},\mathbf{q},\mathbf{p}_2, \mathbf{p}_3, \mathbf{k}) = \Upsilon_2^{\|}(\mathbf{p},\mathbf{q},\mathbf{p}_2, \mathbf{p}_3, \mathbf{k})\frac{\mathbf{k}^j}{k_{\perp}^2} + \Upsilon_2^{\perp}(\mathbf{p},\mathbf{q},\mathbf{p}_2, \mathbf{p}_3, \mathbf{k})\frac{\epsilon^{ij}\mathbf{k}^i}{k_{\perp}^2}. 
\end{equation}
The two functions are 
\begin{equation}
\begin{split}
&\Upsilon_2^{\|}(\mathbf{p},\mathbf{q},\mathbf{p}_2, \mathbf{p}_3, \mathbf{k}) \\
=&-\frac{1}{2}\frac{\mathbf{p}_2\times\mathbf{p}}{|\mathbf{p}-\mathbf{p}_2|^2p_2^2}\frac{\mathbf{q}\times\mathbf{p} }{p_{\perp}^2} \left(\frac{(\mathbf{k}-\mathbf{q})\cdot \mathbf{p}_3}{|\mathbf{k}-\mathbf{q}|^2p_3^2} -\frac{(\mathbf{k}-\mathbf{q}-\mathbf{p}_3) \cdot\mathbf{p}_3}{q_{\perp}^2p_3^2} \right)\\
&\qquad-\frac{1}{2}\frac{\mathbf{p}_2\times\mathbf{p}}{ |\mathbf{p}-\mathbf{p}_2|^2 p_2^2} \frac{(\mathbf{k}-\mathbf{q})\times \mathbf{p}_3}{|\mathbf{k}-\mathbf{q}|^2p_3^2} \frac{\mathbf{q}\times\mathbf{p}}{p_{\perp}^2}\frac{\mathbf{k}\times \mathbf{q}}{q_{\perp}^2}\left(i\frac{\mathbf{k}\cdot\mathbf{q}}{|\mathbf{k}\times \mathbf{q}| }-1\right)\\
&\qquad+\frac{1}{2}\frac{(\mathbf{p}_2\times\mathbf{p})}{ p_2^2|\mathbf{p}-\mathbf{p}_2|^2}\frac{(\mathbf{k}-\mathbf{q}-\mathbf{p}_3)\cdot\mathbf{p}_3}{|\mathbf{k}-\mathbf{q}|^2p_3^2} \frac{(\mathbf{p}\cdot\mathbf{q})}{p_{\perp}^2}\frac{\mathbf{k}\times \mathbf{q}}{q_{\perp}^2} \left(i\frac{\mathbf{k}\cdot(\mathbf{k}-\mathbf{q})}{|\mathbf{k}\times \mathbf{q}|} -1\right)\\
\end{split}
\end{equation}
and 
\begin{equation}
\begin{split}
&\Upsilon_2^{\perp}(\mathbf{p},\mathbf{q},\mathbf{p}_2, \mathbf{p}_3, \mathbf{k})\\
=&\frac{1}{2 }\frac{(\mathbf{p}_2\times\mathbf{p})}{|\mathbf{p}-\mathbf{p}_2|^2p_2^2}\frac{(\mathbf{k}-\mathbf{q})\cdot \mathbf{p}_3}{|\mathbf{k}-\mathbf{q}|^2p_3^2} \frac{(\mathbf{p}\cdot\mathbf{k})}{p_{\perp}^2}+\frac{1}{2 }\frac{\mathbf{p}_2\times\mathbf{p}}{ |\mathbf{p}-\mathbf{p}_2|^2 p_2^2}\frac{(\mathbf{k}-\mathbf{q})\times \mathbf{p}_3}{|\mathbf{k}-\mathbf{q}|^2p_3^2} \frac{\mathbf{k}\cdot(\mathbf{k}-\mathbf{q})}{q_{\perp}^2 }\frac{\mathbf{q}\times\mathbf{p}}{p_{\perp}^2}\\
&\qquad-\frac{1}{2 }\frac{\mathbf{p}_2\times\mathbf{p}}{ |\mathbf{p}-\mathbf{p}_2|^2 p_2^2}\frac{(\mathbf{k}-\mathbf{q}-\mathbf{p}_3)\cdot\mathbf{p}_3}{ |\mathbf{k}-\mathbf{q}|^2p_3^2} \frac{\mathbf{q}\times\mathbf{p}}{p_{\perp}^2}\frac{\mathbf{k}\times \mathbf{q}}{q_{\perp}^2}\left(i\frac{\mathbf{q}\cdot(\mathbf{q}-\mathbf{k})}{|\mathbf{q}\times \mathbf{k}|} +1\right)\\
&\qquad +i\frac{1}{2 }\frac{(\mathbf{p}_2\times\mathbf{p})}{p_2^2|\mathbf{p}-\mathbf{p}_2|^2}\frac{(\mathbf{k}-\mathbf{q})\times\mathbf{p}_3}{p_3^2} \frac{(\mathbf{p}\cdot\mathbf{q})}{p_{\perp}^2 }\frac{(\mathbf{k}\times \mathbf{q})}{q^2_{\perp} |\mathbf{k}-\mathbf{q}|^2}\frac{(k_{\perp}^2+q_{\perp}^2-\mathbf{k}\cdot\mathbf{q})}{|\mathbf{k}\times \mathbf{q}|}.
\end{split}
\end{equation}
See Fig.~\ref{fig:M1M5_squared_2wl} for the momentum assignments. $\mathcal{F}_2$ contains both real part and imaginary parts. The imaginary part is associated with the sign function $\mathrm{sgn}(\mathbf{k}\times \mathbf{q})$. \\

There are nineteen terms in $\mathcal{F}_3$. Six of the terms involve the auxiliary  functions $\mathcal{I}_{1,2,3}$ and $\mathcal{G}_{1,2,3}$ we introduced in  Sec.~\ref{Sect:Bulk}. These functions contain definite angular integrals that no closed form results are known to us. The computations are straightforward but very tedious. The final expression of $\mathcal{F}_3$ can also be expressed as a product  of two effective vertices
\begin{equation}
\mathcal{F}_3(\mathbf{p},\mathbf{q}, \mathbf{p}_1, \mathbf{p}_2, \mathbf{p}_3, \mathbf{p}_4, \mathbf{k}) = \Upsilon^j_3(\mathbf{p},\mathbf{q}, \mathbf{p}_1, \mathbf{p}_2, \mathbf{p}_3,\mathbf{k}) L^j(\mathbf{p}_4, -\mathbf{k})
\end{equation}
The third effective vertex at order-$g^5$, $\Upsilon^j_3$, can be further decomposed into two parts depending whether it is parallel or perpendicular to the momentum $\mathbf{k}$,
\begin{equation}
\Upsilon^j_3(\mathbf{p},\mathbf{q}, \mathbf{p}_1, \mathbf{p}_2, \mathbf{p}_3,\mathbf{k}) = \Upsilon_3^{\|} (\mathbf{p},\mathbf{q}, \mathbf{p}_1, \mathbf{p}_2, \mathbf{p}_3,\mathbf{k}) \frac{\mathbf{k}^j}{k_{\perp}^2} + \Upsilon_3^{\perp}(\mathbf{p},\mathbf{q}, \mathbf{p}_1, \mathbf{p}_2, \mathbf{p}_3,\mathbf{k})  \frac{\epsilon^{ij} \mathbf{k}^i}{k_{\perp}^2}.
\end{equation}
Here the two functions $\Upsilon_3^{\|}$ and $ \Upsilon_3^{\perp}$ are
\begin{equation}
\begin{split}
&\Upsilon_3^{\|} (\mathbf{p},\mathbf{q}, \mathbf{p}_1, \mathbf{p}_2, \mathbf{p}_3,\mathbf{k}) \\
=&\frac{\mathbf{p}\times\mathbf{p}_1}{p_{\perp}^2p_1^2} \frac{(\mathbf{q}-\mathbf{p}-\mathbf{p}_2)\cdot\mathbf{p}_2}{|\mathbf{q}-\mathbf{p}|^2p_2^2} \frac{(\mathbf{k}-\mathbf{q})\times \mathbf{p}_3}{|\mathbf{k}-\mathbf{q}|^2 p_3^2} i\mathcal{I}_1 (\mathbf{p}, \mathbf{q}, \mathbf{k})\\
&+ \frac{(\mathbf{p}-\mathbf{p}_1)\cdot\mathbf{p}_1}{p^2_{\perp}p_1^2} \frac{(\mathbf{q}-\mathbf{p}-\mathbf{p}_2)\cdot\mathbf{p}_2}{|\mathbf{q}-\mathbf{p}|^2p_2^2}\frac{(\mathbf{k}-\mathbf{q}-\mathbf{p}_3)\cdot\mathbf{p}_3}{|\mathbf{k}-\mathbf{q}|^2p_3^2}  i\mathcal{I}_2(\mathbf{p}, \mathbf{q}, \mathbf{k})\\
&+ \frac{\mathbf{p}\times\mathbf{p}_1}{p^2_{\perp}p_1^2} \frac{(\mathbf{q}-\mathbf{p})\times\mathbf{p}_2}{|\mathbf{q}-\mathbf{p}|^2p_2^2}\frac{(\mathbf{k}-\mathbf{q}-\mathbf{p}_3)\cdot\mathbf{p}_3}{|\mathbf{k}-\mathbf{q}|^2p_3^2}   i\mathcal{I}_3(\mathbf{p},\mathbf{q}, \mathbf{k})\\
&+\left[-\frac{\mathbf{p}\times \mathbf{p}_1}{p_1^2}\frac{\mathbf{q}\times\mathbf{p}}{p_{\perp}^2}  +\frac{1}{2}\frac{\mathbf{p}\cdot\mathbf{p}_1}{p_1^2} \frac{\mathbf{q}\cdot\mathbf{p}}{p_{\perp}^2}\right]\frac{(\mathbf{q}-\mathbf{p})\cdot\mathbf{p}_2}{|\mathbf{q}-\mathbf{p}|^2p_2^2} \frac{(\mathbf{k}-\mathbf{q}-\mathbf{p}_3)\cdot\mathbf{p}_3}{q_{\perp}^2p_3^2} \\
&-\frac{(\mathbf{p}-\mathbf{p}_1)\cdot\mathbf{p}_1}{p_1^2} \frac{(\mathbf{q}-\mathbf{p})\cdot\mathbf{p}_2}{ |\mathbf{q}-\mathbf{p}|^2p_2^2} \left[\frac{1}{2}\frac{(\mathbf{k}-\mathbf{q})\cdot \mathbf{p}_3}{|\mathbf{k}-\mathbf{q}|^2p_3^2} +\frac{(\mathbf{k}-\mathbf{q})\times \mathbf{p}_3}{|\mathbf{k}-\mathbf{q}|^2p_3^2}  \frac{\mathbf{k}\times \mathbf{q}}{q_{\perp}^2}\left(i \frac{(\mathbf{k}\cdot \mathbf{q})}{|\mathbf{k}\times \mathbf{q}|}-1\right)\right]\\
&+\left[- \frac{(\mathbf{p}\times \mathbf{p}_1)}{p_1^2} \frac{\mathbf{q}\cdot\mathbf{p}}{p_{\perp}^2} +\frac{1}{2 }\frac{\mathbf{p}\cdot\mathbf{p}_1}{p_1^2} \frac{ (\mathbf{p}\times \mathbf{q})}{p_{\perp}^2} \right]\frac{(\mathbf{q}-\mathbf{p})\cdot\mathbf{p}_2}{|\mathbf{q}-\mathbf{p}|^2p_2^2} \frac{(\mathbf{k}-\mathbf{q}-\mathbf{p}_3)\cdot\mathbf{p}_3}{|\mathbf{k}-\mathbf{q}|^2p_3^2} \frac{\mathbf{k}\times \mathbf{q}}{q_{\perp}^2} \left(i\frac{\mathbf{k}\cdot(\mathbf{k}-\mathbf{q})}{|\mathbf{k}\times \mathbf{q}|} -1  \right) \\
\end{split}
\end{equation}
and 
\begin{equation}
\begin{split}
&\Upsilon_3^{\perp}(\mathbf{p},\mathbf{q}, \mathbf{p}_1, \mathbf{p}_2, \mathbf{p}_3,\mathbf{k}) \\
=&\frac{\mathbf{p}\times \mathbf{p}_1}{p_{\perp}^2p_1^2} \frac{(\mathbf{q}-\mathbf{p}-\mathbf{p}_2)\cdot\mathbf{p}_2}{|\mathbf{q}-\mathbf{p}|^2p_2^2} \frac{(\mathbf{k}-\mathbf{q}-\mathbf{p}_3)\cdot\mathbf{p}_3}{|\mathbf{k}-\mathbf{q}|^2p_3^2} \frac{(\mathbf{k}\times \mathbf{p}_4)}{k^2_{\perp}p_4^2}i\mathcal{G}_1(\mathbf{p}, \mathbf{q}, \mathbf{k})\\
&+\frac{(\mathbf{p}-\mathbf{p}_1)\cdot\mathbf{p}_1}{p_{\perp}^2p_1^2} \frac{(\mathbf{q}-\mathbf{p}-\mathbf{p}_2)\cdot\mathbf{p}_2}{|\mathbf{q}-\mathbf{p}|^2p_2^2} \frac{(\mathbf{k}-\mathbf{q})\times \mathbf{p}_3}{|\mathbf{k}-\mathbf{q}|^2p_3^2}\frac{(\mathbf{k}\times \mathbf{p}_4)}{k^2_{\perp}p_4^2}i\mathcal{G}_2(\mathbf{p}, \mathbf{q}, \mathbf{k})\\
&+\frac{\mathbf{p}\times \mathbf{p}_1}{p_{\perp}^2p_1^2} \frac{(\mathbf{q}-\mathbf{p})\times \mathbf{p}_2}{ |\mathbf{q}-\mathbf{p}|^2p_2^2} \frac{(\mathbf{k}-\mathbf{q})\times \mathbf{p}_3}{|\mathbf{k}-\mathbf{q}|^2p_3^2} \frac{(\mathbf{k}\times \mathbf{p}_4)}{k^2_{\perp}p_4^2} i\mathcal{G}_3(\mathbf{p}, \mathbf{q}, \mathbf{k})\\
&+\left[-\frac{\mathbf{p}\times \mathbf{p}_1}{p_1^2}\frac{\mathbf{q}\times \mathbf{p}}{p_{\perp}^2} +\frac{1}{2}\frac{\mathbf{p}\cdot\mathbf{p}_1}{p_1^2} \frac{\mathbf{q}\cdot\mathbf{p}}{p_{\perp}^2}\right]\frac{(\mathbf{q}-\mathbf{p})\cdot\mathbf{p}_2}{|\mathbf{q}-\mathbf{p}|^2p_2^2}  \frac{(\mathbf{k}-\mathbf{q})\times \mathbf{p}_3}{ |\mathbf{k}-\mathbf{q}|^2p_3^2}\frac{\mathbf{k}\cdot(\mathbf{k}-\mathbf{q})}{q_{\perp}^2}\\
&\quad +\left[-\frac{1}{2 }\frac{\mathbf{p}\times\mathbf{p}_1}{p_1^2}\frac{\mathbf{k}\cdot\mathbf{p}}{p_{\perp}^2} +\frac{1}{3 } \frac{\mathbf{p}\cdot\mathbf{p}_1}{p_1^2} \frac{\mathbf{p}\times \mathbf{k}}{p_{\perp}^2}\right]\frac{(\mathbf{q}-\mathbf{p})\cdot\mathbf{p}_2}{|\mathbf{q}-\mathbf{p}|^2 p_2^2}\frac{(\mathbf{k}-\mathbf{q})\cdot\mathbf{p}_3}{|\mathbf{k}-\mathbf{q}|^2p_3^2}\\
&\quad +i\left[-\frac{\mathbf{p}\times \mathbf{p}_1}{p_1^2} \frac{\mathbf{q}\cdot\mathbf{p}}{p_{\perp}^2} + \frac{1}{2 }\frac{\mathbf{p}\cdot\mathbf{p}_1}{p_1^2}\frac{(\mathbf{p}\times \mathbf{q})}{p_{\perp}^2}\right]\frac{(\mathbf{q}-\mathbf{p})\cdot\mathbf{p}_2}{|\mathbf{q}-\mathbf{p}|^2\mathbf{p}_2^2} \frac{(\mathbf{k}-\mathbf{q})\times \mathbf{p}_3}{|\mathbf{k}-\mathbf{q}|^2p_3^2}  \frac{(\mathbf{k}\times \mathbf{q})}{q^2_{\perp} }\frac{(k_{\perp}^2+q_{\perp}^2-\mathbf{k}\cdot\mathbf{q})}{|\mathbf{k}\times \mathbf{q}|} \\
&\quad -\frac{(\mathbf{p}-\mathbf{p}_1)\cdot\mathbf{p}_1}{p_1^2} \frac{(\mathbf{q}-\mathbf{p})\cdot\mathbf{p}_2}{|\mathbf{q}-\mathbf{p}|^2p_2^2} \frac{(\mathbf{k}-\mathbf{q}-\mathbf{p}_3)\cdot\mathbf{p}_3}{|\mathbf{k}-\mathbf{q}|^2p_3^2}  \frac{\mathbf{k}\times \mathbf{q}}{q_{\perp}^2 }\left(i\frac{\mathbf{q}\cdot(\mathbf{q}-\mathbf{k})}{|\mathbf{q}\times \mathbf{k}|} +1\right).
\end{split}
\end{equation}
See Fig.~\ref{fig:M1M5_squared_4wl} for the momentum assignments. 

In addition to the explicit imaginary part that is associated with the sign function $\mathrm{sgn}(\mathbf{k}\times \mathbf{q})$, the six functions $\mathcal{I}_{1,2,3}$ and $\mathcal{G}_{1,2,3}$ are in general complex. Therefore the effective vertex is complex. 

Inspecting   Eq.~ \eqref{eq:M1M5_final_result}, we conclude that we only need to consider three terms as far as the color structure is concerned. The color structures of the terms containing two, three and four  Wilson lines are shown in Fig.~\ref{fig:M1M5}. Of course, the complication is hidden in the three kinematic functions $\mathcal{F}_1, \mathcal{F}_2, \mathcal{F}_3$. With the help of the effective vertices, these coefficients functions can also be computed.  

Adding Eq.~ \eqref{eq:M3_squared_result} and Eq.~ \eqref{eq:M1M5_final_result}, one obtains the main result of this paper, the first saturation correction to single gluon production 
\begin{equation}\label{eq:FSC_final_result}
\begin{split}
n^{(6)}(\mathbf{k})
=&\frac{1}{\pi}\int_{\mathbf{p}, \mathbf{q}, \mathbf{p}_2, \mathbf{p}_4}\mathcal{H}_1(\mathbf{p}, \mathbf{q}, \mathbf{p}_2, \mathbf{p}_4, \mathbf{k}) f^{ab_1b_2}f^{cb_3b_4}\rho_P^{b_1}(\mathbf{p}-\mathbf{p}_2)\rho_P^{b_2}(\mathbf{p}_2) \rho_P^{b_3}(\mathbf{q}-\mathbf{p}_4)\rho_P^{b_4}(\mathbf{p}_4) \\
&\qquad\qquad \times U^{ae}(\mathbf{k}-\mathbf{p})U^{ce}(-\mathbf{k}-\mathbf{q})\\
&+\frac{1}{\pi}\int_{\mathbf{p}, \mathbf{q}, \mathbf{p}_2, \mathbf{p}_4} \mathcal{F}_1(\mathbf{p}, \mathbf{q}, \mathbf{p}_2, \mathbf{p}_4, \mathbf{k}) f^{dab_3}f^{ab_1b_2} \rho_P^{b_1}(\mathbf{p}-\mathbf{p}_2)\rho_P^{b_2}(\mathbf{p}_2) \rho_P^{b_3}(\mathbf{q}-\mathbf{p})\rho_P^{b_4}(\mathbf{p}_4)\\
&\qquad\qquad \times U^{de}(\mathbf{k}-\mathbf{q})U^{b_4e}(-\mathbf{k}-\mathbf{p}_4)\\
&+\frac{1}{\pi}\int_{\mathbf{p}, \mathbf{q}, \mathbf{p}_1, \mathbf{p}_2, \mathbf{p}_4} \mathcal{H}_2(\mathbf{p},\mathbf{q}, \mathbf{p}_1, \mathbf{p}_2, \mathbf{p}_4, \mathbf{k}) f^{db_3b_4}f^{ebc}\rho^{b_1}_P(\mathbf{p}_1)\rho^{b_2}_P(\mathbf{p}_2)\rho^{b_3}_P(\mathbf{q}-\mathbf{p}_4)\rho^{b_4}_P(\mathbf{p}_4)\\
&\qquad\qquad \times U^{b_1b}(\mathbf{k}-\mathbf{p}-\mathbf{p}_1)U^{b_2c}(\mathbf{p}-\mathbf{p}_2)U^{de}(-\mathbf{k}-\mathbf{q})\\
&+\frac{1}{\pi}\int_{\mathbf{p}, \mathbf{q}, \mathbf{p}_2, \mathbf{p}_3, \mathbf{p}_4} \mathcal{F}_2(\mathbf{p}, \mathbf{q}, \mathbf{p}_2, \mathbf{p}_3, \mathbf{p}_4, \mathbf{k}) f^{bb_1b_2} f^{ade} \rho_P^{b_1}(\mathbf{p}-\mathbf{p_2})\rho_P^{b_2}(\mathbf{p}_2)\rho_P^{b_3}(\mathbf{p}_3)\rho_P^{b_4}(\mathbf{p}_4)\\
&\qquad\qquad \times U^{ba}(\mathbf{q}-\mathbf{p}) U^{b_3 d}(\mathbf{k}-\mathbf{q}-\mathbf{p}_3)U^{b_4 e}(-\mathbf{k}-\mathbf{p}_4)\\
&+\frac{1}{\pi}\int_{\mathbf{p}, \mathbf{q}, \mathbf{p}_1, \mathbf{p}_2, \mathbf{p}_3, \mathbf{p}_4}\mathcal{H}_3(\mathbf{p},\mathbf{q}, \mathbf{p}_1, \mathbf{p}_2, \mathbf{p}_3, \mathbf{p}_4, \mathbf{k})  f^{abc}f^{ade} \rho_P^{b_1}(\mathbf{p}_1)\rho_P^{b_2}(\mathbf{p}_2)\rho_P^{b_3}(\mathbf{p}_3)  \rho_P^{b_4}(\mathbf{p}_4)\\
&\qquad\qquad \times U^{b_1 b}(\mathbf{k}-\mathbf{p}-\mathbf{p}_1)U^{b_2 c}(\mathbf{p}-\mathbf{p}_2) U^{b_3d}(-\mathbf{k}-\mathbf{q}-\mathbf{p}_3) U^{b_4 e}(\mathbf{q}-\mathbf{p}_4)\\
&+\frac{1}{\pi}\int_{\mathbf{p}, \mathbf{q}, \mathbf{p}_1, \mathbf{p}_2, \mathbf{p}_3, \mathbf{p}_4} \mathcal{F}_3(\mathbf{p}, \mathbf{q}, \mathbf{p}_1, \mathbf{p}_2, \mathbf{p}_3, \mathbf{p}_4, \mathbf{k})f^{abc}f^{ade}\rho^{b_1}_P(\mathbf{p}_1)\rho^{b_2}_P(\mathbf{p}_2)\rho^{b_3}_P(\mathbf{p}_3)\rho^{b_4}_P(\mathbf{p}_4)\\
&\qquad\qquad\times U^{b_1b}(\mathbf{p}-\mathbf{p}_1)U^{b_2c}(\mathbf{q}-\mathbf{p}-\mathbf{p}_2) U^{b_3 d}(\mathbf{k}-\mathbf{q}-\mathbf{p}_3) U^{b_4e}(-\mathbf{k}-\mathbf{p}_4)\\
&+c.c.\,.
\end{split}
\end{equation}
There are a few comments about $n^{(6)}(\mathbf{k})$. First of all, it is impossible to relabel the integration momenta to make the color structures of the terms that have equal number of Wilson lines the same. Taking the two terms having four Wilson lines as an example, denoting the eikonally rotated color charge density $\rho_R^b(\mathbf{k})=\rho_P^a(\mathbf{p})U^{ab}(\mathbf{k}-\mathbf{p})$, the two color structures are
\begin{equation}
\begin{split}
&-f^{abc}f^{ade} \rho_R^b(\mathbf{k}-\mathbf{p}) \rho_R^c(\mathbf{p})\rho_R^d(-\mathbf{k}-\mathbf{q}) \rho^e_R(\mathbf{q}) = \Big[\rho_R(\mathbf{k}-\mathbf{p}), \rho_R(\mathbf{p})\Big]^a\Big[\rho_R(-\mathbf{k}-\mathbf{q}), \rho_R(\mathbf{q})\Big]^a,\\
&-f^{abc}f^{ade} \rho_R^{b}(\mathbf{p})\rho_R^c(\mathbf{q}-\mathbf{p}) \rho_R^{d}(\mathbf{k}-\mathbf{q}) \rho_R^e(-\mathbf{k}) = \Big[ \rho_R(\mathbf{p}), \rho_R(\mathbf{q}-\mathbf{p})\Big]^a \Big[\rho_R(\mathbf{k}-\mathbf{q}), \rho_R(-\mathbf{k})\Big]^a.\\
\end{split}
\end{equation}
Here $\mathbf{k}$ is some given external momentum. The integration variables are $\mathbf{p}, \mathbf{q}$.  It is obvious that one can not redefine $\mathbf{p}, \mathbf{q}$ to make these two expressions the same.  On the other hand, if one is interested in the total number of gluons rather than the gluon spectrum, with an additional integration $\int d^2\mathbf{k}$ in the game, one can redefine $\mathbf{q}\rightarrow \mathbf{k}$ and $\mathbf{k}\rightarrow -\mathbf{q}$ in the second expression so that the two color structures are the same.  There is another possibility that after averaging over the projectile color charge configuration, terms with equal number of Wilson lines might be combined. We have explicitly checked this situation using the MV model, however, only partial terms could be combined and no significant cancellations happen.  

The $n^{(6)}(\mathbf{k})$ is expressed as a functional of the color charge densities $\rho_P$ and $\rho_T$ (through the Wilson line U). This event-by-event expression is useful for computing double/multiple gluon productions. To obtain event averaged gluon spectrum one needs to perform ensemble average over all possible color charge configurations of the projectile and the target. One popular statistical distribution for the color charge density used in the literature is the Gaussian distribution. It is straightforward to carry out the average over $\rho_P$.  However, the average over $\rho_T$ contains four adjoint Wilson lines and is in general very complicated albeit possible \cite{Blaizot:2004wv, Dominguez:2012ad, Shi:2017gcq}. There also exists other approximation schemes, see discussions in Refs. \cite{Lappi:2015vta, Kovner:2017ssr, Altinoluk:2018ogz}. Event averaging will be studied in a follow-up paper. 

As far as color structure is concerned, $n^{(6)}(\mathbf{k})$ can be simply computed from the six diagrams in Fig.~\ref{fig:M3_squared} and Fig.~\ref{fig:M1M5}. These six diagrams capture the correct dependence on $\rho_P$ and $\rho_T$. Of course, the realistic coefficients $\mathcal{F}_{1,2,3}$ and $\mathcal{H}_{1,2,3}$ can not be read from the diagrams. However, with the help of the effective vertices, one can directly computing the correct expression for $n^{(6)}(\mathbf{k})$ from the six diagrams . 

Compared with the leading order gluon production $n^{(2)}(\mathbf{k})$, the most important feature of the first saturation correction $n^{(6)}(\mathbf{k})$ is that it includes nontrivial \textit{final state interactions}. This is explicitly shown in figs. \ref{fig:M3_squared_3wl}, \ref{fig:M3_squared_4wl}, \ref{fig:M1M5_squared_3wl} and \ref{fig:M1M5_squared_4wl}. Gluons produced by different color sources merge together after independently scattering on the target nucleus. Figs. \ref{fig:M3_squared_2wl} and \ref{fig:M1M5_squared_2wl} only contain \textit{initial state effects}, no gluon interactions after scattering with the nucleus. 

Finally, as a consistence check, when the target Wilson line $U(\mathbf{x})=1$, Eq.~\eqref{eq:FSC_final_result} vanishes as expected because no gluon will be produced if there are no interactions with the  nucleus.

%\subsubsection{First saturation correction in the  MV model}
%The obtained NNLO term can be averaged over the projectile configurations. To that end, we will use the Gaussian MV model. We explicitly assume that  the color charge density fluctuations of the proton is 
%homogenous in the transverse plane, i.e.  $\mu_P^2$ is independent of  the transverse momentum. The problem is straightforward but tedious due to combinatorics. We present here only the final result

\section{Conclusions}
\label{Sect:Conclusions}
In this tour de force calculations, we obtained the first saturation correction to the single inclusive semi-hard gluon production in high energy proton-nucleus collisions. The main result is given in Eq.~ \eqref{eq:FSC_final_result}.  It is expressed as a functional of the color charge densities of the projectile $\rho_p$ and the target $\rho_T$. Interestingly we found that to obtain the correct functinal dependence on $\rho_{P(T)}$, we only need to compute the six diagrams shown in Fig.~\ref{fig:M3_squared} and Fig.~\ref{fig:M1M5}.  The coefficients of these functional dependence, however, have to be determined by a full analysis of all the contributing diagrams or through the tedious computation using the LSZ reduction formula, as have been done in this paper. To encapsulate our results, we introduced several effective vertices. With these effective vertices, directly computing the six diagrams in Fig.~\ref{fig:M3_squared} and Fig.~\ref{fig:M1M5} leads to the correct result for the first saturation correction to single gluon production. 

The order-$g$ effective vertex $L^j$ is the famous Lipatov vertex. For the two effective vertices at order-$g^3$, $\Gamma_1^j, \Gamma_2^j$ and the three effective vertices at order-$g^5$, $\Upsilon_1^j, \Upsilon^j_2, \Upsilon_3^j$, although they are very convenient quantities to organize the calculations, they  represent summations of many diagrams and it is hard to percieve their physical meanings. 

Eq. \eqref{eq:FSC_final_result} is an event-by-event result, in a follow-up paper, we will averaging over the color configurations. The averaging heavily depends on different modelings of the statistical distributions of $\rho_P$ and $\rho_T$. It is probably more reasonable to model the dilute proton and the dense nucleus differently.  

Our ultimate goal is to evaluate the saturation effects of the dilute object on double gluon productions in high energy dilute-dense scatterings, this includes proton-nucleus collisions as well as light-heavy ion collisions. Since the first saturation effects also contains non trivial final state interactions, we hope to study this effect, work out the phenomenology and compare with data from LHC and RHIC (see early works \cite{Mace:2018vwq, Mace:2018yvl}).

In appendix. \ref{sec:comparison_with_LCPT}, we made comparisons with the order-$g^3$ gluon production amplitude derived using the light-cone perturbation theory in \cite{Chirilli:2015tea}. The odd part under $\mathbf{k}\leftrightarrow -\mathbf{k}$ of the order-$g^3$ amplitude is shown to be exactly the same even though their calculation was done in the light-cone gauge while ours is done in the Fock-Schwinger gauge. Other than that, comparisons at amplitude level at two different gauges would not make sense. Only the gauge invariant gluon production should be compared. 

It is also possible that beyond the leading order, rapidity dependent fluctuations might play an important role in semi-hard gluon productions, see Refs.~\cite{Romatschke:2005pm,  Gelis:2013rba} for their influences on the bulk properties of the Glasma (recall that we assumed boost invariance in this paper). It would be interesting to evaluates its contribution compared with those from the first saturation correction. In this sense, our complete result on the first saturation correction, besides its own importance, also serves as a very useful reference. \\

 \acknowledgments
 
 We thank A. Dumitru, A. Kovner, Yu. Kovchegov, M. Lublinsky,   L. McLerran, and R. Venugopalan for insightful discussions and collaboration on related projects. We thank H. Duan for his contribution in the exploratory stage of this project. 

We acknowledge support by the DOE Office of Nuclear Physics through Grant No. DE-SC0020081.\\

\appendix

\section{Definite integrals involving Bessel functions}
\label{App:BInt}

In order to simplify the readers task to reproduce our results, we list all definite integrals involving Bessel function that were used in the main body of this paper.

\subsection*{Integrals involving one Bessel function}
\begin{equation}
\begin{split}
\int_0^{\infty} d\tau \tau H_0^{(2)}(k_{\perp}\tau) =&\int_0^{\infty} d\tau \tau \left(J_0(k_{\perp}\tau) -iY_0(k_{\perp}\tau)\right) \\
=& \delta(k_{\perp})/k_{\perp} -i\frac{2}{\pi} \frac{1}{k_{\perp}^2}= -i\frac{2}{\pi} \frac{1}{k_{\perp}^2 -i\epsilon}.
\end{split}
\end{equation}
Here we used the following regularization procedure in order to assign the meaning to  the integrals over $J_0$ and $Y_0$:
\begin{itemize}
\item
Taking the limit $p_{\perp} \rightarrow 0$, in the closure relation~\eqref{Eq:JClosure}
\begin{equation}
\int_0^{\infty} d\tau \tau J_0(k_{\perp}\tau)  = \lim_{p\to0} \int_0^{\infty} d\tau \tau J_0(k_{\perp}\tau) J_0(p_{\perp}\tau) = \delta(k_{\perp})/k_{\perp}.
\end{equation}

\item  Additionally  from Eq.~\eqref{Eq:CrossprodClosure}, we have
\begin{equation}
\int_0^{\infty} d\tau \tau Y_0(k_{\perp}\tau) = \lim_{p\to0} \int_0^{\infty} d\tau \tau Y_0(k_{\perp}\tau) J_0(p_{\perp}\tau)
= \frac{2}{\pi} \frac{1}{k_{\perp}^2}.
\end{equation}
\end{itemize}

\subsection*{Integrals involving two Bessel functions}
The closure relation for Bessel functions is
\begin{equation}
\label{Eq:JClosure}
\int_0^{\infty} x J_{\alpha} (ax)J_{\alpha}(bx) dx = \frac{1}{b} \delta(a-b)\,.
\end{equation}
Together with the crossproducts
\begin{align}
&\int_0^{\infty} x J_1(ax) Y_1(bx) dx =\frac{2a}{\pi b} \frac{1}{b^2-a^2},\\
&\int_0^{\infty} xJ_0(ax)Y_0(bx) dx =  \frac{2}{\pi} \frac{1}{b^2-a^2}.
\label{Eq:CrossprodClosure}
\end{align}
the closure relation  leads to
\begin{equation}\label{eq:integral_J1H1}
\begin{split}
\int_0^{\infty} x J_1(ax)H_1^{(2)}(bx) dx =& \frac{1}{b}\delta(a-b) -i \frac{2a}{\pi b} \frac{1}{b^2-a^2}\\
%=&-i\frac{2a}{\pi b}\left(\frac{1}{b^2-a^2} + i\pi \delta(b^2-a^2)\right)\\
=&-i\frac{2a}{\pi b} \frac{1}{b^2-a^2-i\epsilon}
\end{split}
\end{equation}
and
\begin{equation}\label{eq:integral_J0H0}
\begin{split}
\int_0^{\infty}xJ_0(ax) H_0^{(2)}(bx) dx = &\frac{1}{b}\delta(a-b) - i \frac{2}{\pi} \frac{1}{b^2-a^2}\\
%=&-i\frac{2}{\pi} \left(\frac{1}{b^2-a^2} +i\pi \frac{1}{2b}\delta(b-a)\right)\\
=&-i\frac{2}{\pi}  \frac{1}{b^2-a^2-i\epsilon}.
\end{split}
\end{equation}

\subsection*{Integrals involving three Bessel functions}
\label{ap:integral_formulas_three_bessels}

Here we list the most general expressions; they are followed by the specific integrals used in the main body of the paper.

\subsubsection*{General formulae}
\label{ap:integral_formulas_three_bessels}
The following integrals contain products of three Bessel functions in the integrands. (see more in Ref.~\cite{Prudnikov1986}):
\begin{itemize}
\item{Integrand $x J_0(ax)J_1(bx)H_1^{(2)}(cx)$}

Lets consider the real and imaginary part separately, $H^{(2)}_1(cx) = J_1(cx) -iY_1(cx)$.

For positive  $a,b,c>0$, the integral
\begin{equation}
\int_0^{\infty} x J_0(ax)J_1(bx)J_1(cx) dx = \frac{1}{\pi bc} \frac{r_{+}}{\sqrt{1-r_{+}^2}}, \qquad r_{+} =\frac{b^2+c^2-a^2}{2bc}
\end{equation}
is non-zero only if   $|c-b|<a<c+b $.

 The imaginary part of the integrand  involves $Y_1(cx)$. The associated integral is given by
\begin{equation}
\begin{split}
&\int_0^{\infty} x J_0(ax)J_1(bx)Y_1(cx) dx = I,\\
&I= \frac{1}{\pi bc} \left(\pm \frac{r_+}{\sqrt{r_+^2-1}} -1\right), \qquad\qquad   a<\pm (c-b)\\
&I= -\frac{1}{\pi bc},\qquad\qquad \qquad \qquad\qquad \qquad|b-c|<a<b+c \\
&I=\frac{1}{\pi bc} \left(-\frac{r_+}{\sqrt{r_+^2-1}}-1\right). \quad\qquad\qquad a>b+c
\end{split}
\end{equation}

In order to simplify notation it is convenient to introduce
\begin{equation}
L_{011}(a,b,c)\equiv\int_0^{\infty} x J_0(ax)J_1(bx)H^{(2)}_1(cx) dx \,.
\end{equation}

Explicitly
\begin{equation}\label{eq:L011}
\begin{split}
L_{011}(a,b,c)  %=
% &\frac{1}{\pi bc} \Bigg\{-i\Theta(c-b -a)\left(\frac{r_+}{\sqrt{r_+^2-1}} -1\right) \\
%&-i\Big[\Theta(b-c -a)+\Theta(a-(b+c))\Big] \left(-\frac{r_+}{\sqrt{r_+^2-1}}  -1\right)\\
%&+\Theta(b+c-a)\Theta(a-|b-c|)\left(\frac{r_+}{\sqrt{1-r_+^2}} +i\right)\Bigg\}\\
=&\frac{i}{\pi bc}  +\frac{1}{\pi bc}\frac{r_+}{\sqrt{|r_+^2-1|}} \Big( -i\Theta(c-b-a) + i\Theta(b-c-a)+i\Theta(a-b-c) \\
&\qquad+ \Theta(b+c-a)\Theta(a-|b-c|)\Big)\,.
\end{split}
\end{equation}
In particular, for $|c-b|<a<c+b$, the above expression simplifies to
\begin{equation}
L_{011}(a,b,c) = \frac{1}{\pi bc} \left(\frac{r_{+}}{\sqrt{1-r_{+}^2}} +i\right), \qquad\qquad |c-b|<a<c+b\,.
\end{equation}

\item{Integrand $ x J_1(ax)J_1(bx)H_0^{(2)}(cx)$}

For the real part, we quote the formula from \cite{Prudnikov1986} ($a, b, c >0$)
\begin{equation}
\int_0^{\infty} \tau J_1(a\tau) J_1(b\tau)J_0(c\tau) d\tau = \frac{1}{\pi ab} \frac{r_+}{\sqrt{1-r_+^2}}.
\end{equation}
It is nonvanishing only in the parameter region $|a-b|<c<a+b$. Here
\begin{equation}
r_{\pm} = \pm \frac{a^2+b^2-c^2}{2ab}
\end{equation}
and $a, b, c >0$.

The imaginary part of the integrand  involving $Y_0(c\tau)$ deserves more explanation. From  Refs.~\cite{Prudnikov1986,Gervois1984},
\begin{equation}
\begin{split}
&\int_0^{\infty} \tau J_1(a\tau)J_1(b\tau) Y_0(c\tau) d\tau  = I ,\\
&I = i\frac{\sqrt{2}}{\pi^{3/2} ab} (r_+^2-1)^{-1/4} Q^{1/2}_{1/2} (r_+), \qquad \qquad  c<|a-b|  \\
&I = -\frac{\sqrt{2}}{\pi^{3/2} ab} (1-r_+^2)^{-1/4} Q^{1/2}_{1/2}(r_+). \qquad\qquad |a-b|<c<a+b \\
&I = i \frac{\sqrt{2}}{\pi^{3/2} ab} (r_-^2-1)^{-1/4} Q^{1/2}_{1/2}(r_-). \qquad\qquad c>a+b.
\end{split}
\end{equation}
Here $Q^{\mu}_{\nu}(x)$ is the associated Legendre function of second kind. The explicit form is not very transparent ($|z|>1$)
\begin{align}\notag
Q_{\nu}^{\mu} (z) &= \frac{e^{i\mu \pi}\sqrt{\pi}}{2^{\nu+1}} \frac{\Gamma(\mu+\nu+1)}{\nu+\frac{3}{2}} \frac{(z^2-1)^{\mu/2}}{z^{\mu+\nu+1}}
 \\ & \times {}_2F_1\left(\frac{\mu+\nu+1}{2}, \frac{\mu+\nu+2}{2}; \nu+\frac{3}{2}; \frac{1}{z^2} \right)\,.
\end{align}
%Here the hypergeometric function is
%\begin{equation}
%{}_2F_1(a,b;c, z) = \frac{\Gamma(c)}{\Gamma(a)\Gamma(b)} \sum_0^{\infty} \frac{\Gamma(a+n)\Gamma(b+n)}{\Gamma(c+n)} \frac{z^n}{n!}.
%\end{equation}
The integral representation~\cite{abramowitz+stegun}
\begin{equation}
Q^{\mu}_{\nu} (z)= \frac{e^{i\mu\pi}\pi^{1/2} (z^2-1)^{\mu/2}\Gamma(\mu+\nu+1)}{2^{\mu} \Gamma(\mu+\frac{1}{2}) \Gamma(\nu-\mu+1)}\int_0^{\infty} \frac{(\sinh{t})^{2\mu}}{( z+ (z^2-1)^{1/2}\cosh{t})^{\nu+\mu+1}} dt
\end{equation}
enables us to obtain, for $z>1$,
\begin{equation}\label{eq:Qonehalf_offCut}
Q^{1/2}_{1/2}(z) = i\left(\frac{\pi}{2}\right)^{1/2} (z^2-1)^{-1/4} \left(z- \sqrt{z^2-1} \right)\,.
\end{equation}
Note that the associated Legendre functions have a cut located on $(-\infty, 1)$. For a  real $z$ in the range $-1<z<1$, the  analytic continuation can be implemented~\cite{Erdelyi1953}
\begin{equation}\label{eq:Qmunu_onCut}
\begin{split}
&Q_{\nu}^{\mu}(x) = \frac{1}{2} e^{-i\mu\pi}\Big(e^{-i\mu\pi/2} Q_{\nu}^{\mu}(x+i0) + e^{i\mu\pi/2} Q_{\nu}^{\mu}(x-i0)\Big)\,.
\end{split}
\end{equation}
Using the result of Eq.~\eqref{eq:Qonehalf_offCut}
\begin{equation}
\begin{split}
e^{-i\pi/4} Q_{1/2}^{1/2}(x+i0) = &i\left(\frac{\pi}{2}\right)^{1/2} \left[(1-x^2)e^{i\pi}\right]^{-1/4} \Big( x+i0 - ((1-x^2)e^{i\pi})^{1/2}\Big)e^{-i\pi/4}\\
=&\left(\frac{\pi}{2}\right)^{1/2} (1-x^2)^{-1/4} (x+i0 -i(1-x^2)^{1/2})
\end{split}
\end{equation}
\begin{equation}
\begin{split}
e^{i\pi/4} Q_{1/2}^{1/2}(x-i0) = &i\left(\frac{\pi}{2}\right)^{1/2} \left[(1-x^2)e^{-i\pi}\right]^{-1/4} \Big( x-i0 - ((1-x^2)e^{-i\pi})^{1/2}\Big)e^{i\pi/4}\\
=&-\left(\frac{\pi}{2}\right)^{1/2} (1-x^2)^{-1/4} (x-i0 +i(1-x^2)^{1/2})
\end{split}
\end{equation}
and substituting these two expressions into Eq.~\eqref{eq:Qmunu_onCut}, one obtains, for $-1<x<1$
\begin{equation}
Q^{1/2}_{1/2}(x)  =-\sqrt{ \frac{\pi}{2}}(1-x^2)^{\frac{1}{4}}.
\end{equation}
We thus obtain the final result for  the imaginary part of the integrand
\begin{equation}
\begin{split}
&\int_0^{\infty} \tau J_1(a\tau)J_1(b\tau) Y_0(c\tau) d\tau  = I ,\\
&I = -\frac{1}{\pi ab} \left(\frac{r_+}{\sqrt{r_+^2-1}} -1\right), \qquad \qquad \{ c<|a-b|   \}\\
&I = \frac{1}{\pi ab},  \qquad\qquad\qquad\qquad\qquad \{|a-b|<c<a+b\} \\
&I =-\frac{1}{\pi ab} \left(\frac{r_-}{\sqrt{r_-^2-1}} -1\right). \qquad\qquad \{c>a+b\}.
\end{split}
\end{equation}
For convenienence, we introduce the notation
\begin{equation}
 L_{110}(a, b, c) \equiv \int_0^{\infty} \tau J_1(a\tau)J_1(b\tau) H^{(2)}_0(c\tau) d\tau
\end{equation}
with the  explicit expression
\begin{equation}
\begin{split}
L_{110}(a,b,c) = &\frac{i}{\pi ab}\Big\{\Theta( |a-b|-c)\Big(\frac{r_+}{\sqrt{r_+^2-1}} -1\Big) - \Theta(c-a-b) \Big(\frac{r_+}{\sqrt{r_+^2-1}} +1\Big)\Big\}\\
&+\Theta(a+b-c)\Theta(c-|a-b|) \frac{1}{\pi ab} \Big( \frac{r_+}{\sqrt{1-r_+^2}}-i\Big).
\end{split}
\end{equation}
In the particular case $|a-b|<c<a+b$,
\begin{equation}
L_{110}(a,b,c) = \frac{1}{\pi ab} \Big( \frac{r_+}{\sqrt{1-r_+^2}}-i\Big).
\end{equation}

\item{Integrand $xJ_0(ax)J_0(bx)H_0^{(2)}(cx)$}

From Ref.~\cite{Prudnikov1986}, we find that the real part of the integral ($a,b,c>0$)
\begin{equation}
\int_0^{\infty} xJ_0(ax)J_0(bx)J_0(cx) dx =  \frac{1}{\pi ab} \frac{1}{\sqrt{1-r_+^2}}, \qquad r_{+} =  \frac{a^2+b^2-c^2} {2ab}
\end{equation}
is nonzero only for  $|a-b|<c<a+b$.  On the other hand, the imaginary part
\begin{equation}
\begin{split}
\int_0^{\infty} xJ_0(ax)J_0(bx)Y_0(cx) dx =& \pm \frac{1}{\pi ab} \frac{1}{\sqrt{r_+^2-1}}
\end{split}
\end{equation}
is  nonvanishing only in the parameter range $c>a+b$ and $c<|a-b|$.

It is handy to use the notation
\begin{equation}
L_{000}(a,b,c) = \int_0^{\infty} xJ_0(ax)J_0(bx)H^{(2)}_0(cx) dx
\end{equation}
for the explicit form
\begin{equation}
\begin{split}
L_{000}(a,b,c) &= \frac{1}{\pi  ab} \Big(\Theta(a+b-c)\Theta(c-|a-b|) \frac{1}{\sqrt{1-r_+^2}} \\ & +   \Big[\Theta(c-a-b) -\Theta(|a-b|-c)\Big]\frac{-i}{\sqrt{r_+^2-1}}\Big).
\end{split}
\end{equation}
In particular, when $|a-b|<c<a+b$, it simplifies to
\begin{equation}
L_{000}(a,b,c) =  \frac{1}{\pi ab} \frac{1}{\sqrt{1-r_+^2}}\,.
\end{equation}

 \end{itemize}

\subsubsection*{Specific formulae}
 Using the above general forms for the integrals involving three Bessel functions, it is straightforward to establish the following  list of identities
\begin{align}
&\int_0^{\infty} d\tau \tau J_0(p_{\perp}\tau)J_1(|\mathbf{k}-\mathbf{p}|\tau)  H_1^{(2)}(k_{\perp}\tau)  = \frac{1}{\pi k_{\perp}|\mathbf{k}-\mathbf{p}|} \left(\frac{\mathbf{k}\cdot(\mathbf{k}-\mathbf{p})}{|\mathbf{k}\times \mathbf{p}|} +i\right),\\
&\int_0^{\infty} d\tau \tau J_1(p_{\perp}\tau)J_1(|\mathbf{k}-\mathbf{p}|\tau) H_0^{(2)}(k_{\perp}\tau)  = \frac{1}{\pi p_{\perp} |\mathbf{k}-\mathbf{p}|} \left(\frac{\mathbf{p}\cdot(\mathbf{p}-\mathbf{k})}{|\mathbf{p}\times \mathbf{k}|}  - i\right),\\
&\int_0^{\infty} d\tau \tau J_0(p_{\perp}\tau)J_0(|\mathbf{k}-\mathbf{p}|\tau) H_0^{(2)}(k_{\perp}\tau)  = \frac{1}{\pi |\mathbf{p}\times \mathbf{k}|},\\
&\int_0^{\infty} d\tau \tau J_0(|\mathbf{k}-\mathbf{q}|\tau) J_1(q_{\perp}\tau) H_1^{(2)}(k_{\perp}\tau) = \frac{1}{\pi q_{\perp}k_{\perp}} \left(\frac{\mathbf{k}\cdot\mathbf{q}}{|\mathbf{k}\times \mathbf{q}| }+i\right),\\
&\int_0^{\infty} d\tau \tau J_0(|\mathbf{k}-\mathbf{q}|\tau) J_1(w_{\perp}\tau) H_1^{(2)}(k_{\perp}\tau)  = L_{011}(|\mathbf{k}-\mathbf{q}|, w_{\perp}, k_{\perp}),\\
&\int_0^{\infty} d\tau \tau J_0(q_{\perp}\tau) J_1(|\mathbf{k}-\mathbf{q}|\tau) H_1^{(2)}(k_{\perp}\tau) = \frac{1}{\pi |\mathbf{k}-\mathbf{q}|k_{\perp}} \left(\frac{\mathbf{k}\cdot(\mathbf{k}-\mathbf{q})}{|\mathbf{k}\times \mathbf{q}|} + i \right),\\
%&\int_0^{\infty} d\tau \tau J_1(|\mathbf{k}-\mathbf{q}|\tau)H_1^{(2)} (k_{\perp}\tau)
%=-i\frac{ 2|\mathbf{k}-\mathbf{q}|}{\pi k_{\perp}} \frac{1}{\mathbf{q}\cdot(2\mathbf{k}-\mathbf{q}) -i\epsilon},\\
&\int_0^{\infty} d\tau \tau J_0(w_{\perp}\tau)  J_1(|\mathbf{k}-\mathbf{q}|\tau)H_1^{(2)} (k_{\perp}\tau) = L_{011}(w_{\perp},|\mathbf{k}-\mathbf{q}|, k_{\perp}), \\
&\int_0^{\infty} d\tau \tau J_1(q_{\perp}\tau)J_1(|\mathbf{k}-\mathbf{q}|\tau) H_0^{(2)}(k_{\perp}\tau)  =\frac{1}{\pi q_{\perp} |\mathbf{k}-\mathbf{q}|} \left(\frac{\mathbf{q}\cdot(\mathbf{q}-\mathbf{k})}{|\mathbf{q}\times \mathbf{k}|} -i\right) %\equiv L_{110}(q_{\perp}, |\mathbf{k}-\mathbf{q}|,k_{\perp})
, \\
&\int_0^{\infty} d\tau \tau J_1(w_{\perp}\tau)J_1(|\mathbf{k}-\mathbf{q}|\tau) H_0^{(2)}(k_{\perp}\tau)=L_{110}( w_{\perp},|\mathbf{k}-\mathbf{q}|, k_{\perp}) , \\
&\int_0^{\infty}d\tau \tau J_0(q_{\perp}\tau)J_0(|\mathbf{k}-\mathbf{q}|\tau)H_0^{(2)}(k_{\perp}\tau) = \frac{1}{\pi |\mathbf{k}\times \mathbf{q}|}, \\
&\int_0^{\infty}d\tau \tau J_0(w_{\perp}\tau)J_0(|\mathbf{k}-\mathbf{q}|\tau)H_0^{(2)}(k_{\perp}\tau) = L_{000}(w_{\perp}, |\mathbf{k}-\mathbf{q}|, k_{\perp})
%&\int_0^{\infty} d\tau \tau J_0(|\mathbf{k}-\mathbf{q}|\tau) H_0^{(2)}(k_{\perp}\tau) =\frac{-2i}{\pi} \frac{1}{k_{\perp}^2 - |\mathbf{k}-\mathbf{q}|^2 -i\epsilon}
\end{align}

\subsection*{Integrals involving four Bessel Functions}
\label{ap:integral_formulas_four_bessels}

We also need integrals involving  four Bessel Functions.
Our strategy is to reduce the number of Bessel functions in the integrand from four to three, which we have explicit formula do the integration.
\begin{itemize}
\item $\int_0^{\infty}d\tau \tau J_0(p_{\perp}\tau)J_1(|\mathbf{q}-\mathbf{p}|\tau)J_1(|\mathbf{k}-\mathbf{q}|\tau) H_0^{(2)}(k_{\perp}\tau) $:

By applying the Garf's formula  (see Appendix of Ref.~\cite{Li:2021zmf})
\begin{equation}
J_0(p_{\perp}\tau)J_1(|\mathbf{q}-\mathbf{p}|\tau) = \int_{-\pi}^{\pi}\frac{d \theta}{2\pi} e^{i\Psi} J_1(w_{\perp}\tau)
\end{equation}
with
\begin{equation}
\label{Eq:wperp}
w_{\perp} = \sqrt{ p_{\perp}^2 + |\mathbf{q}-\mathbf{p}|^2-2p_{\perp}|\mathbf{q}-\mathbf{p}| \cos{\theta}}
\end{equation} and
\begin{equation}
e^{i\Psi} = \frac{ |\mathbf{q}-\mathbf{p}| -p_{\perp}\cos{\theta}}{w_{\perp}} + i \frac{p_{\perp} \sin{\theta}}{w_{\perp}},
\end{equation}
we perform the desired reduction to get
\begin{equation}
\label{Eq:4iBL110}
\begin{split}
&\int_0^{\infty}d\tau \tau J_0(p_{\perp}\tau)J_1(|\mathbf{q}-\mathbf{p}|\tau)J_1(|\mathbf{k}-\mathbf{q}|\tau) H_0^{(2)}(k_{\perp}\tau) \\
=&\int_{-\pi}^{\pi}\frac{d\theta}{2\pi} e^{i\Psi}  \int_0^{\infty}d\tau \tau J_1(w_{\perp},\tau)J_1(|\mathbf{k}-\mathbf{q}|\tau)H_0^{(2)}(k_{\perp}\tau)\\
=&\int_{-\pi}^{\pi}\frac{d \theta}{2\pi} e^{i\Psi}   L_{110}(w_{\perp}, |\mathbf{k}-\mathbf{q}|, k_{\perp}).
\end{split}
\end{equation}

\item $\int_0^{\infty}d\tau \tau J_0(p_{\perp}\tau)J_0(|\mathbf{q}-\mathbf{p}|\tau)J_0(|\mathbf{k}-\mathbf{q}|\tau) H_0^{(2)}(k_{\perp}\tau)$:

Here we have (see Eq.~\eqref{Eq:wperp} for the definition of $w_\perp$)
\begin{equation}
J_0(p_{\perp}\tau)J_0(|\mathbf{q}-\mathbf{p}|\tau) = \int_{-\pi}^{\pi}\frac{d \theta}{2\pi} J_0(w_{\perp}\tau)\,.
\end{equation}
With this reduction we  obtain
\begin{equation}
\begin{split}
&\int_0^{\infty}d\tau \tau J_0(p_{\perp}\tau)J_0(|\mathbf{q}-\mathbf{p}|\tau)J_0(|\mathbf{k}-\mathbf{q}|\tau) H_0^{(2)}(k_{\perp}\tau) \\
=&\int_{-\pi}^{\pi} \frac{d\theta}{2\pi} \int_0^{\infty}d\tau \tau  J_0(w_{\perp}\tau)J_0(|\mathbf{k}-\mathbf{q}|\tau) H_0^{(2)}(k_{\perp}\tau)\\
=&\int_{-\pi}^{\pi} \frac{d\theta}{2\pi}  L_{000}(w_{\perp}, |\mathbf{k}-\mathbf{q}|, k_{\perp}).
\end{split}
\end{equation}

\item $\int_0^{\infty}d\tau \tau\,  J_1(|\mathbf{q}-\mathbf{p}|\tau)J_0(p_{\perp}\tau) J_0(|\mathbf{k}-\mathbf{q}|\tau) H_1^{(2)} (k_{\perp}\tau)$:

Finally,
\begin{equation}
\begin{split}
&\int_0^{\infty}d\tau \tau\,  J_1(|\mathbf{q}-\mathbf{p}|\tau)J_0(p_{\perp}\tau) J_0(|\mathbf{k}-\mathbf{q}|\tau) H_1^{(2)} (k_{\perp}\tau)\\
=&\int_0^{\infty}d\tau \tau\, \int_{-\pi}^{\pi} \frac{d\theta}{2\pi} e^{i\Psi}J_1(w_{\perp}\tau)J_0(|\mathbf{k}-\mathbf{q}|\tau) H_1^{(2)} (k_{\perp}\tau)\\
=&\int_{-\pi}^{\pi} \frac{d\theta}{2\pi} e^{i\Psi} L_{011}( |\mathbf{k}-\mathbf{q}|,w_{\perp}, k_{\perp})\,.
%=&\int_{-\pi}^{\pi} \frac{d\phi}{2\pi} \frac{|\mathbf{q}-\mathbf{p}|-p_{\perp}\cos\phi}{w_{\perp}} L_{011}( |\mathbf{k}-\mathbf{q}|,w_{\perp}, k_{\perp})\\
\end{split}
\end{equation}
In this expression, one can substitute $\frac{|\mathbf{q}-\mathbf{p}|^2-p_{\perp}^2 + w_{\perp}^2}{2|\mathbf{q}-\mathbf{p}| w_{\perp}} $ for $e^{i\Psi}$, since only even
part of $e^{i\Psi}$ contributes to the final result. The same applies to Eq.~\eqref{Eq:4iBL110}.
\end{itemize}

%The explicit expresions of the function  $L_{011}(a,b,c)$ are given in the Appendix. The results of these integrals depend on the relative magnitudes of $w_{\perp}, |\mathbf{k}-\mathbf{q}|, k_{\perp}$. For fixed $k_{\perp}$, because $\mathbf{p}$ and $\mathbf{q}$ are integration variables and thus can be any values. Their relative magnitudes are not determined.

\section{Relations among $L_{000}(a,b,c)$, $L_{110}(a,b,c)$ and $L_{011}(a,b,c)$}
\label{ap:Ls}
We previously introduced three functions $L_{000}(a,b,c)$, $L_{110}(a,b,c)$ and $L_{011}(a,b,c)$.
It may appear that they are independent.  However, further analysis shows that only one function is independent.
Indeed, using the reduction (see Appendix of Ref.~\cite{Li:2021zmf})
\begin{equation}
J_1(a\tau) J_1(b\tau) = \int_{-\pi}^\pi \frac{d\phi}{2\pi} \cos\phi  J_0(w\tau)
\end{equation}
with $w^2 = a^2+b^2-2ab\cos\phi$   and Eq.~\eqref{eq:integral_J0H0},
we get
\begin{equation}
\begin{split}
L_{110}(a,b,c) \equiv& \int_0^{\infty} \tau J_1(a\tau) J_1(b\tau) H_0^{(2)}(c\tau) d\tau\\
%=& \int_0^{\infty} \tau \int_{-\pi}^{\pi}\frac{d\phi}{2\pi} \cos\phi J_0(w\tau) H_0^{(2)}(c\tau) d\tau\\
=&\int_{-\pi}^{\pi}\frac{d\phi}{2\pi} \cos\phi  \int_0^{\infty} \tau J_0(w\tau) H_0^{(2)}(c\tau) d\tau \\
=&\int_{-\pi}^{\pi}\frac{d\phi}{2\pi} \frac{a^2+b^2-w^2}{2ab} \left(\frac{-2i}{\pi}\right) \frac{1}{c^2- w^2-i\epsilon}\\
%=&\frac{a^2+b^2-c^2 +i\epsilon} {2ab}\int_{-\pi}^{\pi}\frac{d\phi}{2\pi}\left(\frac{-2i}{\pi}\right) \frac{1}{c^2- w^2-i\epsilon}  + \frac{1}{2ab} \frac{-2i}{\pi} \int_{-\pi}^{\pi}\frac{d\phi}{2\pi}\\
=&\frac{a^2+b^2-c^2 +i\epsilon} {2ab}\int_{-\pi}^{\pi}\frac{d\phi}{2\pi}  \int_0^{\infty}\tau J_0(w\tau)H_0^{(2)}(c\tau) d\tau - \frac{i}{\pi ab}\\
%=&\frac{a^2+b^2-c^2 +i\epsilon} {2ab}\int_0^{\infty}\tau J_0(a\tau)J_0(b\tau)H_0^{(2)}(c\tau) d\tau- \frac{i}{\pi ab}\\
=&\frac{a^2+b^2-c^2 } {2ab}L_{000}(a,b,c) - \frac{i}{\pi ab}.
\end{split}
\end{equation}
Note that the final expression is symmetric with respect to $a\leftrightarrow b$. This establishes the connection between $L_{110}(a,b,c)$ and $L_{000}(a,b,c)$.

For $L_{011}(a,b,c)$, we similarly have
\begin{equation}
\begin{split}
L_{011}(a,b,c)\equiv&\int_0^{\infty}\tau J_0(a\tau) J_1(b\tau)H_1^{(2)}(c\tau) d\tau \\
=&\int_0^{\infty}\tau \int_{-\pi}^{\pi}\frac{d\phi}{2\pi} \frac{b-a\cos\phi}{w} J_1(w\tau)H_1^{(2)}(c\tau) d\tau\\
=&\int_{-\pi}^{\pi}\frac{d\phi}{2\pi} \frac{b-a\cos\phi}{w}\left( \frac{-2i}{\pi}\frac{w}{c} \frac{1}{c^2-w^2-i\epsilon} \right)\\
%=&\int_{-\pi}^{\pi}\frac{d\phi}{2\pi} \frac{b^2-a^2+ w^2}{2bc}\left( \frac{-2i}{\pi} \frac{1}{c^2-w^2-i\epsilon} \right)\\
=&\frac{b^2-a^2+c^2 -i\epsilon}{2bc} \int_{-\pi}^{\pi}\frac{d\phi}{2\pi}\left( \frac{-2i}{\pi} \frac{1}{c^2-w^2-i\epsilon} \right) -\frac{-2i}{\pi} \frac{1}{2bc} \int_{-\pi}^{\pi}\frac{d\phi}{2\pi} \\
=&\frac{b^2-a^2+c^2 -i\epsilon}{2bc} \int_{-\pi}^{\pi}\frac{d\phi}{2\pi} \int_0^{\infty}\tau J_0(w\tau)H_0^{(2)}(c\tau) d\tau + \frac{i}{\pi bc}\\
%=&\frac{b^2-a^2+c^2 -i\epsilon}{2bc}  \int_0^{\infty}\tau J_0(a\tau)J_0(b\tau)H_0^{(2)}(c\tau) d\tau + \frac{i}{\pi bc}\\
=&\frac{b^2+c^2-a^2}{2bc} L_{000}(a,b,c) + \frac{i}{\pi bc}.\\
\end{split}
\end{equation}
We have used Eqs.~\eqref{eq:integral_J1H1} and \eqref{eq:integral_J0H0}. This established the connection between $L_{011}(a,b,c)$ and $ L_{000}(a,b,c)$.

For the completeness, we note that $L_{000}(a,b,c)$ have somewhat laconic  integral representation
\begin{equation}
L_{000}(a,b,c) =\frac{-2i}{\pi} \int_{-\pi}^{\pi} \frac{d\phi}{2\pi} \frac{1}{c^2-a^2-b^2+2ab\cos\phi-i\epsilon}.
\end{equation}

In the following, we give an explicit computation of the integral. 
We first change  of variables and introduce $z = e^{i\phi}$; thus the angular integration over $\phi$ from $-\pi$ to $\pi$ becomes a closed contour integral of $|z|=1$ in counter-clockwise direction.
\begin{equation}
\int_{-\pi}^{\pi} \frac{d\phi}{2\pi} \to -\frac{i}{2\pi} \oint_{|z|=1} \frac{dz}{z}
\end{equation}
The  integral becomes (for real $a$, $b$ and with $b>0$)
\begin{equation}
\int_{-\pi}^{\pi} \frac{d\phi}{2\pi} \frac{1}{a+b\cos{\phi} - i\epsilon} 
=-i\frac{1}{\pi b} \oint_{|z|=1} \frac{dz}{(z-z_+)(z-z_-)}
\end{equation}
where
\begin{equation}
\begin{split}
&z_\pm = -\frac{1}{b}(a-i\epsilon) \pm  \sqrt{\frac{1}{b^2}(a-i\epsilon)^2 -1}.
\end{split}
\end{equation}
To proceed, we have to consider specific relations between  $a$ and $b$,
\begin{itemize}
\item $a>b$

When $a>b>0$, the solution $z_-$ lies outside of the unit circle $|z|=1$. Thus  $z_+$ is the only pole inside the unit circle.
Using the residue theorem, one obtains the result
\begin{equation}
\int_{-\pi}^{\pi} \frac{d\phi}{2\pi} \frac{1}{a+b\cos{\phi} - i\epsilon}   = \frac{1}{\sqrt{a^2-b^2}}.
\end{equation}

\item $-b\leq a\leq b$

In this case, the situation is more subtle. We can write
\begin{equation}
\begin{split}
z_\pm=&-\frac{a}{b} + \frac{1}{b}i\epsilon  \pm i\sqrt{ 1- \frac{a^2}{b^2} + \frac{2a}{b^2} i\epsilon}\\
=&-\frac{a}{b} + \frac{1}{b}i\epsilon  \pm i \sqrt{ 1- \frac{a^2}{b^2} } \mp \frac{a}{b\sqrt{b^2-a^2}} \epsilon
\end{split}
\end{equation}
If $\epsilon=0$, the two roots are located  exactly  on the unit circle. We are interested in  the case of  $\epsilon=0^+$.  The root with negative imaginary part is slightly shifted inside the unit circle while the solution with positive imaginary part is slightly shifted outside the unit circle.  Following  the residue theorem, one obtains
\begin{equation}
\int_{-\pi}^{\pi} \frac{d\phi}{2\pi} \frac{1}{a+b\cos{\phi} - i\epsilon}   = \frac{i}{\sqrt{b^2-a^2}}.
\end{equation}

Amusingly, we can apply this formulae to explicitly derive the identity
\begin{equation}
\int_0^{\infty} d\tau \tau J_0(p_{\perp}\tau)J_0(|\mathbf{k}-\mathbf{p}|\tau)H_0^{(2)}(k_{\perp}\tau) =\frac{1}{\pi} \frac{1}{|\mathbf{k}\times \mathbf{p}| }\,.
\end{equation}
Indeed,
\begin{equation}
\begin{split}
&\int_0^{\infty} d\tau \tau J_0(p_{\perp}\tau)J_0(|\mathbf{k}-\mathbf{p}|\tau)H_0^{(2)}(k_{\perp}\tau) \\
=&\int_{-\pi}^{\pi}\frac{d\phi}{2\pi} \int_0^{\infty}d\tau \tau J_0(w_{\perp}\tau) H_0^{(2)}(k_{\perp}\tau) \\
%=&\int_{-\pi}^{\pi}\frac{d\phi}{2\pi}  \frac{-2i}{\pi} \frac{1}{k_{\perp}^2-w_{\perp}^2 -i\epsilon}\\
=&\frac{-2i}{\pi} \int_{-\pi}^{\pi}\frac{d\phi}{2\pi}  \frac{1}{k_{\perp}^2-p_{\perp}^2-|\mathbf{k}-\mathbf{p}|^2 + 2p_{\perp}|\mathbf{k}-\mathbf{p}|\cos{\phi} -i\epsilon}\\
=&\frac{1}{\pi} \frac{1}{|\mathbf{k}\times \mathbf{p}| }
\end{split}
\end{equation}
Note that $p_{\perp}|\mathbf{k}-\mathbf{p}|\leq k_{\perp}^2-p_{\perp}^2-|\mathbf{k}-\mathbf{p}|^2\leq 2p_{\perp}|\mathbf{k}-\mathbf{p}|$ falls into the case under the consideration.

\item $a<-b$

In this case, $z_+$  ($z_-$) is located outside (inside) of the unit circle. Thus
\begin{equation}
\int_{-\pi}^{\pi} \frac{d\phi}{2\pi} \frac{1}{a+b\cos{\phi} - i\epsilon}    = -\frac{1}{\sqrt{a^2-b^2}}.
\end{equation}

\end{itemize}
To sum up, we evaluated the auxiliary  integral to obtain
\begin{equation}\label{eq:auxiliary_integral}
\begin{split}
&\int_{-\pi}^{\pi} \frac{d\phi}{2\pi} \frac{1}{a+b\cos{\phi} - i\epsilon}\\
=&\Big[\Theta(a-b) -\Theta(-b-a)\Big]\frac{1}{\sqrt{a^2-b^2}}+ \Theta(b-|a|) \frac{i}{\sqrt{b^2-a^2}}.\\
\end{split}
\end{equation}
Using Eq.~\eqref{eq:auxiliary_integral}, one can prove all the formulae involving integrals of three Bessel functions given in Appendix~\ref{ap:integral_formulas_three_bessels}.

\section{Analytic structure of the functions ${\cal I}_{1,2,3}$ and $\mathcal{G}_{1,2,3}$}
\label{ap:Is} 
In this appendix we consider the analytic structure of the functions  ${\cal I}_{1,2,3}$ and $\mathcal{G}_{1,2,3}$. In particular we  concentrate on possible singularities.
\begin{itemize}
\item  $\mathcal{I}_1 (\mathbf{p}, \mathbf{q}, \mathbf{k})$:

First for the readers convenience, we  recall the definition
\begin{equation}
\begin{split}
&\mathcal{I}_1 (\mathbf{p}, \mathbf{q}, \mathbf{k})\\
= &\pi k_{\perp}(\mathbf{k}\times \mathbf{q})(\mathbf{q}\times\mathbf{p})\int_{-\pi}^{\pi}\frac{d\phi}{2\pi}\frac{|\mathbf{q}-\mathbf{p}|^2-p_{\perp}^2+w_{\perp}^2}{q_{\perp}^2-w_{\perp}^2}\\
&\qquad \times \Big(\frac{1}{ w_{\perp}}L_{011}(|\mathbf{k}-\mathbf{q}|, w_{\perp},k_{\perp})- \frac{1}{q_{\perp}}L_{011}(|\mathbf{k}-\mathbf{q}|, q_{\perp}, k_{\perp})\Big) \\
&+\frac{1}{4}\pi k_{\perp}(\mathbf{k}-\mathbf{q})\cdot\mathbf{p} \int_{-\pi}^{\pi} \frac{d\phi}{2\pi} \frac{|\mathbf{q}-\mathbf{p}|^2-p_{\perp}^2 + w_{\perp}^2}{w_{\perp}}  L_{011}(|\mathbf{k}-\mathbf{q}|, w_{\perp},k_{\perp})\,. \\
\end{split}
\end{equation}
The behavior of the function $\mathcal{I}_1 (\mathbf{p}, \mathbf{q}, \mathbf{k})$ depends on the input momenta $\mathbf{p}, \mathbf{q}, \mathbf{k}$. Let's consider a few cases.

When $\mathbf{k}\times \mathbf{q}=0$ or $ \mathbf{q}\times\mathbf{p}=0$ or simply $\mathbf{q}=0$, the first term simply vanishes.

When  $p_{\perp} = |\mathbf{q}-\mathbf{p}|$, the angular integration includes the point $w_{\perp}=0$\footnote{From the definition $w_{\perp} =\sqrt{p_{\perp}^2+|\mathbf{q}-\mathbf{p}|^2-2p_{\perp}|\mathbf{q}-\mathbf{p}|\cos\phi}$, it is apparent that $w_{\perp}=0 $ corresponds to $\phi =0$.}. Thus it appears that $\mathcal{I}_1$ may have a singularity in this case. We want to show that this is a removable singularity. Consider the limit  $w_{\perp} \rightarrow 0$ of the ratio
\begin{equation}
\begin{split}
\lim_{w_{\perp} \rightarrow 0}
\frac{1}{w_{\perp}}L_{011}(|\mathbf{k}-\mathbf{q}|, w_{\perp},k_{\perp}) = &
\lim_{w_{\perp} \rightarrow 0}
\int_0^{\infty}\tau d\tau J_0(|\mathbf{k}-\mathbf{q}|\tau) \frac{J_1(w_{\perp}\tau)}{w_{\perp}}H_1^{(2)}(k_{\perp}\tau)\\
 &  = \frac{1}{2}\int_0^{\infty}\tau^2 d\tau J_0(|\mathbf{k}-\mathbf{q}|\tau) H_1^{(2)}(k_{\perp}\tau)\\
\end{split}
\end{equation}
We thus conclude that this ratio is finite.
On the other hand, when  $p_{\perp} = |\mathbf{q}-\mathbf{p}|$ and $w_{\perp}=0$ the numerators in the integrands vanish. This proves that $w_{\perp}=0$ is a removable singularity of the integrand and the value of the integrand is   zero.

When $w_{\perp} = q_{\perp}$, one may suspect the first integrand is singular. Consider the piece:
\begin{equation}
\begin{split}
&\frac{1}{ w_{\perp}}L_{011}(|\mathbf{k}-\mathbf{q}|, w_{\perp},k_{\perp})- \frac{1}{q_{\perp}}L_{011}(|\mathbf{k}-\mathbf{q}|, q_{\perp}, k_{\perp})\\
=&\int_0^{\infty} \tau d\tau J_0(|\mathbf{k}-\mathbf{q}|\tau) \left(\frac{J_1(w_{\perp}\tau)}{w_{\perp}} - \frac{J_1(q_{\perp}\tau)}{q_{\perp}} \right) H_1^{(2)}(k_{\perp}\tau)\\
=&(w_{\perp}-q_{\perp})\int_0^{\infty} \tau d\tau J_0(|\mathbf{k}-\mathbf{q}|\tau) \left( -\tau \frac{J_2(q_{\perp}\tau)}{q_{\perp}} \right) H_1^{(2)}(k_{\perp}\tau) + \mathcal{O}((w_{\perp}-q_{\perp})^2).\\
\end{split}
\end{equation}
We performed Taylor expansion around $w_{\perp}= q_{\perp}$ for the ratio  $J_1(w_{\perp}\tau)/w_{\perp}$ in the last equality.  Thus in the limit  $w_{\perp}\rightarrow q_{\perp}$, the integrand becomes
\begin{equation}\label{eq:apdix_I1}
\begin{split}
&
\lim_{w_{\perp}\rightarrow q_{\perp}}
\frac{\mathbf{q}\cdot(\mathbf{q}-2\mathbf{p})+w_{\perp}^2}{q_{\perp}^2-w_{\perp}^2}  \Big(\frac{1}{ w_{\perp}}L_{011}(|\mathbf{k}-\mathbf{q}|, w_{\perp},k_{\perp})- \frac{1}{q_{\perp}}L_{011}(|\mathbf{k}-\mathbf{q}|, q_{\perp}, k_{\perp})\Big)\\
 & =  -\frac{\mathbf{q}\cdot(\mathbf{q}-\mathbf{p})}{q_{\perp}}\int_0^{\infty} \tau d\tau J_0(|\mathbf{k}-\mathbf{q}|\tau) \left( -\tau \frac{J_2(q_{\perp}\tau)}{q_{\perp}} \right) H_1^{(2)}(k_{\perp}\tau).
\end{split}
\end{equation}
We  are left with an integral of the form
\begin{equation}
\begin{split}
&\int_0^{\infty} \tau d\tau J_0(a\tau) \left( -\tau \frac{J_2(b\tau)}{b} \right) H_1^{(2)}(c\tau)\\
=&\frac{d}{db} \left(\int_0^{\infty} \tau d\tau J_0(a\tau) \frac{J_1(b\tau)}{b} H_1^{(2)}(c\tau)\right)\\
=&\frac{d}{db} \left(\frac{1}{b} L_{011}(a,b,c)\right)\,.
\end{split}
\end{equation}
For the case of interest,  $|b-c|<a<b+c$, one can evaluate $L_{011}(a,b,c)$:
\begin{equation}
L_{011}(a,b,c) = \frac{1}{\pi b c} \left(\frac{b^2+c^2-a^2}{\sqrt{((b+c)^2-a^2)(a^2-(b-c)^2)}} + i\right).
\end{equation}
Using this result it is straightforward to compute
\begin{equation}
\begin{split}
&\frac{d}{db} \left(\frac{1}{b} L_{011}(a,b,c)\right)\\
=&\frac{-2}{\pi b^3 c} \left(\frac{b^2+c^2-a^2}{\sqrt{((b+c)^2-a^2)(a^2-(b-c)^2)}} + i\right) + \frac{4c}{\pi b} \frac{a^2+b^2-c^2}{[((b+c)^2-a^2)(a^2-(b-c)^2)]^{3/2}}\,.\\
\end{split}
\end{equation}
Substituting  $a=|\mathbf{k}-\mathbf{q}|, b= q_{\perp}, c=k_{\perp}$ into the formula, one can compute Eq.~ \eqref{eq:apdix_I1}.
to obtain the final expression for the integrand at  $w_{\perp}=q_{\perp}$
\begin{equation}
-\frac{\mathbf{q}\cdot(\mathbf{q}-\mathbf{p})}{\pi q^4_{\perp} k_{\perp}}\left[-2\left(\frac{\mathbf{q}\cdot\mathbf{k}}{|\mathbf{q}\times \mathbf{k}|} +i \right) + q_{\perp}^2k^2_{\perp} \frac{\mathbf{q}\cdot(\mathbf{q}-\mathbf{k})}{|\mathbf{q}\times \mathbf{k}|^3}\right].
\end{equation}
Recall that from we have implicitly assumed that $\mathbf{q}\times\mathbf{k} \neq 0$. Therefore $w_{\perp}=q_{\perp}$ is a removable pole.

\item $\mathcal{I}_2(\mathbf{p}, \mathbf{q}, \mathbf{k})$
\begin{equation}
\begin{split}
&\mathcal{I}_2(\mathbf{p}, \mathbf{q}, \mathbf{k}) \\
=&\frac{1}{2q_{\perp}^2}(\mathbf{k}\times \mathbf{q})(\mathbf{q}\times \mathbf{p})\pi |\mathbf{k}-\mathbf{q}|k_{\perp}\\
&\quad \times\int_{-\pi}^{\pi}\frac{d\phi}{2\pi} \left(1+\frac{2\mathbf{p}\cdot(\mathbf{p}-\mathbf{q})}{q_{\perp}^2-w_{\perp}^2}\right) \Big(L_{011}(w_{\perp}, |\mathbf{k}-\mathbf{q}|, k_{\perp})-  L_{011}(q_{\perp}, |\mathbf{k}-\mathbf{q}|, k_{\perp})\Big)\\
&+\frac{1}{8q_{\perp}^2}(k_{\perp}^2-|\mathbf{k}-\mathbf{q}|^2)(|\mathbf{q}-\mathbf{p}|^2-p_{\perp}^2)\pi |\mathbf{k}-\mathbf{q}|k_{\perp}\\
&\qquad \times \int_{-\pi}^{\pi} \frac{d\phi}{2\pi}\left(1 -\frac{|\mathbf{q}-\mathbf{p}|^2+p_{\perp}^2}{w_{\perp}^2}\right)\Big(L_{011}(w_{\perp}, |\mathbf{k}-\mathbf{q}|,k_{\perp}) -L_{011}(0,|\mathbf{k}-\mathbf{q}|, k_{\perp}) \Big) \\
\end{split}
\end{equation}
First of all, there is no singularity at $w_{\perp}=0$. The superficial singularity when $w_{\perp} =q_{\perp}$ in computing the first angular integral will be cancelled by
\begin{equation}
\begin{split}
&L_{011}(w_{\perp}, |\mathbf{k}-\mathbf{q}|, k_{\perp})-  L_{011}(q_{\perp}, |\mathbf{k}-\mathbf{q}|, k_{\perp})\\
=&\int_0^{\infty} \tau d\tau \Big(J_0(w_{\perp}\tau)-J_0(q_{\perp}\tau)\Big) J_1(|\mathbf{k}-\mathbf{q}|\tau) H_1^{(2)}(k_{\perp}\tau) \\
=&(w_{\perp}-q_{\perp})\int_0^{\infty} \tau d\tau \Big(-\tau J_1(q_{\perp}\tau)\Big) J_1(|\mathbf{k}-\mathbf{q}|\tau) H_1^{(2)}(k_{\perp}\tau) + \mathcal{O}((w_{\perp}-q_{\perp})^2). \\
\end{split}
\end{equation}
So in the limit $w_{\perp} \rightarrow q_{\perp}$, the integrand becomes
\begin{equation}
\begin{split}
&-\frac{\mathbf{p}\cdot(\mathbf{p}-\mathbf{q})}{q_{\perp}} \int_0^{\infty} \tau d\tau \Big(-\tau J_1(q_{\perp}\tau)\Big) J_1(|\mathbf{k}-\mathbf{q}|\tau) H_1^{(2)}(k_{\perp}\tau)\\
\end{split}
\end{equation}
One needs
\begin{equation}
\begin{split}
&\int_0^{\infty} \tau d\tau \Big(-\tau J_1(a\tau)\Big) J_1(b\tau) H_1^{(2)}(c\tau)\\
=&\frac{d}{da} \int_0^{\infty} \tau d\tau  J_0(a\tau) J_1(b\tau) H_1^{(2)}(c\tau)\\
=&\frac{d}{da} L_{011}(a,b,c)\\
\end{split}
\end{equation}
When $|c-b|<a<c+b$,
\begin{equation}
L_{011}(a,b,c) =  \frac{1}{\pi b c} \left(\frac{b^2+c^2-a^2}{\sqrt{((b+c)^2-a^2)(a^2-(b-c)^2)}} + i\right)
\end{equation}
and
\begin{equation}
\frac{d}{da}L_{011}(a,b,c)
=\frac{-8abc}{\pi} \left(\frac{1}{\sqrt{((b+c)^2-a^2)(a^2-(b-c)^2)}}\right)^3
\end{equation}
With $a=q_{\perp}, b=|\mathbf{k}-\mathbf{q}|, c= k_{\perp}$, 
the integrand now becomes
\begin{equation}
\begin{split}
&-\frac{\mathbf{p}\cdot(\mathbf{p}-\mathbf{q})}{q_{\perp}} \int_0^{\infty} \tau d\tau \Big(-\tau J_1(q_{\perp}\tau)\Big) J_1(|\mathbf{k}-\mathbf{q}|\tau) H_1^{(2)}(k_{\perp}\tau)\\
=&\frac{\mathbf{p}\cdot(\mathbf{p}-\mathbf{q})}{|\mathbf{k}\times\mathbf{q}|^3} \frac{k_{\perp}|\mathbf{k}-\mathbf{q}|}{\pi}.
\end{split}
\end{equation}
Therefore, $w_{\perp}=q_{\perp}$ is a removable singularity in the first angular integral. 
For the second angular integral, the superficial singularity is $w_{\perp}=0$. The integrand when $w_{\perp}=0$ can be analysed similarly.
\begin{equation}
\begin{split}
&L_{011}(w_{\perp}, |\mathbf{k}-\mathbf{q}|,k_{\perp}) -L_{011}(0,|\mathbf{k}-\mathbf{q}|, k_{\perp})\\
=&\int_0^{\infty} \tau d\tau (J_0(w_{\perp}\tau) -J_0(0\tau))J_1(|\mathbf{k}-\mathbf{q}|\tau)H_1^{(2)}(k_{\perp}\tau)\\
=&-\frac{1}{4}w^2_{\perp}\int_0^{\infty} \tau^3 d\tau J_1(|\mathbf{k}-\mathbf{q}|\tau)H_1^{(2)}(k_{\perp}\tau)+\mathcal{O}(w^4_{\perp})
\end{split}
\end{equation}
On the other hand, $w_{\perp}=0$ can only happen when $p_{\perp} = |\mathbf{q}-\mathbf{p}|$ and then in the angular integral  $\phi=0$. However, when $p_{\perp} = |\mathbf{q}-\mathbf{p}|$, the prefactor $\mathbf{q}\cdot(\mathbf{q}-2\mathbf{p})=0$. So there is no singularity at $w_{\perp}=0$.

\item $\mathcal{I}_3(\mathbf{p}, \mathbf{q}, \mathbf{k})$
\begin{equation}
\begin{split}
&\mathcal{I}_3(\mathbf{p}, \mathbf{q}, \mathbf{k}) \\
=& \frac{1}{q_{\perp}^2} (\mathbf{k}\times \mathbf{q})(\mathbf{q}\times \mathbf{p}) (p_{\perp}^2+q_{\perp}^2-\mathbf{p}\cdot\mathbf{q}) \pi|\mathbf{k}-\mathbf{q}|k_{\perp}\\
&\qquad \times \int_{-\pi}^{\pi}\frac{d\phi}{2\pi}  \frac{1}{q_{\perp}^2-w_{\perp}^2}\Big(L_{011}(w_{\perp}, |\mathbf{k}-\mathbf{q}|, k_{\perp})- L_{011}(q_{\perp}, |\mathbf{k}-\mathbf{q}|, k_{\perp})\Big)\\
&+\frac{1}{4q_{\perp}^2} \left(k_{\perp}^2-|\mathbf{k}-\mathbf{q}|^2\right)\left(|\mathbf{q}-\mathbf{p}|^2-p_{\perp}^2\right)\mathbf{p}\cdot(\mathbf{q}-\mathbf{p})\pi |\mathbf{k}-\mathbf{q}| k_{\perp} \\
&\qquad\times\int_{-\pi}^{\pi} \frac{d\phi}{2\pi}\frac{1}{w_{\perp}^2} \Big(L_{011}(w_{\perp}, |\mathbf{k}-\mathbf{q}|, k_{\perp})-L_{011}(0, |\mathbf{k}-\mathbf{q}|, k_{\perp}) \Big). \\
\end{split}
\end{equation}
Again, the superfical singularity at $w_{\perp}=0$ is removalbe. The superfical singularity at $w_{\perp}=q_{\perp}$ is also removable. The integrand when $w_{\perp}=q_{\perp}$ is
\begin{equation}
-\frac{1}{2q_{\perp}}   \frac{-k_{\perp}q_{\perp}|\mathbf{k}-\mathbf{q}|}{\pi} \frac{1}{|\mathbf{k}\times\mathbf{q}|^3} = \frac{1}{2\pi}  \frac{k_{\perp}|\mathbf{k}-\mathbf{q}|}{|\mathbf{k}\times\mathbf{q}|^3}.
\end{equation}

\end{itemize}

%%%%%%%%%%%%%%%%%%%%%%%%%%%%%%%%%%%%%%%%%%%%%%%%%%%%%%%%%%%
%%%%%%%%%%%%%%%%%%%%%%%%%%%%%%%%%%%%%%%%%%%%%%%%%%%%%%%%%
\begin{itemize}
\item  $\mathcal{G}_1(\mathbf{p}, \mathbf{q}, \mathbf{k})$. 
\begin{equation}
\begin{split}
\mathcal{G}_1(\mathbf{p}, \mathbf{q}, \mathbf{k})
=& \pi |\mathbf{k}-\mathbf{q}|(\mathbf{k}\times \mathbf{q})(\mathbf{q}\times \mathbf{p})\int_{-\pi}^{\pi}\frac{d\phi}{2\pi}\left(-1 + \frac{2\mathbf{q}\cdot(\mathbf{q}-\mathbf{p})}{q_{\perp}^2-w_{\perp}^2}\right)\\
&\qquad  \times \left(\frac{1}{w_{\perp}}L_{110}(w_{\perp}, |\mathbf{k}-\mathbf{q}|, k_{\perp})-\frac{1}{q_{\perp}}L_{110}(q_{\perp}, |\mathbf{k}-\mathbf{q}|, k_{\perp})\right)\\
&+\frac{\pi}{4} |\mathbf{k}-\mathbf{q}|(\mathbf{p}\cdot \mathbf{k})\int_{-\pi}^{\pi}\frac{d \phi}{2\pi}\frac{|\mathbf{q}-\mathbf{p}|^2-p_{\perp}^2 + w_{\perp}^2}{w_{\perp}}L_{110}(w_{\perp}, |\mathbf{k}-\mathbf{q}|, k_{\perp})\\
\end{split}
\end{equation}
The superficial singularity at $w_{\perp} =q_{\perp}$ is analysed by
\begin{equation}
\begin{split}
&\frac{1}{w_{\perp}}L_{110}(w_{\perp}, |\mathbf{k}-\mathbf{q}|, k_{\perp})-\frac{1}{q_{\perp}}L_{110}(q_{\perp}, |\mathbf{k}-\mathbf{q}|, k_{\perp})\\
=&\int_0^{\infty} \tau d\tau \left(\frac{1}{w_{\perp}}J_1(w_{\perp}\tau) - \frac{1}{q_{\perp}}J_1(q_{\perp}\tau)\right)J_1(|\mathbf{k}-\mathbf{q}|\tau) H_0^{(2)}(k_{\perp}\tau)\\
=&(w_{\perp}-q_{\perp})\int_0^{\infty} \tau d\tau \left(\frac{-\tau J_2(q_{\perp}\tau)}{q_{\perp}}\right)J_1(|\mathbf{k}-\mathbf{q}|\tau) H_0^{(2)}(k_{\perp}\tau) + \mathcal{O}((w_{\perp}-q_{\perp})^2). \\
\end{split}
\end{equation}
The integrand becomes 
\begin{equation}
-\frac{\mathbf{q}\cdot(\mathbf{q}-\mathbf{p})}{q_{\perp}}\int_0^{\infty} \tau d\tau \left(\frac{-\tau J_2(q_{\perp}\tau)}{q_{\perp}}\right)J_1(|\mathbf{k}-\mathbf{q}|\tau) H_0^{(2)}(k_{\perp}\tau).
\end{equation}
We need to compute integral of the form
\begin{equation}
\begin{split}
&\int_0^{\infty} \tau d\tau \left(\frac{-\tau J_2(a\tau)}{a}\right)J_1(b\tau) H_0^{(2)}(c\tau)\\
=&\frac{d}{da}\left (\frac{1}{a} \int_0^{\infty} \tau d\tau  J_1(a\tau)J_1(b\tau) H_0^{(2)}(c\tau)\right)\\
=&\frac{d}{da}\left (\frac{1}{a} L_{110}(a, b,c)\right).
\end{split}
\end{equation}
When $|a-b|<c<a+b$, 
\begin{equation}
L_{110}(a,b,c) = \frac{1}{\pi ab} \left(\frac{a^2+b^2-c^2}{\sqrt{((a+b)^2-c^2)(c^2-(a-b)^2)}} -i\right).
\end{equation}
\begin{equation}
\begin{split}
&\frac{d}{da}\left (\frac{1}{a} L_{110}(a, b,c)\right)\\
=&\frac{-2}{\pi a^3b} \left(\frac{a^2+b^2-c^2}{\sqrt{((a+b)^2-c^2)(c^2-(a-b)^2)}} -i\right) + \frac{4b}{\pi a} \frac{a^2+c^2-b^2}{[((a+b)^2-c^2)(c^2-(a-b)^2)]^{3/2}}
\end{split}
\end{equation}
with $a=q_{\perp}, b=|\mathbf{k}-\mathbf{q}|, c=k_{\perp}$, 
The integrand becomes 
\begin{equation}
\begin{split}
&-\frac{\mathbf{q}\cdot(\mathbf{q}-\mathbf{p})}{q_{\perp}}\int_0^{\infty} \tau d\tau \left(\frac{-\tau J_2(q_{\perp}\tau)}{q_{\perp}}\right)J_1(|\mathbf{k}-\mathbf{q}|\tau) H_0^{(2)}(k_{\perp}\tau)\\
=&-\frac{\mathbf{q}\cdot(\mathbf{q}-\mathbf{p})}{\pi q^4_{\perp}|\mathbf{k}-\mathbf{q}|}\left[-2 \left(\frac{\mathbf{q}\cdot(\mathbf{q}-\mathbf{k})}{|\mathbf{k}\times \mathbf{q}|} -i\right) + q_{\perp}^2|\mathbf{k}-\mathbf{q}|^2 \frac{\mathbf{k}\cdot\mathbf{q}}{|\mathbf{k}\times \mathbf{q}|^3}\right]\\
\end{split}
\end{equation}

There are no superficial singularities at $w_{\perp}=0$ or $q_{\perp}=0$.

\item $\mathcal{G}_2(\mathbf{p}, \mathbf{q}, \mathbf{k})$
\begin{equation}
\begin{split}
\mathcal{G}_2(\mathbf{p}, \mathbf{q}, \mathbf{k}) =&\frac{1}{q_{\perp}^2}(k_{\perp}^2+q_{\perp}^2-\mathbf{k}\cdot\mathbf{q})(\mathbf{k}\times \mathbf{q})( \mathbf{q}\times \mathbf{p})\frac{\pi}{2}\int_{-\pi}^{\pi}\frac{d\phi}{2\pi}  \frac{1}{q_{\perp}^2-w_{\perp}^2}(-w_{\perp}^2+p_{\perp}^2+|\mathbf{q}-\mathbf{p}|^2)\\
&\times\Big(L_{000}(w_{\perp}, |\mathbf{k}-\mathbf{q}|, k_{\perp}) -L_{000}(q_{\perp}, |\mathbf{k}-\mathbf{q}|, k_{\perp})\Big) \\
&+\frac{1}{4q_{\perp}^2}(k_{\perp}^2-|\mathbf{k}-\mathbf{q}|^2)\,\mathbf{k}\cdot(\mathbf{k}-\mathbf{q})(|\mathbf{q}-\mathbf{p}|^2-p_{\perp}^2) \frac{\pi}{2}\int_{-\pi}^{\pi} \frac{d\phi}{2\pi}\frac{1}{w_{\perp}^2}(w_{\perp}^2-p_{\perp}^2-|\mathbf{q}-\mathbf{p}|^2)\\
&\times \Big(-L_{000}(0, |\mathbf{k}-\mathbf{q}|, k_{\perp})+L_{000}(w_{\perp}, |\mathbf{k}-\mathbf{q}|,k_{\perp}\Big) 
\\
\end{split}
\end{equation}
First of all, there is no singularity at $w_{\perp}=0$. The superficial singularity at $w_{\perp}=q_{\perp}$ can be isolated as 
\begin{equation}
\begin{split}
&L_{000}(w_{\perp}, |\mathbf{k}-\mathbf{q}|, k_{\perp}) -L_{000}(q_{\perp}, |\mathbf{k}-\mathbf{q}|, k_{\perp})\\
=&\int_0^{\infty} \tau d\tau \Big(J_0(w_{\perp}\tau)-J_0(q_{\perp}\tau)\Big)J_0(|\mathbf{k}-\mathbf{q}|\tau)H_0^{(2)}(k_{\perp}\tau)\\
=&(w_{\perp}-q_{\perp})\int_0^{\infty} \tau d\tau \Big(-\tau J_1(q_{\perp}\tau)\Big)J_0(|\mathbf{k}-\mathbf{q}|\tau)H_0^{(2)}(k_{\perp}\tau)+\mathcal{O}((w_{\perp}-q_{\perp})^2)\\
\end{split}
\end{equation}
We need to calculate integral of the form
\begin{equation}
\begin{split}
\int_0^{\infty} \tau d\tau \Big(-\tau J_1(a\tau)\Big)J_0(b\tau)H_0^{(2)}(c\tau) =& \frac{d}{da} \int_0^{\infty} \tau d\tau J_0(a\tau)J_0(b\tau)H_0^{(2)}(c\tau)\\
=&\frac{d}{da} L_{000}(a, b,c)\\
\end{split}
\end{equation}
When $|a-b|<c<a+b$
\begin{equation}
L_{000} (a,b,c)= \frac{2}{\pi} \frac{1}{\sqrt{((a+b)^2-c^2)(c^2-(a-b)^2)}}.
\end{equation}
\begin{equation}
\frac{d}{da} L_{000}(a, b,c) 
=-\frac{4a}{\pi} \frac{(c^2+b^2-a^2)}{[((a+b)^2-c^2)(c^2-(a-b)^2)]^{3/2}}.
\end{equation}
With $a=q_{\perp}, b=|\mathbf{k}-\mathbf{q}|, c=k_{\perp}$,
\begin{equation}
\int_0^{\infty} \tau d\tau \Big(-\tau J_1(q_{\perp}\tau)\Big)J_0(|\mathbf{k}-\mathbf{q}|\tau)H_0^{(2)}(k_{\perp}\tau) =-\frac{q_{\perp}}{\pi} \frac{\mathbf{k}\cdot(\mathbf{k}-\mathbf{q})}{|\mathbf{k}\times \mathbf{q}|^3}.
\end{equation}
The integrand at $w_{\perp}=q_{\perp}$ becomes
\begin{equation}
 \frac{\mathbf{p}\cdot(\mathbf{p}-\mathbf{q})}{\pi} \frac{\mathbf{k}\cdot(\mathbf{k}-\mathbf{q})}{|\mathbf{k}\times \mathbf{q}|^3}.
\end{equation}

\item $\mathcal{G}_3(\mathbf{p}, \mathbf{q}, \mathbf{k})$
\begin{equation}
\begin{split}
\mathcal{G}_3(\mathbf{p}, \mathbf{q}, \mathbf{k})=&\frac{\pi }{q_{\perp}^2}(k_{\perp}^2+q_{\perp}^2-\mathbf{k}\cdot\mathbf{q})(\mathbf{k}\times \mathbf{q})( \mathbf{q}\times \mathbf{p})(-\mathbf{p}\cdot\mathbf{q}+p_{\perp}^2+q_{\perp}^2)\\
&\qquad \times \int_{-\pi}^{\pi}\frac{d\phi}{2\pi}  \frac{1}{q_{\perp}^2-w_{\perp}^2} \Big(L_{000}(w_{\perp}, |\mathbf{k}-\mathbf{q}|, k_{\perp}) - L_{000}(q_{\perp}, |\mathbf{k}-\mathbf{q}|, k_{\perp})\Big) \\
&+ \frac{\pi}{4q_{\perp}^2}(k_{\perp}^2-|\mathbf{k}-\mathbf{q}|^2)\,\mathbf{k}\cdot(\mathbf{k}-\mathbf{q}) (|\mathbf{q}-\mathbf{p}|^2-p_{\perp}^2)\mathbf{p}\cdot(\mathbf{q}-\mathbf{p}) \\
&\qquad \times \int_{-\pi}^{\pi} \frac{d\phi}{2\pi}\frac{1}{w_{\perp}^2} \Big(-L_{000}(0, |\mathbf{k}-\mathbf{q}|, k_{\perp})+ L_{000}(w_{\perp}, |\mathbf{k}-\mathbf{q}|,k_{\perp}\Big) 
\\
&-\frac{1}{2} (\mathbf{k}\times\mathbf{q})(\mathbf{q}\times\mathbf{p})\frac{\pi}{2}\int_{-\pi}^{\pi} \frac{d\phi}{2\pi}  L_{000}(w_{\perp}, |\mathbf{k}-\mathbf{q}|, k_{\perp}).\\
\end{split}
\end{equation}
\end{itemize}
Again, there is no singularity at $w_{\perp}=0$. For the superficial singularity at $w_{\perp}=q_{\perp}$.  Repeating similar analysis as discussed above, the integrand in the first integral becomes
\begin{equation}
\frac{1}{2\pi} \frac{\mathbf{k}\cdot(\mathbf{k}-\mathbf{q})}{|\mathbf{k}\times \mathbf{q}|^3}.
\end{equation}
Thus $w_{\perp}=q_{\perp}$ is a removable singularity.

\section{Comparison with partial results from LCPT}
\label{sec:comparison_with_LCPT}
Previously Chirilli, Kovchegov and Wertepny (CKW) derived the single-gluon production amplitude (in general, a gauge-variant object) at leading-order and  order-$g^{3}$ in Ref.~\cite{Chirilli:2015tea}; the full expression for the first saturation correction to the gluon production cross-section (a gauge-invariant object) was not obtained. 
We thus are forced to compare the leading-order $M_1$ and  order-$g^{3}$ ($M_3$)  amplitudes only. 
However,  in contrast to our approach, these calculations were done in the light cone gauge. Naively, this difference in gauges would not allow us to perform one-to-one comparison of the results on the level of the amplitudes as they are  gauge-variant objects. Deeper inspection shows that the leading order amplitude is gauge invariant as it is the only contribution to the leading order particle production cross-section (order-by-order gauge invariant-object).  At the leading order our calculations are in full agreement with  Ref.~\cite{Chirilli:2015tea} and previous studies. The first saturation correction contribution to single inclusive gluon production involves  leading-order, order-$g^{3}$ and order-$g^{5}$ amplitudes: $M_{3}(k) M^{*}_{3}(k) + M_{1}(k) M^{*}_{5}(k)+ M^{*}_{1}(k) M_{5}(k)$. This is a gauge invariant combination. Since  $M_{5}$ was not computed in  
Ref.~\cite{Chirilli:2015tea} we cannot perform the comparison with our result. Remarkably we can compare the odd part of $M_{3}(k)$: that is $M_{3}(k)-M_{3}(-k)$. 
Indeed, the leading odd contribution to the double inclusive gluon production is given by  $\langle  (n^{(3)} (k) - n^{(3)} (-k))  (n^{(3)} (p) - n^{(3)} (-p))\rangle$,
which in the amplitude notation would correspond to $\langle  M_{1}(k) (M_{3} (k) - M_{3} (-k))  M_{1}(p) (M_{3} (p) - M_{3} (-p))  \rangle$. This is an observable, that is a gauge invariant object. Moreover as we established before, $M_{1}(k)$ is gauge invariant, therefore $(M_{3} (k) - M_{3} (-k))$ is gauge invariant as well. Now the final subtle point is that the calculations in Ref.~\cite{Chirilli:2015tea} were performed not in terms of continuous charge distributions $\rho$, but for projectile's quarks. This would not allow direct  comparison {\it before} averaging of our results with respect to the projectile configurations. Fortunately the odd part of the amplitude   $(M_{3} (k) - M_{3} (-k))$ of Ref.~\cite{Chirilli:2015tea}  contains only single gluon emissions from a quark! Thus by performing the replacement
\begin{equation}
\label{Eq:CGCfromCKW}
 V(\mathbf{b})t^c\longrightarrow \rho^c(\mathbf{b})
 \end{equation}
 in $(M_{3} (k) - M_{3} (-k))$ of  Ref.~\cite{Chirilli:2015tea}   we should be able to reproduce our result. 
 
Here we provide a detailed account of this comparison.  First let's reproduce the order-$g$ gluon production amplitude. In the light cone gauge, see e.g. Ref.~\cite{Chirilli:2015tea}, 
\begin{equation}
M^{(1),a}_{\lambda}( \mathbf{z}, \mathbf{b}) = 2g \int \frac{d^2\mathbf{p}}{(2\pi)^2} e^{i\mathbf{p}\cdot(\mathbf{z}-\mathbf{b})} \frac{\varepsilon_{\lambda}^{\ast} \cdot \mathbf{p}}{p_{\perp}^2} \Big[ U^{ac}(\mathbf{z})-U^{ac}(\mathbf{b})\Big] (V(\mathbf{b})t^c)
\end{equation}
Here the polarization vector $\varepsilon_{\lambda} = -\frac{1}{\sqrt{2}}(\lambda, i)$ with $\lambda = \pm 1$. 
At this order we only have single gluon emission processes in the amplitude, thus we can utilize the prescription Eq.~\eqref{Eq:CGCfromCKW}. 
\begin{equation}
 V(\mathbf{b})t^c\longrightarrow \rho^c(\mathbf{b})
 \end{equation}
We would like to work in momentum space, for that we need to integrate over the impact parameter space  $\int d^2\mathbf{b}$ and Fourier transform 
 $\int d^2\mathbf{z} e^{-i\mathbf{k}\cdot\mathbf{z}}$. 

We have 
\begin{equation}
\begin{split}
M^{(1),a}_{\lambda}(\mathbf{k}) =& \int d^2\mathbf{z} e^{-i\mathbf{k}\cdot\mathbf{z}}\int d^2\mathbf{b} M^{(1),a}_{\lambda}(\mathbf{z}, \mathbf{b})=-2ig \, \varepsilon_{\lambda}^i \mathcal{A}^{i,a} (\mathbf{k})
\end{split}
\end{equation}
where the field 
\begin{equation}
\mathcal{A}_{{\rm CKW}}^{i,a} (\mathbf{k}) = \int \frac{d^2\mathbf{p}}{(2\pi)^2}\left(\frac{i \mathbf{p}^i}{p_{\perp}^2} - \frac{i \mathbf{k}^i}{k_{\perp}^2}\right) U^{ac}(\mathbf{k}-\mathbf{p})\rho^c(\mathbf{p}) \,.
\end{equation}
Using $\sum_{\lambda}\varepsilon_{\lambda}^i \varepsilon_{\lambda}^{j\ast} = \delta^{ij}$,  the amplitude squared can be written in the following form 
\begin{equation}
\sum_{\lambda} |M_{\lambda, {\rm CKW}}(\mathbf{k})|^2 = 4g^2 \mathcal{A}_{a\ {\rm CKW}}^i(\mathbf{k})\mathcal{A}_{a\ {\rm CKW}}^{i\ast}(\mathbf{k})
\end{equation}
Finally gluon production is given by  
\begin{equation}
\frac{d^2N_{\rm CKW} }{d^2\mathbf{k}} = \frac{1}{2(2\pi)^3} \sum_{\lambda} |M_{\lambda\ {\rm CKW}}(\mathbf{k})|^2  = \frac{g^2}{4\pi^3}  \mathcal{A}^i_{a\ {\rm CKW}}(\mathbf{k})\mathcal{A}^{i\ast}_{a\ {\rm CKW}}{(\mathbf{k})}
\end{equation}

In our calculation, i.e.  in the Fock-Schwinger gauge, we work with two components of the gluon  field 
\begin{equation}
b_{\eta}(\mathbf{k}) = -i\mathbf{k}^i \mathcal{A}^i(\mathbf{k}), \qquad b_{\perp}(\mathbf{k}) = -i\epsilon^{ij} \mathbf{k}^i\mathcal{A}^j(\mathbf{k}).
\end{equation}
In other words, 
\begin{equation}
\mathcal{A}^{j,a}(\mathbf{k}) = \frac{i \mathbf{k}^j}{k_{\perp}^2} b_{\eta}(\mathbf{k}) +\frac{i\epsilon^{lj}\mathbf{k}^l}{k_{\perp}^2} b_{\perp}(\mathbf{k}).
\end{equation}
Note that our $\mathcal{A}^{j,a}(\mathbf{k})$ is equal to $\mathcal{A}_{{\rm CKW}}^{j,a}(\mathbf{k})$. 
We have
\begin{equation}
\begin{split}
\frac{d^2N}{d^2\mathbf{k}}  = &\frac{g^2}{(2\pi)^2} \Big(|\mathfrak{S}(\mathbf{k})|^2 + |\mathfrak{B}(\mathbf{k})|^2\Big)\\
=&\frac{g^2}{4\pi^3 k^2_{\perp}} (|b_{\eta}(\mathbf{k})|^2 + |b_{\perp}(\mathbf{k})|^2)\\
=&\frac{g^2}{4\pi^3} \mathcal{A}^i_a(\mathbf{k})\mathcal{A}^{i\ast}_a(\mathbf{k})
\end{split}
\end{equation}
Here we used the identity for Levi-Civita symbol in two-dimensions: $\epsilon^{ij}\epsilon^{mn} = \delta^{im}\delta^{jn}-\delta^{in}\delta^{jm}$.  
We thus fully reproduce the same production cross section in both approaches. 

This exercise enables us to establish the mapping we can use for comparing $(M_{3} (k) - M_{3} (-k))$ between different approaches. 
We first remove the polarization vector $\varepsilon_{\lambda}^i$ and the coupling constant in the expressions given in CKW's paper and
instead denote the remaining by  ${\cal M}^{i,a}(\mathbf{k})$. For example, at order-$g$,
\begin{equation}
{\cal M}^{i,a}_{(1)}(\mathbf{k}) = -2i \mathcal{A}^{i,a}(\mathbf{k})
\end{equation}
Then we project out the amplitude for the $\eta$ polarization and the $\perp$ polarizations 
to establish the following equivalence 
\begin{align}
\label{Eq:Equiv1}
\mathfrak{S}_{\eta}(\mathbf{k})+\mathfrak{B}_{\eta}(\mathbf{k}) &\iff \frac{1}{2\sqrt{\pi}k_{\perp}} \mathbf{k}^i {\cal M}^{i,a}(\mathbf{k}) \\ 
\label{Eq:Equiv2}
\mathfrak{S}_{\perp}(\mathbf{k})+\mathfrak{B}_{\perp}(\mathbf{k}) &\iff \frac{1}{2\sqrt{\pi}k_{\perp}} \epsilon^{ij}\mathbf{k}^i {\cal M}^{j,a}(\mathbf{k}) 
\end{align}

Here we quote only the {\it relevant} part (which gives contribution to $M_{(3)} (k) - M_{(3)} (-k)$ ) of the full expression  for the  order-$g^3$ gluon production amplitude from CKW paper, see Eq.~(32):
\begin{equation}
\begin{split}
&{\cal M}^{i,a}_{(3), {\rm odd}}(\mathbf{z}, \mathbf{b}_1, \mathbf{b}_2)\\
=& -2i\int d^2\mathbf{x}_1 d^2\mathbf{x}_2 \int \frac{d^2\mathbf{p}}{(2\pi)^2} \frac{d^2\mathbf{l}}{(2\pi)^2} \frac{d^2\mathbf{q}_1}{(2\pi)^2} \frac{d^2\mathbf{q}_2}{(2\pi)^2} e^{i\mathbf{q}_1\cdot(\mathbf{x}_1-\mathbf{b}_1)}e^{i\mathbf{q}_2\cdot(\mathbf{x}_2-\mathbf{b}_2)}e^{i\mathbf{l}\cdot(\mathbf{x}_2-\mathbf{x}_1)} e^{i\mathbf{p}\cdot(\mathbf{z}-\mathbf{x}_2)}\\
&\qquad \times \frac{1}{q_1^2q_2^2}\left(-\mathbf{q}_1\cdot\mathbf{q}_2 \frac{\epsilon^{ij}\mathbf{p}^j}{p_{\perp}^2 } + \mathbf{q}_1^i \frac{\mathbf{q}_2\times(\mathbf{p}-\mathbf{l})}{|\mathbf{p}-\mathbf{l}|^2}+\mathbf{q}_2^i \frac{\mathbf{q}_1\times \mathbf{l}}{l_{\perp}^2}\right)\mathrm{sgn}(\mathbf{p}\times \mathbf{l}) \\
&\qquad \times f^{abc}[U^{bd}(\mathbf{x}_1)-U^{bd}(\mathbf{b}_1)]\rho^{d}(\mathbf{b}_1) [U^{ce}(\mathbf{x}_2)-U^{ce}(\mathbf{b}_2)]\rho^e(\mathbf{b}_2)\\
\end{split}
\end{equation}
Performing  transformation to the momentum space, we obtain 
\begin{equation}
\begin{split}
&{\cal M}^{i,a}_{(3), {\rm odd}}(\mathbf{k}) = 
\int d^2\mathbf{z} e^{-i\mathbf{k}\cdot\mathbf{z}} \int d^2\mathbf{b}_1d^2\mathbf{b}_2  {\cal M}^{i,a}_{(3), {\rm odd}}(\mathbf{z}, \mathbf{b}_1, \mathbf{b}_2) \\ & = 
-2i \int_{\mathbf{l}, \mathbf{q}_1, \mathbf{q}_2}\Bigg(\frac{-\epsilon^{ij}\mathbf{k}^j}{k_{\perp}^2}\left(\frac{\mathbf{q}_1}{q_1^2} - \frac{\mathbf{l}}{l_{\perp}^2}\right)\cdot \left(\frac{\mathbf{q}_2}{q_2^2} - \frac{(\mathbf{k}-\mathbf{l})}{|\mathbf{k}-\mathbf{l}|^2}\right)+ \left(\frac{\mathbf{q}_1^i}{q_1^2} - \frac{\mathbf{l}^i}{l_{\perp}^2}\right)\frac{\mathbf{q}_2\times(\mathbf{k}-\mathbf{l})}{q_2^2|\mathbf{k}-\mathbf{l}|^2}\\
&\qquad + \left(\frac{\mathbf{q}_2^i}{q_2^2} - \frac{(\mathbf{k}-\mathbf{l})^i}{|\mathbf{k}-\mathbf{l}|^2}\right)\frac{\mathbf{q}_1\times\mathbf{l}}{q_1^2l_{\perp}^2}\Bigg)\mathrm{sgn}(\mathbf{k}\times \mathbf{l}) f^{abc}  U^{bd}(\mathbf{l}-\mathbf{q}_1)\rho^d(\mathbf{q}_1)U^{ce}(\mathbf{k}-\mathbf{l}-\mathbf{q}_2)\rho^e(\mathbf{q}_2)\\
\end{split}
\end{equation}
This expression can be rewritten in terms of functions $\mathcal{A}^i, b_{\eta}, b_{\perp}$
\begin{equation}
\begin{split}
{\cal M}^{i,a}_{(3), {\rm odd}}(\mathbf{k}) 
=&-2i \int_{\mathbf{l}} \mathrm{sgn}(\mathbf{k}\times \mathbf{l}) \Big(\frac{-\epsilon^{ij} \mathbf{k}^j}{k_{\perp}^2}i\Big[\mathcal{A}^{m}(\mathbf{l}),\mathcal{A}^{m}(\mathbf{k}-\mathbf{l})\Big]^a +\Big [\mathcal{A}^{i}(\mathbf{l}), b_{\perp}(\mathbf{k}-\mathbf{l})\Big]^a\frac{1}{|\mathbf{k}-\mathbf{l}|^2} \\
&\qquad -\Big[\mathcal{A}^{i}(\mathbf{k}-\mathbf{l}), b_{\perp}(\mathbf{l})\Big]^a \frac{1}{l_{\perp}^2}\Big)\\
\end{split}
\end{equation}

To compare with our result we need to project out the longitudinal and the transverse parts: 
\begin{equation}
\begin{split}
&\mathbf{k}^i {\cal M}_{(3), {\rm odd}}^{i,a} =2 \int_{\mathbf{l}}\mathrm{sgn}(\mathbf{k}\times \mathbf{l}) \frac{2\mathbf{k}\cdot(\mathbf{k} - \mathbf{l})}{|\mathbf{k}-\mathbf{l}|^2 l_{\perp}^2} [ b_{\perp}(\mathbf{l}), b_{\eta}(\mathbf{k}-\mathbf{l})]\\
\end{split}
\end{equation}
\begin{equation}
\begin{split}
&\epsilon^{n i}\mathbf{k}^n {\cal M}_{(3), {\rm odd}}^{i,a}  =2  \int_{\mathbf{l}}\mathrm{sgn}(\mathbf{k}\times \mathbf{l})\Big(-\frac{\mathbf{l}\cdot(\mathbf{k}-\mathbf{l})}{l^2_{\perp}|\mathbf{k}-\mathbf{l}|^2}[b_{\eta}(\mathbf{l}), b_{\eta}(\mathbf{k}-\mathbf{l})]-\frac{\mathbf{l}\cdot(\mathbf{k}-\mathbf{l}) -k_{\perp}^2}{l^2_{\perp}|\mathbf{k}-\mathbf{l}|^2} [b_{\perp}(\mathbf{l}), b_{\perp}(\mathbf{k}-\mathbf{l})] \Big)\\
\end{split}
\end{equation}
The odd part of 
$\mathfrak{S}_{\eta, \perp}(\mathbf{k})+\mathfrak{B}_{\eta, \perp}(\mathbf{k})$  is equal to the odd part of the bulk terms alone 
$\mathfrak{B}_{\eta, \perp}(\mathbf{k})$, as the surface terms is an even function of the momentum $\mathfrak{S}_{\eta, \perp}(\mathbf{k}) = \mathfrak{S}_{\eta, \perp}(-\mathbf{k})$. Starting from here and using Eqs.~\eqref{Eq:Bulk3},  it is simple to show that 
\begin{equation} 
\mathfrak{B}_{\eta}(\mathbf{k}) - \mathfrak{B}_{\eta}(-\mathbf{k}) =  \frac{1}{2\sqrt{\pi}k_{\perp}}  \mathbf{k}^i {\cal M}_{(3), {\rm odd}}^{i,a} 
\end{equation} 
and 
\begin{equation} 
\mathfrak{B}_{\perp}(\mathbf{k}) - \mathfrak{B}_{\perp}(-\mathbf{k}) =  \frac{1}{2\sqrt{\pi}k_{\perp}}  \epsilon^{n i}\mathbf{k}^n {\cal M}_{(3), {\rm odd}}^{i,a} \,. 
\end{equation} 
This proofs the equivalence between the odd contributions to the order-$g^{3}$ amplitude between our results and the CKW paper.  The even part of the order-$g^{3}$ amplitude is not gauge invariant and we cannot perform the direct comparison between our result and that of the CKW paper.

\newpage
\bibliography{spires}

\providecommand{\href}[2]{#2}\begingroup\raggedright\begin{thebibliography}{10}

\bibitem{Balitsky:2004rr}
I.~Balitsky, \emph{{Scattering of shock waves in QCD}},
  \href{https://doi.org/10.1103/PhysRevD.70.114030}{\emph{Phys. Rev. D}
  {\bfseries 70} (2004) 114030}
  [\href{https://arxiv.org/abs/hep-ph/0409314}{{\ttfamily hep-ph/0409314}}].

\bibitem{Chirilli:2015tea}
G.A.~Chirilli, Y.V.~Kovchegov and D.E.~Wertepny, \emph{{Classical Gluon
  Production Amplitude for Nucleus-Nucleus Collisions: First Saturation
  Correction in the Projectile}},
  \href{https://doi.org/10.1007/JHEP03(2015)015}{\emph{JHEP} {\bfseries 03}
  (2015) 015} [\href{https://arxiv.org/abs/1501.03106}{{\ttfamily
  1501.03106}}].

\bibitem{Li:2021zmf}
M.~Li and V.V.~Skokov, \emph{{First Saturation Correction in High Energy
  Proton-Nucleus Collisions: I. Time evolution of classical Yang-Mills fields
  beyond leading order}},  \href{https://arxiv.org/abs/2102.01594}{{\ttfamily
  2102.01594}}.

\bibitem{McLerran:2016snu}
L.~McLerran and V.~Skokov, \emph{{Odd Azimuthal Anisotropy of the Glasma for pA
  Scattering}},
  \href{https://doi.org/10.1016/j.nuclphysa.2016.12.011}{\emph{Nucl. Phys. A}
  {\bfseries 959} (2017) 83}
  [\href{https://arxiv.org/abs/1611.09870}{{\ttfamily 1611.09870}}].

\bibitem{Kovchegov:2018jun}
Y.V.~Kovchegov and V.V.~Skokov, \emph{{How classical gluon fields generate odd
  azimuthal harmonics for the two-gluon correlation function in high-energy
  collisions}}, \href{https://doi.org/10.1103/PhysRevD.97.094021}{\emph{Phys.
  Rev. D} {\bfseries 97} (2018) 094021}
  [\href{https://arxiv.org/abs/1802.08166}{{\ttfamily 1802.08166}}].

\bibitem{Romatschke:2005pm}
P.~Romatschke and R.~Venugopalan, \emph{{Collective non-Abelian instabilities
  in a melting color glass condensate}},
  \href{https://doi.org/10.1103/PhysRevLett.96.062302}{\emph{Phys. Rev. Lett.}
  {\bfseries 96} (2006) 062302}
  [\href{https://arxiv.org/abs/hep-ph/0510121}{{\ttfamily hep-ph/0510121}}].

\bibitem{Gelis:2013rba}
T.~Epelbaum and F.~Gelis, \emph{{Pressure isotropization in high energy heavy
  ion collisions}},
  \href{https://doi.org/10.1103/PhysRevLett.111.232301}{\emph{Phys. Rev. Lett.}
  {\bfseries 111} (2013) 232301}
  [\href{https://arxiv.org/abs/1307.2214}{{\ttfamily 1307.2214}}].

\bibitem{Berges:2013eia}
J.~Berges, K.~Boguslavski, S.~Schlichting and R.~Venugopalan, \emph{{Turbulent
  thermalization process in heavy-ion collisions at ultrarelativistic
  energies}}, \href{https://doi.org/10.1103/PhysRevD.89.074011}{\emph{Phys.
  Rev. D} {\bfseries 89} (2014) 074011}
  [\href{https://arxiv.org/abs/1303.5650}{{\ttfamily 1303.5650}}].

\bibitem{Kovchegov:1997pc}
Y.V.~Kovchegov, \emph{{Quantum structure of the nonAbelian Weizsacker-Williams
  field for a very large nucleus}},
  \href{https://doi.org/10.1103/PhysRevD.55.5445}{\emph{Phys. Rev. D}
  {\bfseries 55} (1997) 5445}
  [\href{https://arxiv.org/abs/hep-ph/9701229}{{\ttfamily hep-ph/9701229}}].

\bibitem{Lappi:2007ku}
T.~Lappi, \emph{{Wilson line correlator in the MV model: Relating the glasma to
  deep inelastic scattering}},
  \href{https://doi.org/10.1140/epjc/s10052-008-0588-4}{\emph{Eur. Phys. J. C}
  {\bfseries 55} (2008) 285} [\href{https://arxiv.org/abs/0711.3039}{{\ttfamily
  0711.3039}}].

\bibitem{Kovchegov:1998bi}
Y.V.~Kovchegov and A.H.~Mueller, \emph{{Gluon production in current nucleus and
  nucleon - nucleus collisions in a quasiclassical approximation}},
  \href{https://doi.org/10.1016/S0550-3213(98)00384-8}{\emph{Nucl. Phys. B}
  {\bfseries 529} (1998) 451}
  [\href{https://arxiv.org/abs/hep-ph/9802440}{{\ttfamily hep-ph/9802440}}].

\bibitem{Dumitru:2001ux}
A.~Dumitru and L.D.~McLerran, \emph{{How protons shatter colored glass}},
  \href{https://doi.org/10.1016/S0375-9474(01)01301-X}{\emph{Nucl. Phys. A}
  {\bfseries 700} (2002) 492}
  [\href{https://arxiv.org/abs/hep-ph/0105268}{{\ttfamily hep-ph/0105268}}].

\bibitem{Chen:2015wia}
G.~Chen, R.J.~Fries, J.I.~Kapusta and Y.~Li, \emph{{Early Time Dynamics of
  Gluon Fields in High Energy Nuclear Collisions}},
  \href{https://doi.org/10.1103/PhysRevC.92.064912}{\emph{Phys. Rev. C}
  {\bfseries 92} (2015) 064912}
  [\href{https://arxiv.org/abs/1507.03524}{{\ttfamily 1507.03524}}].

\bibitem{Li:2016eqr}
M.~Li and J.I.~Kapusta, \emph{{Analytic calculation of the energy-momentum
  tensor in heavy ion collisions from color glass condensate}},
  \href{https://doi.org/10.1103/PhysRevC.94.024908}{\emph{Phys. Rev. C}
  {\bfseries 94} (2016) 024908}
  [\href{https://arxiv.org/abs/1602.09060}{{\ttfamily 1602.09060}}].

\bibitem{Krasnitz:1999wc}
A.~Krasnitz and R.~Venugopalan, \emph{{The Initial energy density of gluons
  produced in very high-energy nuclear collisions}},
  \href{https://doi.org/10.1103/PhysRevLett.84.4309}{\emph{Phys. Rev. Lett.}
  {\bfseries 84} (2000) 4309}
  [\href{https://arxiv.org/abs/hep-ph/9909203}{{\ttfamily hep-ph/9909203}}].

\bibitem{Krasnitz:2000gz}
A.~Krasnitz and R.~Venugopalan, \emph{{The Initial gluon multiplicity in heavy
  ion collisions}},
  \href{https://doi.org/10.1103/PhysRevLett.86.1717}{\emph{Phys. Rev. Lett.}
  {\bfseries 86} (2001) 1717}
  [\href{https://arxiv.org/abs/hep-ph/0007108}{{\ttfamily hep-ph/0007108}}].

\bibitem{Schenke:2015aqa}
B.~Schenke, S.~Schlichting and R.~Venugopalan, \emph{{Azimuthal anisotropies in
  p$+$Pb collisions from classical Yang\textendash{}Mills dynamics}},
  \href{https://doi.org/10.1016/j.physletb.2015.05.051}{\emph{Phys. Lett. B}
  {\bfseries 747} (2015) 76}
  [\href{https://arxiv.org/abs/1502.01331}{{\ttfamily 1502.01331}}].

\bibitem{Schlichting:2019bvy}
S.~Schlichting and V.~Skokov, \emph{{Saturation corrections to dilute-dense
  particle production and azimuthal correlations in the Color Glass
  Condensate}},
  \href{https://doi.org/10.1016/j.physletb.2020.135511}{\emph{Phys. Lett. B}
  {\bfseries 806} (2020) 135511}
  [\href{https://arxiv.org/abs/1910.12496}{{\ttfamily 1910.12496}}].

\bibitem{McLerran:1993ni}
L.D.~McLerran and R.~Venugopalan, \emph{{Computing quark and gluon distribution
  functions for very large nuclei}},
  \href{https://doi.org/10.1103/PhysRevD.49.2233}{\emph{Phys. Rev. D}
  {\bfseries 49} (1994) 2233}
  [\href{https://arxiv.org/abs/hep-ph/9309289}{{\ttfamily hep-ph/9309289}}].

\bibitem{McLerran:1993ka}
L.D.~McLerran and R.~Venugopalan, \emph{{Gluon distribution functions for very
  large nuclei at small transverse momentum}},
  \href{https://doi.org/10.1103/PhysRevD.49.3352}{\emph{Phys. Rev. D}
  {\bfseries 49} (1994) 3352}
  [\href{https://arxiv.org/abs/hep-ph/9311205}{{\ttfamily hep-ph/9311205}}].

\bibitem{Jeon:2005cf}
S.~Jeon and R.~Venugopalan, \emph{{A Classical Odderon in QCD at high
  energies}}, \href{https://doi.org/10.1103/PhysRevD.71.125003}{\emph{Phys.
  Rev. D} {\bfseries 71} (2005) 125003}
  [\href{https://arxiv.org/abs/hep-ph/0503219}{{\ttfamily hep-ph/0503219}}].

\bibitem{Dumitru:2020fdh}
A.~Dumitru, V.~Skokov and T.~Stebel, \emph{{Subfemtometer scale color charge
  correlations in the proton}},
  \href{https://doi.org/10.1103/PhysRevD.101.054004}{\emph{Phys. Rev. D}
  {\bfseries 101} (2020) 054004}
  [\href{https://arxiv.org/abs/2001.04516}{{\ttfamily 2001.04516}}].

\bibitem{Dumitru:2021tvw}
A.~Dumitru, H.~M\"antysaari and R.~Paatelainen, \emph{{Color charge
  correlations in the proton at NLO: beyond geometry based intuition}},
  \href{https://arxiv.org/abs/2103.11682}{{\ttfamily 2103.11682}}.

\bibitem{Gelis:2008rw}
F.~Gelis, T.~Lappi and R.~Venugopalan, \emph{{High energy factorization in
  nucleus-nucleus collisions}},
  \href{https://doi.org/10.1103/PhysRevD.78.054019}{\emph{Phys. Rev. D}
  {\bfseries 78} (2008) 054019}
  [\href{https://arxiv.org/abs/0804.2630}{{\ttfamily 0804.2630}}].

\bibitem{Gelis:2008ad}
F.~Gelis, T.~Lappi and R.~Venugopalan, \emph{{High energy factorization in
  nucleus-nucleus collisions. II. Multigluon correlations}},
  \href{https://doi.org/10.1103/PhysRevD.78.054020}{\emph{Phys. Rev. D}
  {\bfseries 78} (2008) 054020}
  [\href{https://arxiv.org/abs/0807.1306}{{\ttfamily 0807.1306}}].

\bibitem{Gelis:2008sz}
F.~Gelis, T.~Lappi and R.~Venugopalan, \emph{{High energy factorization in
  nucleus-nucleus collisions. 3. Long range rapidity correlations}},
  \href{https://doi.org/10.1103/PhysRevD.79.094017}{\emph{Phys. Rev. D}
  {\bfseries 79} (2009) 094017}
  [\href{https://arxiv.org/abs/0810.4829}{{\ttfamily 0810.4829}}].

\bibitem{Kovner:2010xk}
A.~Kovner and M.~Lublinsky, \emph{{Angular Correlations in Gluon Production at
  High Energy}}, \href{https://doi.org/10.1103/PhysRevD.83.034017}{\emph{Phys.
  Rev. D} {\bfseries 83} (2011) 034017}
  [\href{https://arxiv.org/abs/1012.3398}{{\ttfamily 1012.3398}}].

\bibitem{Kovner:2016jfp}
A.~Kovner, M.~Lublinsky and V.~Skokov, \emph{{Exploring correlations in the CGC
  wave function: odd azimuthal anisotropy}},
  \href{https://doi.org/10.1103/PhysRevD.96.016010}{\emph{Phys. Rev. D}
  {\bfseries 96} (2017) 016010}
  [\href{https://arxiv.org/abs/1612.07790}{{\ttfamily 1612.07790}}].

\bibitem{Kovchegov:1996ty}
Y.V.~Kovchegov, \emph{{NonAbelian Weizsacker-Williams field and a
  two-dimensional effective color charge density for a very large nucleus}},
  \href{https://doi.org/10.1103/PhysRevD.54.5463}{\emph{Phys. Rev. D}
  {\bfseries 54} (1996) 5463}
  [\href{https://arxiv.org/abs/hep-ph/9605446}{{\ttfamily hep-ph/9605446}}].

\bibitem{Kovner:1995ja}
A.~Kovner, L.D.~McLerran and H.~Weigert, \emph{{Gluon production from
  nonAbelian Weizsacker-Williams fields in nucleus-nucleus collisions}},
  \href{https://doi.org/10.1103/PhysRevD.52.6231}{\emph{Phys. Rev. D}
  {\bfseries 52} (1995) 6231}
  [\href{https://arxiv.org/abs/hep-ph/9502289}{{\ttfamily hep-ph/9502289}}].

\bibitem{Gyulassy:1997vt}
M.~Gyulassy and L.D.~McLerran, \emph{{Yang-Mills radiation in ultrarelativistic
  nuclear collisions}},
  \href{https://doi.org/10.1103/PhysRevC.56.2219}{\emph{Phys. Rev. C}
  {\bfseries 56} (1997) 2219}
  [\href{https://arxiv.org/abs/nucl-th/9704034}{{\ttfamily nucl-th/9704034}}].

\bibitem{Blaizot:2010kh}
J.P.~Blaizot, T.~Lappi and Y.~Mehtar-Tani, \emph{{On the gluon spectrum in the
  glasma}}, \href{https://doi.org/10.1016/j.nuclphysa.2010.06.009}{\emph{Nucl.
  Phys. A} {\bfseries 846} (2010) 63}
  [\href{https://arxiv.org/abs/1005.0955}{{\ttfamily 1005.0955}}].

\bibitem{Berges:2013fga}
J.~Berges, K.~Boguslavski, S.~Schlichting and R.~Venugopalan, \emph{{Universal
  attractor in a highly occupied non-Abelian plasma}},
  \href{https://doi.org/10.1103/PhysRevD.89.114007}{\emph{Phys. Rev. D}
  {\bfseries 89} (2014) 114007}
  [\href{https://arxiv.org/abs/1311.3005}{{\ttfamily 1311.3005}}].

\bibitem{Itzykson:1980rh}
C.~Itzykson and J.B.~Zuber, \emph{{Quantum Field Theory}}, International Series
  In Pure and Applied Physics, McGraw-Hill, New York (1980).

\bibitem{abramowitz+stegun}
M.~Abramowitz and I.A.~Stegun, \emph{Handbook of Mathematical Functions with
  Formulas, Graphs, and Mathematical Tables}, Dover, New York (1964).

\bibitem{Blaizot:2004wu}
J.P.~Blaizot, F.~Gelis and R.~Venugopalan, \emph{{High-energy pA collisions in
  the color glass condensate approach. 1. Gluon production and the Cronin
  effect}}, \href{https://doi.org/10.1016/j.nuclphysa.2004.07.005}{\emph{Nucl.
  Phys. A} {\bfseries 743} (2004) 13}
  [\href{https://arxiv.org/abs/hep-ph/0402256}{{\ttfamily hep-ph/0402256}}].

\bibitem{Blaizot:2008yb}
J.-P.~Blaizot and Y.~Mehtar-Tani, \emph{{The Classical field created in early
  stages of high energy nucleus-nucleus collisions}},
  \href{https://doi.org/10.1016/j.nuclphysa.2008.11.010}{\emph{Nucl. Phys. A}
  {\bfseries 818} (2009) 97} [\href{https://arxiv.org/abs/0806.1422}{{\ttfamily
  0806.1422}}].

\bibitem{Fadin:1975cb}
V.S.~Fadin, E.A.~Kuraev and L.N.~Lipatov, \emph{{On the Pomeranchuk Singularity
  in Asymptotically Free Theories}},
  \href{https://doi.org/10.1016/0370-2693(75)90524-9}{\emph{Phys. Lett. B}
  {\bfseries 60} (1975) 50}.

\bibitem{Kuraev:1977fs}
E.A.~Kuraev, L.N.~Lipatov and V.S.~Fadin, \emph{{The Pomeranchuk Singularity in
  Nonabelian Gauge Theories}}, {\emph{Sov. Phys. JETP} {\bfseries 45} (1977)
  199}.

\bibitem{Kovchegov:2012mbw}
Y.V.~Kovchegov and E.~Levin, \emph{{Quantum chromodynamics at high energy}},
  vol.~33, Cambridge University Press (8, 2012),
  \href{https://doi.org/10.1017/CBO9781139022187}{10.1017/CBO9781139022187}.

\bibitem{Blaizot:2004wv}
J.P.~Blaizot, F.~Gelis and R.~Venugopalan, \emph{{High-energy pA collisions in
  the color glass condensate approach. 2. Quark production}},
  \href{https://doi.org/10.1016/j.nuclphysa.2004.07.006}{\emph{Nucl. Phys. A}
  {\bfseries 743} (2004) 57}
  [\href{https://arxiv.org/abs/hep-ph/0402257}{{\ttfamily hep-ph/0402257}}].

\bibitem{Dominguez:2012ad}
F.~Dominguez, C.~Marquet, A.M.~Stasto and B.-W.~Xiao, \emph{{Universality of
  multiparticle production in QCD at high energies}},
  \href{https://doi.org/10.1103/PhysRevD.87.034007}{\emph{Phys. Rev. D}
  {\bfseries 87} (2013) 034007}
  [\href{https://arxiv.org/abs/1210.1141}{{\ttfamily 1210.1141}}].

\bibitem{Shi:2017gcq}
Y.~Shi, C.~Zhang and E.~Wang, \emph{{Multipole scattering amplitudes in the
  Color Glass Condensate formalism}},
  \href{https://doi.org/10.1103/PhysRevD.95.116014}{\emph{Phys. Rev. D}
  {\bfseries 95} (2017) 116014}
  [\href{https://arxiv.org/abs/1704.00266}{{\ttfamily 1704.00266}}].

\bibitem{Lappi:2015vta}
T.~Lappi, B.~Schenke, S.~Schlichting and R.~Venugopalan, \emph{{Tracing the
  origin of azimuthal gluon correlations in the color glass condensate}},
  \href{https://doi.org/10.1007/JHEP01(2016)061}{\emph{JHEP} {\bfseries 01}
  (2016) 061} [\href{https://arxiv.org/abs/1509.03499}{{\ttfamily
  1509.03499}}].

\bibitem{Kovner:2017ssr}
A.~Kovner and A.H.~Rezaeian, \emph{{Double parton scattering in the CGC: Double
  quark production and effects of quantum statistics}},
  \href{https://doi.org/10.1103/PhysRevD.96.074018}{\emph{Phys. Rev. D}
  {\bfseries 96} (2017) 074018}
  [\href{https://arxiv.org/abs/1707.06985}{{\ttfamily 1707.06985}}].

\bibitem{Altinoluk:2018ogz}
T.~Altinoluk, N.~Armesto, A.~Kovner and M.~Lublinsky, \emph{{Double and triple
  inclusive gluon production at mid rapidity: quantum interference in p-A
  scattering}},
  \href{https://doi.org/10.1140/epjc/s10052-018-6186-1}{\emph{Eur. Phys. J. C}
  {\bfseries 78} (2018) 702}
  [\href{https://arxiv.org/abs/1805.07739}{{\ttfamily 1805.07739}}].

\bibitem{Mace:2018vwq}
M.~Mace, V.V.~Skokov, P.~Tribedy and R.~Venugopalan, \emph{{Hierarchy of
  Azimuthal Anisotropy Harmonics in Collisions of Small Systems from the Color
  Glass Condensate}},
  \href{https://doi.org/10.1103/PhysRevLett.121.052301}{\emph{Phys. Rev. Lett.}
  {\bfseries 121} (2018) 052301}
  [\href{https://arxiv.org/abs/1805.09342}{{\ttfamily 1805.09342}}].

\bibitem{Mace:2018yvl}
M.~Mace, V.V.~Skokov, P.~Tribedy and R.~Venugopalan, \emph{{Systematics of
  azimuthal anisotropy harmonics in proton\textendash{}nucleus collisions at
  the LHC from the Color Glass Condensate}},
  \href{https://doi.org/10.1016/j.physletb.2018.09.064}{\emph{Phys. Lett. B}
  {\bfseries 788} (2019) 161}
  [\href{https://arxiv.org/abs/1807.00825}{{\ttfamily 1807.00825}}].

\bibitem{Prudnikov1986}
A.P.~{Prudnikov}, Y.A.~{Brychkov} and O.I.~{Marichev}, \emph{{Integrals and
  series. Vol. 1: Elementary functions. Vol. 2: Special functions. Transl. from
  the Russian by N. M. Queen}}, New York etc.: Gordon \&| Breach Science
  Publishers (1986).

\bibitem{Gervois1984}
A.~Gervois and H.~Navelet, \emph{{Some integrals involving three modified
  Bessel functions. 1.}}, \href{https://doi.org/10.1063/1.527169}{\emph{J.
  Math. Phys.} {\bfseries 27} (1986) 682}.

\bibitem{Erdelyi1953}
H.~Bateman and A.~Erd{\'e}lyi, \emph{Higher Transcendental Functions},
  McGraw-Hill (1953).

\end{thebibliography}\endgroup

\end{document}